\documentclass[12pt]{article}

\usepackage{axodraw}

\makeatletter

\def\paragraph{\@startsection{paragraph}{4}{\z@}{+2.00ex plus
 +1ex minus +.2ex}{1.5ex plus .2ex}{\it\normalsize}}

\def\section{\@startsection {section}{1}{\z@}{+3.0ex plus +1ex minus
  +.2ex}{2.3ex plus .2ex}{\normalsize\bf\boldmath}}
\def\subsection{\@startsection{subsection}{2}{\z@}{+2.5ex plus +1ex
minus +.2ex}{1.5ex plus .2ex}{\normalsize\bf\boldmath}}
\def\subsubsection{\@startsection{subsubsection}{3}{\z@}{+3.25ex plus
 +1ex minus +.2ex}{1.5ex plus .2ex}{\normalsize\it}}

\expandafter\ifx\csname mathrm\endcsname\relax\def\mathrm#1{{\rm #1}}\fi

\newcounter{saveeqn}

\@addtoreset{equation}{section}

\newcount\@tempcntc
\def\@citex[#1]#2{\if@filesw\immediate\write\@auxout{\string\citation{#2}}\fi
  \@tempcnta\z@\@tempcntb\m@ne\def\@citea{}\@cite{\@for\@citeb:=#2\do
    {\@ifundefined
       {b@\@citeb}{\@citeo\@tempcntb\m@ne\@citea
        \def\@citea{,\penalty\@m\ }{\bf ?}\@warning
       {Citation `\@citeb' on page \thepage \space undefined}}%
    {\setbox\z@\hbox{\global\@tempcntc0\csname
b@\@citeb\endcsname\relax}%
     \ifnum\@tempcntc=\z@ \@citeo\@tempcntb\m@ne
       \@citea\def\@citea{,\penalty\@m}
       \hbox{\csname b@\@citeb\endcsname}%
     \else
      \advance\@tempcntb\@ne
      \ifnum\@tempcntb=\@tempcntc
      \else\advance\@tempcntb\m@ne\@citeo
      \@tempcnta\@tempcntc\@tempcntb\@tempcntc\fi\fi}}\@citeo}{#1}}

\def\@citeo{\ifnum\@tempcnta>\@tempcntb\else\@citea
  \def\@citea{,\penalty\@m}%
  \ifnum\@tempcnta=\@tempcntb\the\@tempcnta\else
   {\advance\@tempcnta\@ne\ifnum\@tempcnta=\@tempcntb \else
\def\@citea{--}\fi
    \advance\@tempcnta\m@ne\the\@tempcnta\@citea\the\@tempcntb}\fi\fi}

\def\nl{\nonumber\\}
\def\nls{\nonumber\\[1ex]}

\def\nln{\nl*[-1ex]}

\def\asymp#1%
{\mathrel{\raisebox{-.4em}{$\widetilde{\scriptstyle #1}$}}}

\def\Nequal#1%
{\mathrel{\raisebox{-.5em}{$\stackrel{=}{\scriptstyle\rm#1}$}}}
\newcommand{\dsl}[1]{\not \hspace{-0.7mm}#1}
\def\dsl{\mathpalette\make@slash}
\def\make@slash#1#2{\setbox\z@\hbox{$#1#2$}%
  \hbox to 0pt{\hss$#1/$\hss\kern-\wd0}\box0}

\def\beq{\begin{equation}}
\def\eeq{\end{equation}}
\def\beqar{\begin{eqnarray}}
\def\eeqar{\end{eqnarray}}
\def\barr#1{\begin{array}{#1}}
\def\earr{\end{array}}
\def\bfi{\begin{figure}}
\def\efi{\end{figure}}
\def\btab{\begin{table}}
\def\etab{\end{table}}
\def\bce{\begin{center}}
\def\ece{\end{center}}
\def\nn{\nonumber}
\def\disp{\displaystyle}
\def\text{\textstyle}

\def\al{\alpha}
\def\be{\beta}

\def\de{\delta}
\def\De{\Delta}
\def\eps{\epsilon}

\def\La{\Lambda}
\def\la{\lambda}

\def\si{\sigma}

\def\refeq#1{\mbox{(\ref{#1})}}
\def\refeqs#1{\mbox{(\ref{#1})}}

\def\reffi#1{\mbox{Figure~\ref{#1}}}

\def\refse#1{\mbox{Section~\ref{#1}}}
\def\refses#1{\mbox{Sections~\ref{#1}}}
\def\refapp#1{\mbox{App.~\ref{#1}}}
\def\refapps#1{\mbox{Apps.~\ref{#1}}}
\def\citere#1{\mbox{Ref.~\cite{#1}}}
\def\citeres#1{\mbox{Refs.~\cite{#1}}}

\newcommand{\ri}{{\mathrm{i}}}
\newcommand{\rd}{{\mathrm{d}}}

\newcommand{\rT}{{\mathrm{T}}}

\newcommand{\ord}{\mathswitch{{\cal{O}}}}

\def\mathswitchr#1{\relax\ifmmode{\mathrm{#1}}\else$\mathrm{#1}$\fi}

\newcommand{\PW}{\mathswitchr W}

\newcommand{\PH}{\mathswitchr H}

\newcommand{\Pd}{\mathswitchr d}

\newcommand{\Pu}{\mathswitchr u}

\newcommand{\Pep}{\mathswitchr {e^+}}
\newcommand{\Pem}{\mathswitchr {e^-}}

\def\mathswitch#1{\relax\ifmmode#1\else$#1$\fi}

\newcommand{\MW}{\mathswitch {M_\PW}}

\newcommand{\GW}{\Gamma_{\PW}}

\def\ie{i.e.,\ }
\def\eg{e.g.\ }

\def\Li{\mathop{\mathrm{Li}_2}\nolimits}

\def\sgn{\mathop{\mathrm{sgn}}\nolimits}

\newcommand{\textfrac}[2]{{\textstyle\frac{#1}{#2}}}
\newcommand{\D}{{\cal{D}}}

\newcommand{\Ctilde}{{\tilde{C}}}
\newcommand{\Dtilde}{{\tilde{D}}}
\newcommand{\Etilde}{{\tilde{E}}}
\newcommand{\TNtilde}{{\tilde{T}^N}}
\newcommand{\Ntilde}{{\tilde{N}}}

\newcommand{\debar}{\bar\delta}
\newcommand{\detY}[1]{\eta_{#1}}

\newcommand{\Ecomb}{E}
\newcommand{\Dcomb}{D}
\newcommand{\Ccomb}{C}

\newcommand{\Shat}{\hat S}
\newcommand{\Zadj}{\tilde Z}
\newcommand{\Zadjadj}{\,\smash{\tilde{\!\tilde Z}}\vphantom{\tilde Z}}
\newcommand{\Ymod}{X}
\newcommand{\Ymodadj}{\tilde{\Ymod}}
\newcommand{\Ymodadjadj}{\,\smash{\tilde{\!\Ymodadj}}\vphantom{\Ymodadj}}

\newcommand{\sst}{\scriptstyle}

\newcommand{\ina}{i_1}
\newcommand{\inb}{i_2}
\newcommand{\inc}{i_3}
\newcommand{\ind}{i_4}
\newcommand{\ine}{i_5}
\newcommand{\ing}{i_6}
\newcommand{\sina}{n}

\newcommand{\sinN}{\sina}

\newcommand{\np}{n_p}
\newcommand{\Ebar}{\bar{E}}
\newcommand{\Fbar}{\bar{F}}

\newcommand{\indexunderbrace}[1]{\underbrace{\sst #1}}

\def\draftdate{\relax}
\def\mda{\relax}
\def\mua{\relax}
\def\mla{\relax}
\def\draft{
\def\thtystars{******************************}
\def\sixtystars{\thtystars\thtystars}
\typeout{}
\typeout{\sixtystars**}
\typeout{* Draft mode!
         For final version remove \protect\draft\space in source file *}
\typeout{\sixtystars**}
\typeout{}
\def\draftdate{\today}
\def\mua{\marginpar[\boldmath\hfil$\uparrow$]%
                   {\boldmath$\uparrow$\hfil}%
                    \typeout{marginpar: $\uparrow$}\ignorespaces}
\def\mda{\marginpar[\boldmath\hfil$\downarrow$]%
                   {\boldmath$\downarrow$\hfil}%
                    \typeout{marginpar: $\downarrow$}\ignorespaces}
\def\mla{\marginpar[\boldmath\hfil$\rightarrow$]%
                   {\boldmath$\leftarrow $\hfil}%
                    \typeout{marginpar: $\leftrightarrow$}\ignorespaces}
\def\Mua{\marginpar[\boldmath\hfil$\Uparrow$]%
                   {\boldmath$\Uparrow$\hfil}%
                    \typeout{marginpar: $\uparrow$}\ignorespaces}
\def\Mda{\marginpar[\boldmath\hfil$\Downarrow$]%
                   {\boldmath$\Downarrow$\hfil}%
                    \typeout{marginpar: $\downarrow$}\ignorespaces}
\def\Mla{\marginpar[\boldmath\hfil$\Rightarrow$]%
                   {\boldmath$\Leftarrow $\hfil}%
                    \typeout{marginpar: $\leftrightarrow$}\ignorespaces}
\overfullrule 5pt
\oddsidemargin -15mm
\marginparwidth 29mm
}

\def\stars{\strut\leaders\hbox{*}\hfill\strut}
\def\starline{\hfil\strut\hfil\hbox to \textwidth {\stars}\hfil}

\begin{document}
\thispagestyle{empty}
\def\thefootnote{\fnsymbol{footnote}}
\setcounter{footnote}{1}
\null
\draftdate
\strut\hfill MPP-2005-84\\
\strut\hfill PSI-PR-05-08\\
\strut\hfill hep-ph/0509141
\vfill
\begin{center}
{\large \bf\boldmath
Reduction schemes for one-loop tensor integrals 
\par} \vskip 1em
\vspace{1cm}
{\large
{\sc A.\ Denner$^1$ and S.\ Dittmaier$^2$} }
\\[.8cm]
$^1$ {\it Paul Scherrer Institut, W\"urenlingen und Villigen\\
CH-5232 Villigen PSI, Switzerland} \\[0.5cm]
$^2$ {\it Max-Planck-Institut f\"ur Physik 
(Werner-Heisenberg-Institut) \\
D-80805 M\"unchen, Germany} 
\\[0.5cm]
\par 
\end{center}\par
\vfill \vskip 2.0cm {\bf Abstract:} \par 
We present new methods for the evaluation of one-loop tensor integrals
which have been used in the calculation of the complete electroweak
one-loop corrections to $\Pep\Pem\to4\,$fermions. The described
methods for 3-point and 4-point integrals are, in particular,
applicable in the case where the conventional Passarino--Veltman
reduction breaks down owing to the appearance of Gram determinants in
the denominator.  One method consists of different variants for
expanding tensor coefficients about limits of vanishing Gram
determinants or other kinematical determinants, thereby reducing all
tensor coefficients to the usual scalar integrals. In a second method
a specific tensor coefficient with a logarithmic integrand is
evaluated numerically, and the remaining coefficients as well as the
standard scalar integral are algebraically derived from this
coefficient.  For 5-point tensor integrals, we give explicit formulas
that reduce the corresponding tensor coefficients to coefficients of
4-point integrals with tensor rank reduced by one.  Similar formulas
are provided for 6-point functions, and the generalization to
functions with more internal propagators is straightforward.  All the
presented methods are also applicable if infrared (soft or collinear)
divergences are treated in dimensional regularization or if mass
parameters (for unstable particles) become complex.
\par
\vskip 1cm
\noindent
September 2005  
\null
\setcounter{page}{0}
\clearpage
\def\thefootnote{\arabic{footnote}}
\setcounter{footnote}{0}

\section{Introduction}
\label{se:intro}

Future high-energy colliders, such as the LHC and the ILC, will allow
us to search for new physics and to test the Standard Model of the
electroweak and strong interaction with high precision.  Various
interesting processes naturally involve many particles in the final
state, where ``many'' means three, four, or more particles. Such
processes often proceed via one or more resonances that subsequently
decay, or they represent an irreducible background to such resonance
processes.  In order to exhaust the potential of future colliders,
precise theoretical predictions including strong and electroweak
corrections to many-particle processes are mandatory.

The calculation of radiative corrections to complicated processes
poses a number of problems. Besides the huge amount of algebra, the
appearance of unstable particles, and the integration of the
multi-dimensional phase space, a numerically stable evaluation of the
loop integrals is an important ingredient. In this paper we are
concerned with the calculation of one-loop integrals, including those
with five and six external legs. The generalization from six to more 
external legs is straightforward.

Pioneering work in the calculation of one-loop integrals was performed
by Veltman and collaborators. Together with `t~Hooft, he provided
compact explicit expressions for the basic one-loop integrals, the
scalar 1-point, 2-point, 3-point, and 4-point integrals
\cite{'tHooft:1978xw}, which have been completed later by other
authors \cite{Beenakker:1990jr}.  Elaborating on an idea of Brown and
Feynman \cite{Brown:1952eu}, together with Passarino he provided
systematic formulas that allow to reduce all tensor integrals with up
to four internal propagators to the basic scalar integrals
\cite{Passarino:1978jh}.  These methods are basically sufficient for
the calculation of radiative corrections to processes with four
external particles for non-exceptional configurations. Nevertheless,
in the sequel some improvements and additions have been worked out.
Van Oldenborgh and Vermaseren constructed a different tensor basis
that allows to concentrate some of the numerical instabilities into a
number of determinants \cite{vanOldenborgh:1989wn}. Ezawa et.\ al
performed the reduction using an orthonormal tensor basis
\cite{Ezawa:1990dh}.  A reduction in Feynman-parameter space, which is
equivalent to the Passarino--Veltman scheme, is used in the GRACE
package \cite{Belanger:2003sd}.

The main drawback of the Passarino--Veltman reduction and variants
thereof is the appearance of Gram determinants in the denominator,
which spoil the numerical stability if they become small. In processes
with up to four external particles this happens usually only near the
edge of phase space, \eg for forward scattering or on thresholds. For
the special cases where a Gram determinant is identically zero,
alternative reduction procedures have been devised by Stuart and
collaborators \cite{Stuart:1987tt,Devaraj:1997es} (see also
\citere{Boudjema:2005hb}). However, in
processes with more than four external particles, Gram determinants
also vanish within phase space, and methods for the calculation of
tensor integrals are needed where Gram determinants are small but not
exactly zero.  In \citere{Campbell:1996zw} such a method has been
devised by constructing combinations of $N$-point and $(N-1)$-point
scalar integrals that are finite in the limit of vanishing Gram
determinants and using this limit if the Gram determinant becomes
small.

On the other hand, alternative tensor reduction schemes have been
developed using different sets of master integrals. Davydychev could
relate the coefficients of one-loop tensor integrals to scalar
integrals in a different number of space-time dimensions
\cite{Davydychev:1991va}, and Tarasov found recursion relations
between these integrals \cite{Tarasov:1996br}. These methods have been
further elaborated by different groups
\cite{Bern:1992em,Binoth:1999sp,Duplancic:2003tv,Giele:2004iy,Giele:2004ub}.
In this approach all one-loop tensor integrals can be reduced to
finite 4-point integrals in $(D+2)$ dimensions and divergent 3-point
and 2-point integrals in $D$ dimensions.  Numerical instabilities in
this reduction, which are also due to small Gram or other kinematical
determinants, have been investigated in \citere{Giele:2004ub} for the
massless case, and a systematic improvement by an iteration technique
has been proposed.  While numerically stable analytic expression for
the basic integrals are available for the massless case, these turn
out to be hard to construct for the massive case. Therefore, one
typically reduces these basic integrals to the usual scalar integrals
or, in particular for vanishing Gram determinants, calculates them by
numerical integration \cite{Binoth:2005ff}.

Other algorithms, which are based on recursion relations similar to
Passarino--Veltman reduction and applicable irrespective of the number
of external legs, have been presented in
\citeres{delAguila:2004nf,vanHameren:2005ed}.  These algorithms do not
completely avoid the appearance of inverse Gram determinants.

It was realized already in the sixties by Melrose that scalar
integrals with more than four lines in the loop, \ie 5-point and
higher-point scalar integrals, can be reduced to scalar integrals with
less internal propagators in four dimensions \cite{Me65}. These
methods were subsequently extended and improved by several authors
\cite{vanOldenborgh:1989wn,Campbell:1996zw,Davydychev:1991va,%
  Bern:1992em,Binoth:1999sp,Duplancic:2003tv,vanNeerven:1983vr,%
  Denner:1993kt,Suzuki:2002js,Denner:2002ii,Belanger:2002ik} 
and generalized to dimensional regularization in
\citeres{Bern:1992em,Beenakker:2002nc,Dittmaier:2003bc}.  In
\citere{Denner:2002ii}, a method for the reduction of 5-point
integrals that completely avoids inverse leading Gram determinants has
been worked out. Recently, a similar reduction has been found that
even reduces 5-point tensor integrals to 4-point integrals with rank
reduced by one \cite{Binoth:2005ff}.  In all these approaches 5- and
higher-point tensor integrals are reduced to 
tensor integrals with less internal propagators.

Various approaches have been proposed that use numerical integration
of loop integrals and are, thus, complementary to most of the methods
mentioned so far.  In the approach of \citere{Ferroglia:2002mz}, which
has been elaborated for general one-loop integrals with up to six
external legs, the Feynman-parameter integrals are rewritten in such a
way that they can be numerically integrated in a stable way.  A fully
numerical approach to calculate loop integrals by contour integration
was proposed in \citere{Kurihara:2005ja}. A semi-numerical approach
that relies on the subtraction of UV and infrared divergences has been
advocated in \citere{Nagy:2003qn}. A different semi-numerical method
makes use of the fact that all tensor one-loop integrals can be
expressed in terms of one- and two-dimensional parameter integrals
which are suitable for numerical integration \cite{Binoth:2002xh}.
A numerical method based on multi-dimensional contour deformation has
been proposed in \citere{Binoth:2005ff}.
Finally, Feynman-parameter integrals have been numerically performed
with a small but finite ``$\ri\eps$'' from the propagator denominators
and a subsequent extrapolation $\eps\to0$ in \citere{deDoncker:2004bf}.
So far, none of these methods
has proven their performance in calculations of higher-order
corrections for processes with more than four external particles.  In
practice, one can still expect problems with the numerical stability
of the algebraic reduction to standard forms in specific regions of
phase space and with the speed of the underlying numerical integration
of the basic loop integrals.

In this paper we describe methods that have actually been used in the
calculation of the electroweak corrections to $\Pep\Pem\to4\,$fermions
\cite{Denner:2005es}, \ie in the first established calculation of the
complete one-loop electroweak corrections to a process with six
external particles.%
\footnote{The GRACE-loop collaboration has recently reported on
  progress towards one-loop calculations for $2\to4$ particle
  processes.  Using the methods described in
  \citeres{Belanger:2003sd,Belanger:2002ik}, first results on
  $\Pep\Pem\to\nu\bar\nu\PH\PH$ have been shown at conferences
  \cite{gracetalks}, and a status report on
  $\Pep\Pem\to\mu^-\bar\nu_\mu\Pu\bar\Pd$ has been given in
  \citere{Boudjema:2004id}.}  In this approach, 6-point integrals are
directly expressed in terms of six 5-point functions, and the 5-point
integrals are written in terms of five 4-point functions. While we
used the methods described in \citeres{Me65,Denner:1993kt} and
\citere{Denner:2002ii} in the original calculation
\cite{Denner:2005es}, in this paper we describe improved methods for
the reduction of 6-point and 5-point integrals which have meanwhile
been implemented in the code for the electroweak corrections to
$\Pep\Pep\to4$ fermions and which further improve its performance in
numerical stability and CPU time.  The 3-point and 4-point tensor
integrals are algebraically reduced to the (standard) scalar 1-point,
2-point, 3-point, and 4-point functions as described below.  For
1-point and 2-point integrals explicit numerically stable results are
used.

In more detail, the 3-point and 4-point functions are reduced to
scalar integrals according to the Passarino--Veltman algorithm if no
small Gram determinants appear. This is the case for most points in
parameter space. If a small Gram determinant appears, the reduction of
4-point to 3-point or 3-point to 2-point functions is done
differently.  Here we have worked out two alternative calculational
methods (referred to as ``rescue systems'' in \citere{Denner:2005es}).
One method makes use of suitable expansions of the tensor coefficients
about the limit of vanishing Gram determinants. This is achieved in an
iterative way and requires to calculate $(N-1)$-point functions of
higher degree compared to the usual Passarino--Veltman reduction.%
\footnote{A similar idea, where tensor coefficients are iteratively determined
from higher rank tensors has been described in \citere{Giele:2004ub}.}
Finally, again all tensor coefficients can be expressed in terms of
the standard scalar 1-point, 2-point, 3-point, and 4-point functions.
In practice, we use the first two to three terms in the expansions and
we have to introduce different expansions for different regions of
parameter space. In the second, alternative method we evaluate a
specific tensor coefficient, the integrand of which is logarithmic in
Feynman parametrization, by numerical integration. Then the remaining
coefficients as well as the standard scalar integral are algebraically
derived from this coefficient. This reduction again involves no
inverse Gram determinants; instead inverse modified Cayley
determinants appear. In this approach, the set of master integrals is
not given by the standard scalar integrals anymore.  For some specific
3-point integrals, where the modified Cayley determinant vanishes
exactly, analytical results have been worked out that allow for a
stable numerical evaluation.

The paper is organized as follows.  We summarize our conventions and
useful definitions in \refse{se:conventions}. The evaluation of
1-point and 2-point tensor integrals is summarized in \refses{se:1pf}
and \ref{se:2pf}, respectively. In \refse{se:4pf}, we provide several
methods for the reduction of 3-point and 4-point tensor coefficients
and describe their actual application to $\Pep\Pem\to4f$ in
\refse{se:ee4f}.  In \refse{se:divs} we consider UV and infrared
divergences in detail and conclude that the proposed methods are valid
independent of method for infrared regularization.  The reduction of
5-point and 6-point tensor integrals to integrals with smaller rank
and smaller number of propagators is detailed in \refses{se:5pf} and
\ref{se:6pf}, respectively. In \refapp{se:tendiv}, we list the
UV-divergent parts of one-loop integrals that enter the reduction
formulas. Appendix~\ref{app:Csing} describes a treatment of singular
3-point integrals based on analytical methods.  Finally, we discuss
alternative reductions of 5- and 6-point tensor integrals in
\refapps{se:5pf-alt} and \ref{se:6pf-alt}, respectively.

\section{Conventions and notation}
\label{se:conventions}

One-loop tensor $N$-point integrals have the general form
\beq
\label{tensorint}
T^{N,\mu_{1}\ldots\mu_{P}}(p_{1},\ldots,p_{N-1},m_{0},\ldots,m_{N-1})=
\displaystyle{\frac{(2\pi\mu)^{4-D}}{\ri\pi^{2}}\int \rd^{D}q\,
\frac{q^{\mu_{1}}\cdots q^{\mu_{P}}}
{N_0N_1\ldots N_{N-1}}}
\eeq
with the denominator factors
\beq \label{D0Di}
N_{k}= (q+p_{k})^{2}-m_{k}^{2}+\ri\epsilon, \qquad k=0,\ldots,N-1 ,
\qquad p_0=0,
\eeq
where $\ri\epsilon$ $(\eps>0)$ is an infinitesimally small imaginary
part.  For $P=0$, \ie no integration momenta in the numerator of the
loop integral, \refeq{tensorint} defines the scalar $N$-point integral
$T^N_0$.  Following the notation of \citere{'tHooft:1978xw} we set
$T^{1}= A,$ $T^{2}= B,$ $T^{3}= C,$ $T^{4}= D$, $T^{5}= E,$ $T^{6}=
F$.  Throughout we use the conventions of
\citeres{Denner:1993kt,Denner:2002ii} to decompose the tensor
integrals into Lorentz-covariant structures.

In order to be able to write down the tensor decompositions in a
concise way we use a notation (similar to the one of
\citere{Passarino:1978jh}) in which curly braces denote symmetrization
with respect to Lorentz indices in such a way that all non-equivalent
permutations of the Lorentz indices on metric tensors $g$ and a
generic momentum $p$ contribute with weight one and that in covariants
with $\np$ momenta $p_{i_j}^{\mu_j}$ $(j=1,\dots,\np$) only one
representative out of the $\np!$ permutations of the indices $i_j$ is
kept.  Thus, we have for example
\beqar
\{p\ldots p\}^{\mu_1\ldots\mu_P}_{i_1\ldots i_P} &=& 
p_{i_1}^{\mu_1}\ldots p_{i_P}^{\mu_P},
\nn\\[.3em]
\{g p\}_{i_1}^{\mu\nu\rho} &=& 
g^{\mu\nu}p_{i_1}^{\rho}+g^{\nu\rho}p_{i_1}^{\mu}+g^{\rho\mu}p_{i_1}^{\nu},
\nn\\[.3em]
\{g pp \}^{\mu\nu\rho\si}_{i_1 i_2} &=& 
 g^{\mu\nu}p_{i_1}^{\rho}p_{i_2}^{\si}
+g^{\mu\rho}p_{i_1}^{\si}p_{i_2}^{\nu}
+g^{\mu\si}p_{i_1}^{\nu}p_{i_2}^{\rho}
+g^{\nu\rho}p_{i_1}^{\si}p_{i_2}^{\mu}
+g^{\rho\si}p_{i_1}^{\nu}p_{i_2}^{\mu}
+g^{\si\nu}p_{i_1}^{\rho}p_{i_2}^{\mu},
\nn\\[.3em]
\{gg\}^{\mu\nu\rho\si} &=& g^{\mu\nu}g^{\rho\si}+g^{\nu\rho}g^{\mu\si}
+g^{\rho\mu}g^{\nu\si}.
\label{eq:covbrace}
\eeqar
This definition is unique up to the selection of the representative
permutations of the momenta. For our calculation this does not matter,
since the covariants are always contracted with quantities that are
totally symmetric in the indices $i_j$. In fact in our calculation the
definition is equivalent to a normalization of the sum of the $\np!$
covariants with a factor $1/\np!$; in this case the third line of
\refeq{eq:covbrace} would contain 12 instead of 6 terms on the r.h.s.

We decompose the general tensor integral into Lorentz-covariant
structures as
\beqar
T^{N,\mu_1\ldots\mu_P} &=& 
\sum_{n=0}^{\left[\frac{P}{2}\right]} \,
\sum_{i_{2n+1},\ldots,i_P=1}^{N-1} \,
\{\underbrace{g \ldots g}_n  p\ldots p\}^{\mu_1\ldots\mu_P}_{i_{2n+1}\ldots i_P}
\, T^N_{\underbrace{\sst 0\ldots0}_{2n} i_{2n+1}\ldots i_{P}}
\nn\\
&=&\sum_{i_1,\ldots,i_P=1}^{N-1} 
p^{\mu_1}_{i_1}\ldots p^{\mu_P}_{i_P} T^N_{i_1\ldots i_P}
+ \sum_{i_3,\ldots,i_{P}=1}^{N-1} 
\{g p\ldots p\}^{\mu_1\ldots\mu_P}_{i_3\ldots i_P} T^N_{00i_3\ldots i_{P}}
\nl&&{}
+ \sum_{i_5,\ldots,i_{P}=1}^{N-1} 
\{g g p\ldots p\}^{\mu_1\ldots\mu_P}_{i_5 \ldots i_P} T^N_{0000i_5\ldots i_{P}}
+\ldots
\nl&&{}+
\left\{
\barr{cc}\disp
\sum_{i_{P}=1}^{N-1} 
\{g \ldots g p\}_{i_P}^{\mu_1\ldots\mu_P} 
T^N_{\underbrace{\sst0\ldots 0}_{P-1} i_{P}},
& \mbox{ for } P\ \mathrm{odd,} \\
\{g \ldots g \}^{\mu_1\ldots\mu_P} 
T^N_{\underbrace{\sst0\ldots 0}_{P}},
& \mbox{ for } P\ \mathrm{even,} 
\earr\right.
\eeqar
where $\left[{P}/{2}\right]$ is the largest natural number smaller 
or equal to ${P}/{2}$.  For each metric tensor in the Lorentz
covariant the corresponding coefficient carries an index pair ``00''
and for each momentum $p_{i_r}$ it carries the corresponding index
$i_r$.

For tensor integrals up to rank five the decompositions more explicitly read
\beqar
T^{N,\mu}&=&\sum_{\ina=1}^{N-1} p_{\ina}^{\mu}T^N_{\ina}, \qquad
T^{N,\mu\nu}=\sum_{\ina,\inb=1}^{N-1} p_{\ina}^{\mu}p_{\inb}^{\nu}T^N_{\ina\inb}
+g^{\mu\nu}T^N_{00},\nls
T^{N,\mu\nu\rho}&=&\sum_{\ina,\inb,\inc=1}^{N-1} p_{\ina}^{\mu}p_{\inb}^{\nu}p_{\inc}^{\rho}T^N_{\ina\inb\inc}
+\sum_{\ina=1}^{N-1}\{g p\}_{\ina}^{\mu\nu\rho} T^N_{00\ina}, 
\nl
T^{N,\mu\nu\rho\si} &=& 
\sum_{\ina,\inb,\inc,\ind=1}^{N-1} p_{\ina}^{\mu}p_{\inb}^{\nu}p_{\inc}^{\rho}p_{\ind}^\si T^N_{\ina\inb\inc\ind}
+\sum_{\ina,\inb=1}^{N-1}
\{g pp\}_{\ina\inb}^{\mu\nu\rho\si}T^N_{00\ina\inb}
+\{g g\}^{\mu\nu\rho\si} T^N_{0000},\nl
T^{N,\mu\nu\rho\si\tau} &=& 
\sum_{\ina,\inb,\inc,\ind,\ine=1}^{N-1}
p_{\ina}^{\mu}p_{\inb}^{\nu}p_{\inc}^{\rho}p_{\ind}^\si p_{\ine}^\tau T^N_{\ina
\inb\inc\ind\ine}
+\sum_{\ina,\inb,\inc=1}^{N-1}
\{g ppp\}_{\ina\inb\inc}^{\mu\nu\rho\si\tau}
T^N_{00\ina\inb\inc}
\nl&&{}
+\sum_{\ina=1}^{N-1} 
\{g g p\}_{\ina}^{\mu\nu\rho\si\tau} 
T^N_{0000\ina}.
\eeqar
Because of the symmetry of the tensor $T^{N}_{\mu_{1}\ldots\mu_{P}}$
all coefficients $T^N_{i_1\ldots i_P}$ are symmetric under permutation
of all indices.  For convenience we assume this symmetry also for
indices that are zero.

When reducing a tensor integral $T^{N+1}_{\mu_1\ldots\mu_P}$, one
encounters tensor integrals that are obtained by omitting the $k$th
denominator $N_k$; we denote such integrals
$T^{N}_{\mu_1\ldots\mu_P}(k)$.  In the decomposition of
$T^{N}_{\mu_1\ldots\mu_P}(k)$, $k=1,\dots,N$, shifted indices appear
which we denote as
\beq\label{eq:defji}
 (i_r)_k=\left\{\barr{lll} i_r & \mathrm{\ for\ }&  k>i_r,\\
                     i_r-1 & \mathrm{\ for\ }& k < i_r.\earr \right.
\eeq
After cancelling the denominator $N_0$ the resulting tensor integrals
are not in the standard form  but can be expressed in terms of standard
integrals by shifting the integration momentum. We choose to perform
the shift $q\to q-p_1$, so that the following $N$-point integrals appear:
\beqar \label{Dshifted}
\tilde {T}^{N,\mu_1\ldots\mu_P}(0)
&=&
\frac{(2\pi\mu)^{(4-D)}}{\ri\pi^{2}}\int\!\rd^{D}q\,
\frac{q^{\mu_1}\cdots q^{\mu_P}}{\Ntilde_1 \cdots \Ntilde_N},
\nl
\Ntilde_{k}&=& (q+p_{k}-p_{1})^{2}-m_{k}^{2}+\ri\epsilon, \qquad k=1,\ldots,N .
\eeqar
Note that the scalar integral $T^N_{0}\equiv T^N$ and the tensor
coefficients $T^N_{00}$, $T^N_{0000}$, $\ldots$ are invariant under this
shift. The other coefficients of $T^{N}_{\mu_1\ldots\mu_P}(k)$ can be
recursively obtained as
\beqar\label{TNaux}
T^N_{\underbrace{\sst 0\ldots0}_{2n} i_{2n+1}\ldots i_{P}}(0) &=& 
\TNtilde_{\underbrace{\sst 0\ldots0}_{2n} i_{2n+1}-1,\ldots, i_P-1}(0), \quad
i_{2n+1},\ldots,i_P>1, \nl
T^N_{\underbrace{\sst 0\ldots0}_{2n} 1 i_{2n+2} \ldots i_P}(0) &=& 
-T^N_{\underbrace{\sst 0\ldots0}_{2n} i_{2n+2}\ldots i_P}(0)
-\sum_{r=2}^{N} T^N_{\underbrace{\sst 0\ldots0}_{2n} r i_{2n+2} \ldots
  i_P}(0), \quad i_{2n+2},\ldots,i_P>0.
\nln
\eeqar
The recursion is solved by
\beqar
T^N_{\underbrace{\sst 0\ldots0}_{2n} \underbrace{\sst 1\ldots1}_{k}
i_{2n+k+1} \ldots i_P}(0) 
&=& (-1)^k\sum_{l=0}^k \pmatrix{k \cr l} \sum_{i_1,\ldots,i_l=1}^{N-1}
\tilde T^N_{\underbrace{\sst 0\ldots0}_{2n} i_1\ldots i_l,
i_{2n+k+1}-1,\ldots,i_P-1}(0), 
\nn\\
&&
i_{2n+k+1},\ldots,i_P > 1.
\label{TNaux2}
\eeqar 

We also use the notation $\debar_{ij} = 1-\de_{ij}$, \ie $\sum_i
\debar_{ij}(\dots) = \sum_{i\ne j}(\dots)$, and employ the caret
``$\hat{\phantom{x}}$'' to indicate indices that are omitted, \ie
\beq
T^N_{i_1\ldots\hat i_r \ldots i_P} \equiv T^N_{i_1\ldots i_{r-1}  i_{r+1}
  \ldots i_P}.
\eeq

In the reduction formulas for the $(N+1)$-point functions the Gram matrix
\beq\label{matrixZ}
Z^{(N)} = \left(\barr{ccc}
2 p_1 p_1     & \ldots & 2 p_1 p_{N} \\
\vdots        & \ddots & \vdots      \\
2 p_{N} p_1 & \ldots & 2 p_{N}p_{N} 
\end{array} \right)
\eeq
appears.  Its determinant, the Gram determinant, is denoted by
\beq
\Delta^{(N)} = \det Z^{(N)},
\eeq
and its inverse can be written as
\beq\label{Zinv}
(Z^{(N)})^{-1}_{ij} = \frac{1}{\Delta^{(N)}} \Zadj^{(N)}_{ji},
\eeq
where $\Zadj^{(N)}_{ij}$ is the adjoint of $Z^{(N)}_{ij}$, which can
be calculated as
\beq\label{Zadj}
\Zadj^{(N)}_{ij} = (-1)^{i+j}
\left|\barr{cccccc}
2 p_1 p_1 &  \ldots & 2 p_1 p_{j-1} & 2 p_1 p_{j+1} &  \ldots & 2 p_1 p_N \\
\vdots    &  \ddots & \vdots   & \vdots    & \ddots  &  \vdots   \\
2p_{i-1}p_1 & \ldots & 2p_{i-1}p_{j-1} & 2p_{i-1}p_{j+1} & \ldots & 2p_{i-1}p_N \\
2p_{i+1}p_1 & \ldots & 2p_{i+1}p_{j-1} & 2p_{i+1}p_{j+1} & \ldots & 2p_{i+1}p_N \\
\vdots    &  \ddots  & \vdots    & \vdots   &  \ddots  &  \vdots   \\
2 p_N p_1 &  \ldots & 2 p_N p_{j-1} & 2 p_N p_{j+1} & \ldots & 2 p_N p_N 
\earr \right|,
\eeq
\ie from a reduced determinant of $Z^{(N)}$ where the $i$th row and
the $j$th column have been discarded.

We introduce a generalization of the adjoint by
\beqar\label{defZadjadj}
\lefteqn{\Zadjadj^{(N)}_{(ik)(jl)} = (-1)^{i+j+k+l}\sgn(i-k)\sgn(l-j)}\qquad\\*
&&\times\left|\barr{ccccccccc}
2 p_1 p_1 &  \ldots & 2 p_1 p_{j-1} & 2 p_1 p_{j+1} &  \ldots  & 
2 p_1 p_{l-1} & 2 p_1 p_{l+1} &  \ldots & 2 p_1 p_N \\
\vdots    &  \ddots  & \vdots  & \vdots   &  \ddots  &  \vdots  &  \vdots   & \ddots &  \vdots  \\
2p_{i-1}p_1 & \ldots & 2p_{i-1}p_{j-1} & 2p_{i-1}p_{j+1} & \ldots & 
2p_{i-1}p_{l-1} & 2p_{i-1}p_{l+1} & \ldots & 2p_{i-1}p_N \\
2p_{i+1}p_1 & \ldots & 2p_{i+1}p_{j-1} & 2p_{i+1}p_{j+1} & \ldots & 
2p_{i+1}p_{l-1} & 2p_{i+1}p_{l+1} & \ldots &  2p_{i+1}p_N \\
\vdots    &  \ddots  & \vdots  & \vdots   &  \ddots  &  \vdots  &  \vdots   & \ddots &  \vdots  \\
2p_{k-1}p_1 & \ldots & 2p_{k-1}p_{j-1} & 2p_{k-1}p_{j+1} & \ldots & 
2p_{k-1}p_{l-1} & 2p_{k-1}p_{l+1} & \ldots & 2p_{k-1}p_N \\
2p_{k+1}p_1 & \ldots & 2p_{k+1}p_{j-1} & 2p_{k+1}p_{j+1} & \ldots & 
2p_{k+1}p_{l-1} & 2p_{k+1}p_{l+1} & \ldots &  2p_{k+1}p_N \\
\vdots    &  \ddots  & \vdots  & \vdots   &  \ddots  &  \vdots  &  \vdots  &  \ddots &  \vdots   \\
2 p_N p_1 &  \ldots & 2 p_N p_{i-1} & 2 p_N p_{j+1} & \ldots 
& 2 p_N p_{l-1} & 2 p_N p_{l+1} & \ldots& 2 p_N p_N 
\earr \right|,\nn
\eeqar
\ie it is defined from a reduced determinant of $Z^{(N)}$ where the
$i$th and $k$th rows and the $j$th and $l$th columns have been
discarded.  Moreover, it is defined to vanish for $i=k$ or $j=l$. 
For the case $N=2$, it is given by
\beq
\Zadjadj^{(2)}_{(ik)(jl)} = \de_{il}\de_{kj}-\de_{ij}\de_{kl}.
\eeq

Expanding the determinant of $Z^{(N)}$ along the $k$th row or the
$l$th column, respectively, it can be written as
\beq\label{columnandrowexp}
\Delta^{(N)} 
=\sum_{m=1}^N Z^{(N)}_{km} \Zadj^{(N)}_{km}
=\sum_{m=1}^N Z^{(N)}_{ml} \Zadj^{(N)}_{ml},
\eeq
where $k$ and $l$ are not summed. 
Expanding the determinant in \refeq{Zadj} with the $l$th
column replaced by $(Z_{1k},\ldots,Z_{Nk})^{\rT}$ along the $l$th
  column yields the relation
\beq\label{Zadjadjrelation0}
\sum_{m=1}^N \Zadjadj^{(N)}_{(im)(jl)}  Z^{(N)}_{mk} =
 \Zadj^{(N)}_{il} \de_{jk} -   \Zadj^{(N)}_{ij} \de_{lk},
\eeq
and analogously
\beq\label{Zadjadjrelation0a}
\sum_{m=1}^N \Zadjadj^{(N)}_{(ik)(jm)}  Z^{(N)}_{lm} =
 \Zadj^{(N)}_{kj} \de_{il} -   \Zadj^{(N)}_{ij} \de_{kl}.
\eeq
These imply the equations
\beqar\label{Zadjadjrelation1}
 \Zadjadj^{(N)}_{(ik)(jl)} &=&
 (Z^{(N)})^{-1}_{jk} \Zadj^{(N)}_{il} -   (Z^{(N)})^{-1}_{lk}
 \Zadj^{(N)}_{ij}
= \left[ \Zadj^{(N)}_{il} \Zadj^{(N)}_{kj} -   \Zadj^{(N)}_{ij}
\Zadj^{(N)}_{kl} \right] / \Delta^{(N)}
\eeqar
and
\beqar\label{Zadjadjrelation2}
\sum_{m,n=1}^N \Zadjadj^{(N)}_{(im)(jn)}  Z^{(N)}_{mn} &=&
 \Zadj^{(N)}_{ij} (1-N),\nl
\sum_{m,n=1}^N \Zadjadj^{(N)}_{(im)(jn)}  Z^{(N)}_{mk}  Z^{(N)}_{ln} &=&
\Delta^{(N)} \de_{il} \de_{jk} -   \Zadj^{(N)}_{ij}  Z^{(N)}_{lk}.
\eeqar
An important special case of the last relation is
\beq
\Delta^{(N)} = Z^{(N)}_{lk} \Zadj^{(N)}_{lk}
+\sum_{m,n=1}^N \Zadjadj^{(N)}_{(lm)(kn)} Z^{(N)}_{mk}  Z^{(N)}_{ln}.
\label{Zadjadjrelation3}
\eeq
The relations \refeq{Zinv}--\refeq{Zadjadjrelation3} are valid for any
(not necessarily symmetric)
$N\times N$ matrix $Z^{(N)}$ with determinant $\De^{(N)}$.

We further introduce the $(N+1)\times(N+1)$ matrix
\beq\label{defYmod}
\Ymod^{(N)}=\left(\barr{cccc}
2m_0^2 & f_1 & \ldots & f_{N} \\
f_1    & 2p_1p_1 & \ldots & 2p_1p_{N} \\
\vdots & \vdots  & \ddots & \vdots \\
f_{N}    & 2p_{N}p_1 & \ldots & 2p_{N}p_{N} 
\earr\right) 
\eeq
with
\beq\label{def-fi}
f_k=p_k^2-m_k^2+m_0^2, \qquad k=1,\ldots,N.
\eeq
Its determinant is given by
\beqar\label{detCayley}
\det(\Ymod^{(N)}) = 2 m_0^2 \Delta^{(N)} 
- \sum_{n,m=1}^{N}f_nf_m\Zadj^{(N)}_{nm}
=\left|\barr{cccc}
Y_{00}   & Y_{01}   & \ldots  &  Y_{0N} \\  
Y_{10}   & Y_{11}   & \ldots  &  Y_{1N} \\  
\vdots  & \vdots    & \ddots  &  \vdots \\  
Y_{N0}   & Y_{N1}   & \ldots  &  Y_{NN}
\earr \right| = \det(Y),
\eeqar
where 
\beq
Y_{ij}=m_i^2+m_j^2-(p_i-p_j)^2,
\qquad i,j=0,\ldots,N.  
\label{eq:Y}
\eeq
The matrix $Y=(Y_{ij})$ is sometimes called modified Cayley matrix and its
determinant the modified Cayley determinant \cite{Me65}.
Its elements are related to those of the Gram matrix via
\beq
Y_{ij}= Z^{(N)}_{ij}-f_i-f_j+2m_0^2,\qquad
Y_{0i}= Y_{i0}= -f_i+2m_0^2, 
\qquad i,j=1,\ldots,N.  
\eeq
The vanishing of $\det(\Ymod^{(N)})$ is a necessary condition for the
appearance of leading Landau singularities \cite{Ed66}.  The adjoint
of $\Ymod^{(N)}_{ij}$, $i,j=0,\ldots,N$ can be expressed as
\beqar\label{Ymodadj}
\Ymodadj^{(N)}_{00} &=& \Delta^{(N)},\nl
\Ymodadj^{(N)}_{0i} &=& \Ymodadj^{(N)}_{i0} =
-\sum_{n=1}^{N} \Zadj^{(N)}_{in} f_n, \nl
\Ymodadj^{(N)}_{ij} &=&  2m_0^2\Zadj^{(N)}_{ij} + \sum_{n,m=1}^{N}
\Zadjadj^{(N)}_{(in)(jm)} f_n f_m, \qquad i,j=1,\ldots,N.
\eeqar
For later use we also consider the generalized adjoint of
$\Ymod^{(N)}$. The relevant part of it is given by
\beqar
\Ymodadjadj^{(N)}_{(0i)(0j)} &=& -\Zadj^{(N)}_{ij}, \qquad
i,j=1,\ldots,N,
\nl
\Ymodadjadj^{(N)}_{(0i)(jk)} &=& 
\Ymodadjadj^{(N)}_{(jk)(0i)} = -\sum_{n=1}^N f_n \Zadjadj^{(N)}_{(ni)(jk)},
\qquad i,j,k=1,\ldots,N.
\label{eq:Ymodadjadj}
\eeqar
These relations together with \refeq{Zadjadjrelation1} imply
\beqar
\det(\Ymod^{(N)})\Zadj^{(N)}_{ij} &=&  
\De^{(N)} \Ymodadj^{(N)}_{ij} -\Ymodadj^{(N)}_{i0}\Ymodadj^{(N)}_{0j},
\nn\\
\det(\Ymod^{(N)}) 
\Ymodadjadj^{(N)}_{(0i)(jk)}
&=&
\Ymodadj^{(N)}_{0k}\Ymodadj^{(N)}_{ij} - \Ymodadj^{(N)}_{ik}\Ymodadj^{(N)}_{0j}.
\label{eq:Ymodadjadj_rels}
\eeqar

\section{Evaluation of 1-point functions}
\label{se:1pf}

The scalar 1-point integral for an arbitrary complex mass $m_0$ is
given by
\beq
A_0(m_0) = m_0^2\left[\Delta+\ln\left(\frac{\mu^2}{m_0^2}\right)+1\right],
\eeq
where $\Delta$ is the standard one-loop divergence
\beq
\Delta = \frac{2}{4-D} - \gamma_{\mathrm{E}} + \ln(4\pi)
\eeq
in $D$ space--time dimensions with $\gamma_{\mathrm{E}}$ denoting
Euler's constant.  The tensor integrals of rank~$2n$ $(n=1,2,\ldots)$
are given by
\beq
A^{\mu_1\ldots\mu_{2n}} =
\{\underbrace{g\ldots g}_{n}\}^{\mu_1\ldots\mu_{2n}} \,
A_{\underbrace{\sst 0\ldots0}_{2n}},
\eeq
where the tensor coefficients are easily evaluated to
\beq
A_{\underbrace{\sst 0\ldots0}_{2n}} = \frac{m_0^{2n}}{2^n(n+1)!}
\left[A_0(m_0)+m_0^2\sum_{k=1}^n\frac{1}{k+1}\right].
\label{A0dots0}
\eeq
Because of Lorentz invariance obviously all tensors of odd rank vanish.

\section{Evaluation of 2-point functions}
\label{se:2pf}

In the following we assume that at least one of the parameters
$p_1^2$, $m_0$, $m_1$ is different from zero; otherwise the
2-point integrals identically vanish in dimensional regularization,
\beq
B_{\dots}(0,0,0) \equiv 0,
\eeq
where the dots stand for any Lorentz index or any index of a tensor
coefficient.

Up to rank~3 the 2-point tensor integrals are decomposed as
\beqar
B^\mu &=& p^\mu_1 B_1, \qquad
B^{\mu\nu} = p^\mu_1 p^\nu_1 B_{11}+g^{\mu\nu}B_{00}, \qquad
B^{\mu\nu\rho} = p^\mu_1 p^\nu_1 p^\rho_1 B_{111}
+\{g p\}_1^{\mu\nu\rho}B_{001}.
\hspace{2em}
\eeqar
The tensor coefficients can be algebraically reduced to scalar 1- and
2-point integrals, $A_0$ and $B_0$, with the Passarino--Veltman
algorithm \cite{Passarino:1978jh} as more generally described in the
next section. The corresponding results for tensors up to rank~3 are,
e.g., given in the appendix of \citere{Denner:2002ii}. The algebraic
reduction for the coefficients $B_{00 i_3 i_4\ldots}$, which
correspond to covariants involving the metric tensor,
\beqar
B_{00} &=& \frac{1}{6}\left[A_0(0)+f_1 B_1+2m_0^2 B_0+m_0^2+m_1^2
-\textfrac{1}{3}p_1^2\right], \nn\\[.1em]
B_{001} 
        &=& \frac{1}{8}\left[ -A_0(0)+f_1 B_{11} +2m_0^2B_1
-\textfrac{1}{6}(2m_0^2+4m_1^2-p_1^2)\right], \qquad \mbox{etc.,}
\eeqar
are numerically well behaved.  However, the reduction formulas for the
coefficients $B_{1\ldots1}$ corresponding to the covariant
$p_1^{\mu_1}\cdots p_1^{\mu_P}$ involve a factor $1/p_1^2$ in each
reduction step, so that these reduction formulas become numerically
unstable for small $p_1^2$.  Owing to the simplicity of 2-point
integrals it is, however, possible to derive closed expressions for
these coefficients that are numerically stable for all values of
$p_1^2$. Such a derivation is described below.  Assuming the knowledge
of the coefficients $B_{1\ldots1}$, the remaining coefficients
$B_{0\ldots01\ldots1}$ can be obtained from the recurrence relations
\beqar\label{Bred1}
B_{\underbrace{\sst 0\ldots0}_{2n+2} \underbrace{\sst 1\ldots1}_{P-2n-2}} 
&=& -\frac{1}{2(P-2n-1)}\left[ 
A_{\underbrace{\sst 0\ldots0}_{2n} \underbrace{\sst 1\ldots1}_{P-2n-1}}(0)
+f_1 B_{\underbrace{\sst 0\ldots0}_{2n} \underbrace{\sst 1\ldots1}_{P-2n-1}}
+2p_1^2 B_{\underbrace{\sst 0\ldots0}_{2n} \underbrace{\sst 1\ldots1}_{P-2n}}
\right], 
\nn\\
&& n=0,\ldots,\left[\frac{P-2}{2}\right].
\eeqar
or
\beqar\label{Bred2}
B_{\underbrace{\sst 0\ldots0}_{2n+2} \underbrace{\sst 1\ldots1}_{P-2n-2}} 
&=& \frac{1}{2(P+1)}\left[ 
A_{\underbrace{\sst 0\ldots0}_{2n} \underbrace{\sst 1\ldots1}_{P-2n-2}}(0)
+2m_0^2 B_{\underbrace{\sst 0\ldots0}_{2n} \underbrace{\sst 1\ldots1}_{P-2n-2}}
+f_1 B_{\underbrace{\sst 0\ldots0}_{2n} \underbrace{\sst 1\ldots1}_{P-2n-1}}
\right.
\nl&&\left.{}
-2(D-4)B_{\underbrace{\sst 0\ldots0}_{2n+2} \underbrace{\sst 1\ldots1}_{P-2n-2}}
\right], 
\quad
n=0,\ldots,\left[\frac{P-2}{2}\right].
\eeqar
The coefficients $A_{0\ldots01\ldots1}(0)$ are given by
\beq
A_{\underbrace{\sst 0\ldots0}_{2n} \underbrace{\sst 1\ldots1}_{P-2n-1}}(0)
= (-1)^{P-2n-1} \tilde A_{\underbrace{\sst 0\ldots0}_{2n}}(0),
\eeq
where $\tilde A_{0\ldots0}(0)$ can be obtained from \refeq{A0dots0}.
The finite polynomial quantities $(D-4)B_{00\ldots}$ can easily be
derived by exploiting \refeq{Bred2} for the UV-singular parts;
explicit results for tensors up to rank~5 are summarized in
\refapp{se:tendiv}.

We derive the expressions for $B_{1\ldots1}$
by explicitly solving the Feynman-parameter integral
\beq
B_{\underbrace{\sst 1\ldots1}_n} = \int_0^1\rd x\, (-x)^n
\left\{\Delta+\ln\mu^2
-\ln\left[-p_1^2 x(1-x)+m_0^2(1-x)+m_1^2 x-\ri\eps\right]\right\}.
\eeq
In the following result we support complex mass parameters; more
precisely, the real parts of $m_i^2$ must be non-negative, the
imaginary parts negative or zero.  The final results are conveniently
written as
\beq
B_{\underbrace{\sst 1\ldots1}_n} = \frac{(-1)^n}{n+1}
\left\{\Delta+\ln\left(\frac{\mu^2}{m_0^2}\right)-\sum_{k=1}^2f_n(x_k)\right\}
\label{eq:b1}
\eeq
with $x_k$ denoting the solutions of the quadratic equation
\beq
0=-p_1^2 x(1-x)+m_0^2(1-x)+m_1^2 x-\ri\eps.
\eeq
For $p_1^2=0$ one of the $x_k$ is formally $\infty$.
The auxiliary functions
\beq
f_n(x) \equiv (n+1)\int_0^1\rd t\,t^n\ln\left(1-\frac{t}{x}\right)
\eeq
can be evaluated in a numerically stable way by choosing one of the two
representations
\beqar
f_n(x) &=& \left(1-x^{n+1}\right)\ln\left(\frac{x-1}{x}\right)
-\sum_{l=0}^n \frac{x^{n-l}}{l+1}
\nn\\
&=& \ln\left(1-\frac{1}{x}\right)
+\sum_{l=n+1}^\infty \frac{x^{n-l}}{l+1}.
\label{eq:fn}
\eeqar
The first form is numerically stable for intermediate values of
$|x|\ne0$. For $x\to0$, $f_n(x)$ develops a true logarithmic
singularity; for $x\to1$ the logarithm $\ln(1-1/x)$ is suppressed
because of its prefactor. The second equality in \refeq{eq:fn} yields
numerically stable results for large $|x|$. In practice, we take the
first form for $|x|<10$ and the second otherwise.  The case where one
of the $x_k$ is zero corresponds to $m_0=0$ and can be easily obtained
via taking the limit $m_0\to0$,
\beq
B_{\underbrace{\sst 1\ldots1}_n}(p_1^2,0,m_1) = \frac{(-1)^n}{n+1}
\left\{\Delta+\ln\left(\frac{\mu^2}{m_1^2-p_1^2-\ri\eps}\right)
+\frac{1}{n+1}-f_n\left(1-\frac{m_1^2-\ri\eps}{p_1^2}\right)\right\}.
\eeq
For $p_1^2=m_1^2$ this further simplifies to
\beq
B_{\underbrace{\sst 1\ldots1}_n}(m_1^2,0,m_1) = \frac{(-1)^n}{n+1}
\left\{\Delta+\ln\left(\frac{\mu^2}{m_1^2}\right)
+\frac{2}{n+1}\right\}.
\eeq
In the vicinity of the last two special cases one of
the $x_k$ becomes small, so that the leading (logarithmic) term
in $f_n(x_k)$ cancels against the explicit logarithm in \refeq{eq:b1}.
Although this somewhat worsens the precision of the evaluation, we
did not find problems with this approach in practice.
Nevertheless we have additionally implemented a more sophisticated
representation of $B_{\sst 1\ldots1}$ with more
branches where such cancellations are avoided.

In the above derivation we essentially followed the approach described
in the appendix of \citere{Passarino:1978jh}; the results given there
are, however, not applicable to the general case of complex masses.

\section{Reduction of 3-point and 4-point functions}
\label{se:4pf}

The tensor decompositions of 3-point tensor integrals up to rank 4 and
4-point tensor integrals up to rank 5 read explicitly
\beqar
C^{\mu}&=&\sum_{\ina=1}^{2} p_{\ina}^{\mu}C_{\ina}, \qquad
C^{\mu\nu}=\sum_{\ina,\inb=1}^{2} p_{\ina}^{\mu}p_{\inb}^{\nu}C_{\ina\inb}
+g^{\mu\nu}C_{00},\nls
C^{\mu\nu\rho}&=&\sum_{\ina,\inb,\inc=1}^{2} p_{\ina}^{\mu}p_{\inb}^{\nu}p_{\inc}^{\rho}C_{\ina\inb\inc}
+\sum_{\ina=1}^{2}\{g p\}_{\ina}^{\mu\nu\rho} C_{00\ina}, 
\nl
C^{\mu\nu\rho\si} &=& 
\sum_{\ina,\inb,\inc,\ind=1}^{2} p_{\ina}^{\mu}p_{\inb}^{\nu}p_{\inc}^{\rho}p_{\ind}^\si C_{\ina\inb\inc\ind}
+\sum_{\ina,\inb=1}^{2}
\{g pp\}_{\ina\inb}^{\mu\nu\rho\si}C_{00\ina\inb}
+\{g g\}^{\mu\nu\rho\si} C_{0000},
\nl
\\
D^{\mu}&=&\sum_{\ina=1}^{3} p_{\ina}^{\mu}D_{\ina}, \qquad
D^{\mu\nu}=\sum_{\ina,\inb=1}^{3} p_{\ina}^{\mu}p_{\inb}^{\nu}D_{\ina\inb}
+g^{\mu\nu}D_{00},\nls
D^{\mu\nu\rho}&=&\sum_{\ina,\inb,\inc=1}^{3} p_{\ina}^{\mu}p_{\inb}^{\nu}p_{\inc}^{\rho}D_{\ina\inb\inc}
+\sum_{\ina=1}^{3}\{g p\}_{\ina}^{\mu\nu\rho} D_{00\ina}, 
\nl
D^{\mu\nu\rho\si} &=& 
\sum_{\ina,\inb,\inc,\ind=1}^{3} p_{\ina}^{\mu}p_{\inb}^{\nu}p_{\inc}^{\rho}p_{\ind}^\si D_{\ina\inb\inc\ind}
+\sum_{\ina,\inb=1}^{3}
\{g pp\}_{\ina\inb}^{\mu\nu\rho\si}D_{00\ina\inb}
+\{g g\}^{\mu\nu\rho\si} D_{0000},\nl
D^{\mu\nu\rho\si\tau} &=& 
\sum_{\ina,\inb,\inc,\ind,\ine=1}^{3}
p_{\ina}^{\mu}p_{\inb}^{\nu}p_{\inc}^{\rho}p_{\ind}^\si p_{\ine}^\tau D_{\ina
\inb\inc\ind\ine}
+\sum_{\ina,\inb,\inc=1}^{3}
\{g ppp\}_{\ina\inb\inc}^{\mu\nu\rho\si\tau}
D_{00\ina\inb\inc}
\nl&&{}
+\sum_{\ina=1}^{3} 
\{g g p\}_{\ina}^{\mu\nu\rho\si\tau} 
D_{0000\ina}.
\label{eq:cdtensor}
\eeqar
Because of the symmetry of the tensor $T^{N}_{\mu_{1}\ldots\mu_{P}}$
all coefficients $C_{i_1\ldots i_P}$, and $D_{i_1\ldots i_P}$ are
symmetric under permutation of all indices.  To be specific, in the
following we give the reduction formulas for the 4-point functions,
i.e.\ $N=4$. To obtain the corresponding results for 3-point functions
one has to perform the substitutions
\beq
C_{\ldots} \to B_{\ldots} , \quad D_{\ldots}  \to C_{\ldots} , \quad
Z^{(3)} \to Z^{(2)}, \quad \De^{(3)} 
\to \De^{(2)}, \quad \Ymod^{(3)} \to \Ymod^{(2)}, \quad N\to 3,
\label{eq:N3subst}
\eeq
and similar obvious substitutions. 

\subsection{Conventional Passarino--Veltman reduction}
\label{se:PV}

The one-loop tensor integrals can be reduced to scalar integrals
recursively by inversion of systems of linear equations
\cite{Passarino:1978jh}.  The inhomogeneity of these equations
consists of coefficients of lower rank. The equations of this system
are obtained by contracting $T^N_{\mu_1\ldots\mu_P}$ with the $(N-1)$
external momenta $p_k^{\mu_1}$ and for $P\ge 2$ also by contraction
with the metric $g^{\mu_1\mu_2}$.  Contracting \refeq{tensorint} with
$p_k^{\mu_1}$ and using
\beq
2p_k q = N_k-N_0-f_k ,
\label{contrtensor}
\eeq
each of the first two terms on the r.h.s.\ of (\ref{contrtensor})
cancels exactly one propagator denominator of
$p_k^{\mu_{1}}T^{N}_{\mu_{1}\ldots\mu_{P}}$ and the third term is
proportional to $T^{N}_{\mu_{2}\ldots\mu_{P}}$.  Likewise the
contraction of \refeq{tensorint} with $g^{\mu_1\mu_2}$ yields a factor
$q^2$ in the numerator of
$g^{\mu_1\mu_2}T^{N}_{\mu_{1}\ldots\mu_{P}}$, which can be written as
\beq
q^2 = N_0 + m_0^2.  
\eeq
The $N_0$ term cancels the first propagator, the second term leads to
the tensor $T^{N}_{\mu_{3}\ldots\mu_{P}}$.
This yields
\beqar\label{recurtensorp}
2p_k^{\mu_{1}}T^{N}_{\mu_{1}\ldots\mu_{P}} &=&
T^{N-1}_{\mu_{2}\ldots\mu_{P}}(k)-T^{N-1}_{\mu_{2}\ldots\mu_{P}}(0)-
f_k T^{N}_{\mu_{2}\ldots\mu_{P}}, \\[.2em]
g^{\mu_1\mu_2}T^{N}_{\mu_{1}\mu_{2}\ldots\mu_{P}} &=&
T^{N-1}_{\mu_{3}\ldots\mu_{P}}(0)+m_0^2T^{N}_{\mu_{3}\ldots\mu_{P}}.
\label{recurtensorg}
\eeqar
Note that for $T^{N-1}_{\mu_{2,3}\ldots\mu_{P}}(0)$ a shift of the
integration momentum $q^\mu\to q^\mu-p_1^\mu$ has to be done in order
to achieve the standard form (\ref{tensorint}). The tensor integrals
with shifted momenta $\tilde T^{N-1}_{\mu_{1}\ldots\mu_{P}}(0)$ are
defined in \refeq{Dshifted}. Inserting the Lorentz decompositions
\refeq{eq:cdtensor} into \refeq{recurtensorp} and
\refeq{recurtensorg}, the desired recurrence relations can be read off
by comparing coefficients.

From \refeq{recurtensorp} we obtain
\beqar\label{recurDp}
S^P_{ki_2\ldots i_P} &\equiv& C_{(i_2)_k\ldots (i_P)_k}(k)
\debar_{ki_2}\ldots \debar_{ki_P}
- C_{i_2\ldots i_P}(0) - f_k D_{i_2\ldots i_P} \nl
&=& \sum_{m=1}^{N-1} Z^{(3)}_{km} D_{mi_2\ldots i_P} + 2 \sum_{r=2}^P
\de_{ki_r} D_{00i_2\ldots \hat i_r\ldots i_P},
\nl
&& k=1,\ldots,N-1, \qquad i_2,\ldots,i_P=0,\ldots, N-1,
\eeqar
and from \refeq{recurtensorg}
\beqar\label{recurDg}
S^P_{00i_3\ldots i_P} &\equiv&
2 C_{i_3\ldots i_P}(0) + 2 m_0^2 D_{i_3\ldots i_P} \nl
&=& \sum_{n,m=1}^{N-1} Z^{(3)}_{nm} D_{nmi_3\ldots i_P} + 2 
\Bigl(D+P-2 + \sum_{r=3}^P \bar\de_{i_r 0} \Bigr)
D_{00i_3\ldots i_P},
\nl
&& i_3,\ldots,i_P=0,\ldots, N-1,
\eeqar
where the matrix $Z^{(3)}$ is defined in \refeq{matrixZ}.
Equations \refeq{recurDp} and \refeq{recurDg} can be solved for 
the coefficients of $D^{\mu_1\ldots \mu_P}$ as
\beqar\label{PVrecursionD1}
D_{00i_3\ldots i_P} &=& \frac{1}{2(3+P-N)}\biggl[
 -2(D-4)D_{00i_3\ldots i_P} 
+ C_{i_3\ldots i_P}(0) 
+ 2m_0^2  D_{i_3\ldots i_P}  
\nl&&{}\qquad\qquad\qquad {}
+ \sum_{n=1}^{N-1} f_n D_{ni_3\ldots i_P} \biggr],
\\\label{PVrecursionD2}
 D_{i_1\ldots i_P} &=&  \sum_{n=1}^{N-1}(Z^{(3)})^{-1}_{i_1n}
\left( S^P_{ni_2\ldots i_P} -  2 \sum_{r=2}^P
\de_{ni_r} D_{00i_2\ldots \hat i_r\ldots i_P}\right), \quad i_1\ne0.
\eeqar

The relations \refeq{PVrecursionD1} and \refeq{PVrecursionD2}
determine $ D_{i_1\ldots i_P}$ in terms of $D_{i_1\ldots i_{P-1}}$ and
3-point functions. Using these relations recursively, all coefficients
of 4-point functions can be expressed in terms of 3-point functions
and the scalar 4-point function $D_0$. The finite polynomial
quantities $(D-4)D_{00i_3\ldots i_P}$ can easily be derived by
exploiting \refeq{PVrecursionD1} for the UV-singular parts; explicit
results for tensors up to rank~7 are summarized in \refapp{se:tendiv}.
As explained in \refse{se:divs}, IR divergences do not occur in
$D_{00i_3\ldots i_P}$.  More explicit formulas for all tensor
functions up to rank 5 are given in the appendix of
\citere{Denner:2002ii}.

Figure~\ref{fig:PV} illustrates the Passarino--Veltman reduction
scheme for 4-point integrals in a plane of tensor coefficients where
the rank of the tensor increases by going down in the rows and the
number of index pairs ``00'' increases by going to the right in the
columns.
\bfi
\centerline{\unitlength 1.2pt
\begin{picture}(280,190)(70,10)%
\SetScale{1.2}
\put(82,190){\small{\fbox{$D_0$} \quad basis integral}}%
\put(85,150){\small $D_{\ina}$}%
\put(85,110){\small $D_{\ina\inb}$}%
\put(205,110){\small $D_{00}$}%
\put(85, 70){\small $D_{\ina\inb\inc}$}%
\put(205, 70){\small $D_{00\ina}$}%
\put(85, 30){\small $D_{\ina\inb\inc\ind}$}%
\put(205, 30){\small $D_{00\ina\inb}$}%
\put(305, 30){\small $D_{0000}$}%
\put(95, 10){\small \vdots}%
\put(215, 10){\small \vdots}%
\put(315, 10){\small \vdots}%
\LongArrow(90,182)(90,162)%
\put(93,172){\small$1$}
\LongArrow(100,185)(198,120)%
\LongArrow(100,150)(198,115)%
\put(190,130){\small$2a$}
\put(160,131){\small$2a$}
\LongArrow(90,140)(90,122)%
\LongArrow(198,110)(107,113)%
\put(93,130){\small$2b$}
\put(120,116){\small$2b$}
\LongArrow(100,145)(198, 80)%
\LongArrow(105,109)(198, 75)
\put(190, 90){\small$3a$}
\put(160, 91){\small$3a$}
\LongArrow(90,100)(90, 82)%
\LongArrow(200, 70)(113, 70)%
\put(93, 90){\small$3b$}
\put(120,75){\small$3b$}
\LongArrow(222,105)(298, 40)%
\LongArrow(227, 65)(298, 35)%
\put(290, 50){\small$4c$}
\put(270, 50){\small$4c$}
\LongArrow(105,105)(198, 40)%
\LongArrow(111, 65)(198, 35)%
\put(190, 50){\small$4a$}
\put(160, 50){\small$4a$}
\LongArrow(200, 30)(119, 30)%
\LongArrow(90, 60)(90, 42)%
\put(93, 50){\small$4b$}
\put(125,35){\small$4b$}
\end{picture}%
}
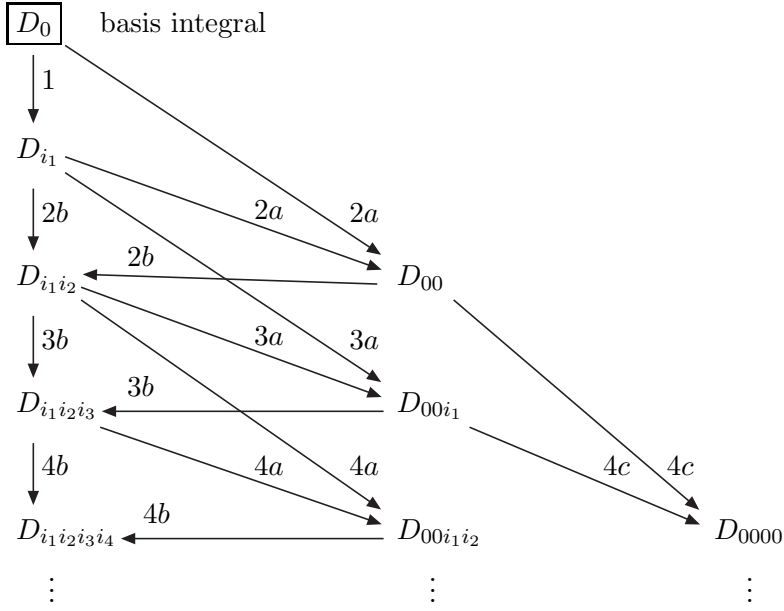
\caption{Schematic illustration of conventional Passarino--Veltman reduction.}
\label{fig:PV}
\efi
The steps in the algorithm are indicated by arrows that show which
coefficient is deduced from previously calculated ones. The numbers
close to the arrows correspond to the step number which is identical
to the rank of the tensor coefficients to be calculated; the
labels ``$a$'', ``$b$'', etc.\ give the order in which the 
coefficients within a step are calculated.

Equation \refeq{PVrecursionD2} becomes numerically unstable
if $Z^{(3)}$ is nearly singular, \ie if the Gram determinant
$\Delta^{(3)}$
is close to zero. Reduction schemes for this case are described in
\refses{se:Cayley}--\ref{se:smallmomenta}.

\subsection{Alternative Passarino--Veltman-like reduction}
\label{se:PValt}

An alternative to the conventional Passarino--Veltman reduction can be
obtained as follows.  Equations \refeq {recurDp} and
\refeq{PVrecursionD1} can be written as
\beqar\label{PValt}
\lefteqn{
\left(\barr{cccc}
2m_0^2 & f_1 & f_2 & f_3 \\
f_1    & 2p_1p_1   & 2p_1p_2 & 2p_1p_{3} \\
f_2    & 2p_{2}p_1 & 2p_2p_2 & 2p_2p_{3}  \\
f_3    & 2p_{3}p_1 & 2p_3p_2 & 2p_{3}p_{3} \\
\earr\right) 
\left(\barr{c}
D_{i_2\ldots i_P} \\
D_{1i_2\ldots i_P} \\
D_{2i_2\ldots i_P} \\
D_{3i_2\ldots i_P} \\
\earr\right) 
=
\Ymod^{(3)}
\left(\barr{c}
D_{i_2\ldots i_P} \\
D_{1i_2\ldots i_P} \\
D_{2i_2\ldots i_P} \\
D_{3i_2\ldots i_P} \\
\earr\right) 
} \hspace{2em}
\nl&&
= \left(\barr{cccc}
2(D+P-N) D_{00i_2\ldots i_P} -  C_{i_2\ldots i_P}(0)\\ 
\Shat^{P}_{1i_2\ldots i_P} -  2 \sum_{r=2}^P
\de_{1i_r} D_{00i_2\ldots \hat i_r\ldots i_P} \\
\Shat^{P}_{2i_2\ldots i_P} -  2 \sum_{r=2}^P
\de_{2i_r} D_{00i_2\ldots \hat i_r\ldots i_P} \\
\Shat^{P}_{3i_2\ldots i_P} -  2 \sum_{r=2}^P
\de_{3i_r} D_{00i_2\ldots \hat i_r\ldots i_P} \\
\earr\right),
\qquad i_2,\ldots,i_P=0,\ldots,N-1,
\hspace{2em}
\eeqar
where on the r.h.s.\ the matrix $\Ymod^{(3)}$ defined in
\refeq{defYmod} appears and the following abbreviations are
introduced,
\beq\label{defShat}
\Shat^{P}_{ki_2\ldots i_P} =  C_{(i_2)_k\ldots (i_P)_k}(k)
\debar_{ki_2}\ldots \debar_{ki_P} - C_{i_2\ldots i_P}(0)
= S^{P}_{ki_2\ldots i_P} + f_k D_{i_2\ldots i_P}.
\eeq

Multiplying \refeq{PValt} with the matrix $\Ymodadj^{(3)}$
from the left, we obtain 
\beqar\label{PValtrec1}
\det(\Ymod^{(3)})D_{i_2\ldots i_P} &=& \Delta^{(3)} \Big[
2(4+P-N) D_{00i_2\ldots i_P} +2(D-4) D_{00i_2\ldots i_P} 
-  C_{i_2\ldots i_P}(0)\Big] \nl
&&{}
+\sum_{n=1}^{N-1} \Ymodadj^{(3)}_{0n}
\left[\Shat^{P}_{ni_2\ldots i_P} -  2 \sum_{r=2}^P
\de_{ni_r} D_{00i_2\ldots \hat i_r\ldots i_P}\right]
\eeqar
and
\beqar\label{PValtrec2}
\det(\Ymod^{(3)})D_{i_1i_2\ldots i_P} &=& 
 \Ymodadj^{(3)}_{i_10}
\Big[2(4+P-N)D_{00i_2\ldots i_P} + 2(D-4)D_{00i_2\ldots i_P}
-  C_{i_2\ldots i_P}(0)\Big] \nl
&&{} +\sum_{n=1}^{N-1}
\Ymodadj^{(3)}_{i_1n}
\left[\Shat^{P}_{ni_2\ldots i_P} -  2 \sum_{r=2}^P
\de_{ni_r} D_{00i_2\ldots \hat i_r\ldots i_P}\right], \qquad i_1\ne 0.
\eeqar
Equation \refeq{PValtrec1} yields $D_{00i_2\ldots i_P}$ in terms of
$D_{00i_2\ldots\hat i_r\ldots i_{P}}$, $D_{i_2\ldots i_{P}}$, and
3-point functions, 
\beqar\label{PValtrec1a}
\lefteqn{
  2(4+P-N) \Delta^{(3)} D_{00i_2\ldots i_P}=
 -2 \Delta^{(3)} (D-4) D_{00i_2\ldots i_P}
+ \Delta^{(3)} C_{i_2\ldots i_P}(0)}\qquad \nl
&&{}
+ \det(\Ymod^{(3)})D_{i_2\ldots i_P} 
-\sum_{n=1}^{N-1} \Ymodadj^{(3)}_{0n}
\left[\Shat^{P}_{ni_2\ldots i_P} -  2 \sum_{r=2}^P
\de_{ni_r} D_{00i_2\ldots \hat i_r\ldots i_P}\right],
\eeqar
and thereafter \refeq{PValtrec2} yields $D_{i_1\ldots i_{P}}$.  Using
these relations recursively, all coefficients of 4-point functions can
be expressed in terms of 3-point functions and the scalar 4-point
function $D_0$.  While the final results are of course identical to
those of the usual Passarino--Veltman reduction, the order in which
the different coefficients are calculated is different. As a
consequence, this recursion can, in some cases, be numerically more
stable than the conventional Passarino--Veltman reduction, in
particular, if all the quantities $\Delta^{(3)}$,
$\Ymodadj_{k0}^{(3)}$, and $\Ymodadj_{kl}^{(3)}$ become small.

For the tensor coefficients up to rank 3 the reduction formulas
explicitly read
\beqar
  2(5-N) \Delta^{(3)} D_{00} &=&
\! -2 \Delta^{(3)} (D-4) D_{00} +
 \Delta^{(3)} C_{0}(0)+ \det(\Ymod^{(3)})D_{0}
-\sum_{n=1}^{N-1} \Ymodadj^{(3)}_{0n}\Shat^{1}_{n},
\\
\label{DiPValt}
\det(\Ymod^{(3)})D_{i_1} &=&
 \Ymodadj^{(3)}_{i_10}
[2(5-N)D_{00} + 2(D-4)D_{00} -  C_{0}(0)]
+\sum_{n=1}^{N-1} \Ymodadj^{(3)}_{i_1n}
\Shat^{1}_{n},
\\
  2(6-N) \Delta^{(3)} D_{00i_2} &=&
 -2 \Delta^{(3)} (D-4) D_{00i_2} +
 \Delta^{(3)} C_{i_2}(0)+ \det(\Ymod^{(3)})D_{i_2} 
\nl
&&{}
-\sum_{n=1}^{N-1} \Ymodadj^{(3)}_{0n}
\left[\Shat^{2}_{ni_2} -  2 \de_{ni_2} D_{00}\right],
\\
\label{DijPValt}
\det(\Ymod^{(3)})D_{i_1i_2} &=&
 \Ymodadj^{(3)}_{i_10}
[2(6-N)D_{00i_2} + 2(D-4)D_{00i_2} -  C_{i_2}(0)]
\nl
&&{} +\sum_{n=1}^{N-1}
\Ymodadj^{(3)}_{i_1n}
\left[\Shat^{2}_{ni_2} - 2 \de_{ni_2} D_{00}\right], \qquad i_1,i_2\ne 0,
\\
\label{D00ijPValt}
  2(7-N) \Delta^{(3)} D_{00i_2i_3} &=&
 -2 \Delta^{(3)} (D-4) D_{00i_2i_3} +
 \Delta^{(3)} C_{i_2i_3}(0)+ \det(\Ymod^{(3)})D_{i_2i_3} 
\nl
&&{}
-\sum_{n=1}^{N-1} \Ymodadj^{(3)}_{0n}
\left[\Shat^{3}_{ni_2i_3} -  2 
\de_{ni_2} D_{00i_3} - 2 \de_{ni_3} D_{00i_2}\right],
\\
\label{DijkPValt}
\det(\Ymod^{(3)})D_{i_1i_2i_3} &=& 
 \Ymodadj^{(3)}_{i_10}
[2(7-N)D_{00i_2i_3} + 2(D-4)D_{00i_2i_3} -  C_{i_2i_3}(0)]
\nl
&&{} +\sum_{n=1}^{N-1}
\Ymodadj^{(3)}_{i_1n}
\left[\Shat^{3}_{ni_2i_3} - 2 \de_{ni_2} D_{00i_3}
- 2 \de_{ni_3} D_{00i_2}\right], \!\quad i_1,i_2,i_3\ne 0.\hspace{3em}
\eeqar
Note that \refeq{D00ijPValt} holds also for $i_2=i_3=0$.

The 3-point tensor coefficients that result from omitting $N_0$ in the
4-point integrals are defined according to \refeq{TNaux} or more
explicitly
\beqar\label{auxC}
\Ccomb_{\ina}(0) &=& \Ctilde_{\ina-1}(0), \quad \ina=2,\ldots,N-1,
\nl
\Ccomb_{1}(0) &=& -\sum_{\sina=2}^{N-1}C_{\sina}(0)-C_{0}(0),
\\
\Ccomb_{\ina\inb}(0) &=& \Ctilde_{\ina-1,\inb-1}(0),
\quad \ina,\inb=2,\ldots,N-1,
\nl
\Ccomb_{1\ina}(0) &=& -\sum_{\sina=2}^{N-1}C_{\sina\ina}(0)-C_{\ina}(0),
\quad \ina=1,\ldots,N-1, 
\\
\Ccomb_{\ina\inb\inc}(0) &=& \Ctilde_{\ina-1,\inb-1,\inc-1}(0),
\quad \ina,\inb,\inc=2,\ldots,N-1,
\nl
\Ccomb_{1\ina\inb}(0) &=&  -\sum_{\sina=2}^{N-1}C_{\sina\ina\inb}(0)-C_{\ina\inb}(0),
\quad \ina,\inb=1,\ldots,N-1.
\eeqar

Figure~\ref{fig:altPV} illustrates the alternative Passarino--Veltman reduction
scheme for 4-point integrals in the plane of tensor coefficients
similarly to \reffi{fig:PV} of the previous section for the conventional
variant.
\bfi
\centerline{\unitlength 1.2pt
\begin{picture}(280,190)(70,10)%
\SetScale{1.2}
\put(82,190){\small{\fbox{$D_0$} \quad basis integral}}%
\put(85,150){\small $D_{\ina}$}%
\put(85,110){\small $D_{\ina\inb}$}%
\put(205,110){\small $D_{00}$}%
\put(85, 70){\small $D_{\ina\inb\inc}$}%
\put(205, 70){\small $D_{00\ina}$}%
\put(85, 30){\small $D_{\ina\inb\inc\ind}$}%
\put(205, 30){\small $D_{00\ina\inb}$}%
\put(305, 30){\small $D_{0000}$}%
\put(95, 10){\small \vdots}%
\put(215, 10){\small \vdots}%
\put(315, 10){\small \vdots}%
\LongArrow(100,185)(198,120)%
\put(190,130){\small$1a$}
\LongArrow(200,115)(102,150)%
\put(160,131){\small$1b$}
\LongArrow(100,145)(198, 80)%
\LongArrow(210,100)(210, 82)%
\put(190, 90){\small$2a$}
\put(215, 90){\small$2a$}
\LongArrow(200,112)(108,112)%
\LongArrow(200, 75)(108,107)
\put(160, 91){\small$2b$}
\put(120,115){\small$2b$}
\LongArrow(106,102)(198, 40)%
\LongArrow(210, 60)(210, 42)%
\put(190, 50){\small$3a$}
\put(215, 50){\small$3a$}
\LongArrow(200, 70)(113, 70)%
\LongArrow(200, 35)(113, 65)
\put(160, 50){\small$3b$}
\put(120, 75){\small$3b$}
\LongArrow(225,105)(300, 40)%
\put(292, 50){\small$3c$}
\end{picture}%
}
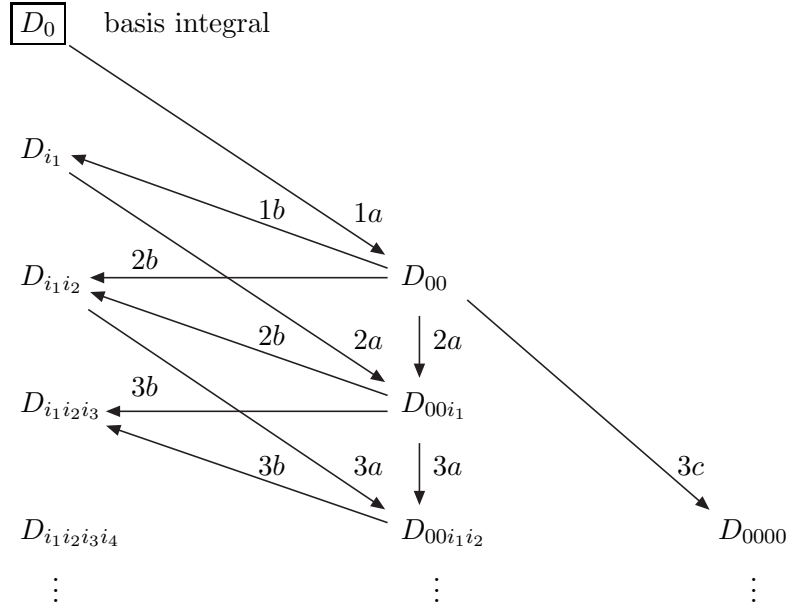
\caption{Schematic illustration of alternative Passarino--Veltman reduction.}
\label{fig:altPV}
\efi

\subsection{Reduction with modified Cayley determinants}
\label{se:Cayley}

Equation \refeq{PValt} can also be exploited directly to calculate
tensor coefficients of lower-rank from higher-rank tensors.
Specifically, the coefficients $D_{i_1\ldots i_P}$ with $i_1\ne0$ for
tensors of rank $P$ are expressed in terms of the coefficients
$D_{00i_2\ldots i_P}$ for tensors of rank $(P+1)$. This means,
\refeq{PValt} recursively expresses tensor coefficients $D_{i_1\ldots
  i_P}$ in terms of $C$ functions and of a single coefficient
$D_{0\ldots0}$ which results from $D_{i_1\ldots i_P}$ upon replacing
all non-zero indices $i_k$ by ``00''.  For sufficiently high tensor
rank $P$, viz.\ $P\ge 2N-4$, the integrand of the Feynman parameter
integral of $D_{0\ldots0}$ involves only polynomials and logarithms of
the integration parameters $x_l$.  Such integrals are numerically well
behaved, because singularities appearing in logarithms can be safely
treated numerically.  The explicit form of the Feynman-parameter
integral for the general coefficient $T^N_{0\ldots0}$ with $P\ge 2N-4$
is given below.

In summary, equation \refeq{PValt} provides a method for deducing all
tensor coefficients $D_{i_1\ldots i_P}$ (including the standard scalar
integral $D_0$) from $C$ functions and the numerically evaluated
coefficient $D_{0\ldots0}$ of tensor rank $2P$.  This procedure does
not involve the inverse of the Gram determinant $\Delta^{(3)}$, as it
is the case in the two versions of Passarino--Veltman reduction
described in the previous sections.  However, the method involves the
inverse of the modified Cayley determinant $\det(\Ymod^{(3)})$, so
that it becomes unstable if $\det(\Ymod^{(3)})$ becomes small.  It is
also interesting to note that the numerically evaluated coefficient
$D_{0\ldots0}$ enters this reduction with a prefactor $\Delta^{(3)}$.
Thus, this method becomes particularly precise if $\Delta^{(3)}$ is
small, where Passarino--Veltman reduction is unstable, because the
error in the numerical calculation of $D_{0\ldots0}$ is suppressed in
this case.  Note, however, that both the reduction of this section and
Passarino--Veltman reduction become problematic if both $\Delta^{(3)}$
and $\det(\Ymod^{(3)})$ are small.

For tensor coefficients up to rank 3 the reduction formulas
explicitly read
\beqar
\det(\Ymod^{(3)})D_{0000} &=& \Delta^{(3)} [2(9-N)
D_{000000} + 2(D-4)D_{000000} -  C_{0000}(0)] 
+\sum_{n=1}^{N-1} \Ymodadj^{(3)}_{0n} \Shat^{5}_{n0000},
\nl
\\
\det(\Ymod^{(3)})D_{00} &=& \Delta^{(3)} [2(7-N)D_{0000} + 2(D-4)
D_{0000} -  C_{00}(0)] 
+\sum_{n=1}^{N-1} \Ymodadj^{(3)}_{0n} \Shat^{3}_{n00},
\\
\det(\Ymod^{(3)})D_{0} &=& \Delta^{(3)} [2(5-N)D_{00} + 2(D-4)D_{00}-  C_{0}(0)] 
+\sum_{n=1}^{N-1} \Ymodadj^{(3)}_{0n} \Shat^{1}_{n},
\label{eq:D0_cayley}
\\
\det(\Ymod^{(3)})D_{0000 i_1} &=& 
\Ymodadj^{(3)}_{i_10} [2(9-N)D_{000000} + 2(D-4)D_{000000} -  C_{0000}(0)] 
+\sum_{n=1}^{N-1} \Ymodadj^{(3)}_{i_1n} \Shat^{5}_{n0000}, 
\nl
\\
\det(\Ymod^{(3)})D_{00 i_1}
&=& \Ymodadj^{(3)}_{i_10} [2(7-N)D_{0000} + 2(D-4)D_{0000} -  C_{00}(0)] 
+\sum_{n=1}^{N-1} \Ymodadj^{(3)}_{i_1n} \Shat^{3}_{n00}, 
\\
\det(\Ymod^{(3)})D_{00i_1i_2} &=& 
 \Ymodadj^{(3)}_{i_10}
[2(8-N)D_{0000i_2} + 2(D-4)D_{0000i_2} -  C_{00i_2}(0)] \nl
&&{} +\sum_{n=1}^{N-1}
\Ymodadj^{(3)}_{i_1n}
\left[\Shat^{4}_{n00i_2} -  2\de_{ni_2} D_{0000}\right], \quad i_1,i_2\ne0,
\eeqar
Finally, $D_{i_1}$, $D_{i_1i_2}$, and $D_{i_1i_2i_3}$ are obtained from
\refeqs{DiPValt}, \refeqs{DijPValt}, and \refeqs{DijkPValt}, respectively.
Thus, all 4-point tensor coefficients up to tensor rank 3 can be recursively
deduced from $D_{000000}$ and 3-point coefficients.

Figure~\ref{fig:mC} illustrates the reduction scheme for 4-point
integrals up to rank~3 in the plane of tensor coefficients similar to
the previous sections.
\bfi
\centerline{\unitlength 1.2pt
\begin{picture}(380,270)(70,10)%
\SetScale{1.2}
\put( 85,270){\small $D_0$}%
\put( 85,230){\small $D_{\ina}$}%
\put( 85,190){\small $D_{\ina\inb}$}%
\put( 85,150){\small $D_{\ina\inb\inc}$}%
\put( 85,110){\small $D_{\ina\inb\inc\ind}$}%
\put( 85, 70){\small $D_{\ina\inb\inc\ind\ine}$}%
\put( 85, 30){\small $D_{\ina\inb\inc\ind\ine\ing}$}%
\put(205,190){\small $D_{00}$}%
\put(205,150){\small $D_{00\ina}$}%
\put(205,110){\small $D_{00\ina\inb}$}%
\put(205, 70){\small $D_{00\ina\inb\inc}$}%
\put(205, 30){\small $D_{00\ina\inb\inc\ind}$}%
\put(305,110){\small{$D_{0000}$}}%
\put(305, 70){\small{$D_{0000\ina}$}}%
\put(305, 30){\small{$D_{0000\ina\inb}$}}%
\put(405, 30){\small{\fbox{$D_{000000}$}}}%
\put(410, 48){\small{\parbox{1.5cm}{basis \\[-.3em] integral}}}%
\put(95, 10){\small \vdots}%
\put(215, 10){\small \vdots}%
\put(315, 10){\small \vdots}%
\put(415, 10){\small \vdots}%
\LongArrow(198,200)(104,265)%
\LongArrow(198,195)(104,225)%
\put(190,210){\small$3a$}
\put(160,210){\small$3b$}
\LongArrow(198,190)(107,190)%
\LongArrow(198,155)(107,185)
\put(160,193){\small$3c$}
\put(160,170){\small$3c$}
\LongArrow(198,150)(113,150)%
\LongArrow(198,115)(113,145)
\put(160,153){\small$3d$}
\put(160,131){\small$3d$}
\LongArrow(298,120)(224,185)%
\LongArrow(298,115)(228,145)%
\put(290,130){\small$2a$}
\put(270,130){\small$2b$}
\LongArrow(398, 40)(330,105)%
\LongArrow(398, 35)(334, 65)%
\put(390, 50){\small$1a$}
\put(370, 50){\small$1b$}
\LongArrow(298,110)(237,110)%
\LongArrow(298, 75)(237,105)
\put(270,113){\small$2c$}
\put(270, 92){\small$2c$}
\end{picture}%
}
\caption{Schematic illustration of the reduction with modified 
  Cayley determinants.}
\label{fig:mC}
\efi
The steps of the reduction now proceed from right to left, starting
with a basis integral $D_{0\ldots0}$ with as many index pairs ``00''
as the finally aimed tensor rank, \ie for rank~3 with $D_{000000}$.
In each step we get all coefficients of at least one rank lower with
one index pair ``00'' less than in the previous steps.

Generically the Feynman-parameter integral for $T^N_{0\ldots0}$ reads
\beqar
T^N_{\underbrace{\sst 0\ldots0}_{2k}} &=& 
\frac{1}{2^k(2+k-N)!}
\left(\prod_{j=0}^{N-1}\int_0^\infty\rd x_j\right)
\delta\left(1-\sum_{l=0}^{N-1}\alpha_l x_l\right)
\left(\sum_{m=0}^{N-1} x_m\right)^{N-4-2k} A^{2+k-N}
\nn\\
&& {} \times
\left[\Delta+\sum_{n=1}^{2+k-N}\frac{1}{n}
-\ln\left(\frac{A-\ri\eps}{\mu^2}\right)
+2\ln\left(\sum_{m=0}^{N-1} x_m\right)
\right], \qquad k\ge N-2,
\label{eq:T02k}
\eeqar
with the shorthand
\beq
A = A(x_0,\ldots,x_{N-1}) = \left(\sum_{l=0}^{N-1}x_l p_l\right)^2
-\left(\sum_{m=0}^{N-1} x_m\right)
\left(\sum_{n=0}^{N-1} x_n(p_n^2-m_n^2)\right).
\eeq
The real parameters $\alpha_l$ appearing in \refeq{eq:T02k} are widely
arbitrary; they only have to fulfil the constraints
$\alpha_l\ge0$ and $\sum_{l=0}^{N-1}\alpha_l>0$.
For the numerical evaluation of the Feynman-parameter integral it is
convenient to take the uniform choice $\alpha_l=1$, in which case
the integral runs over the $(N-1)$-dimensional unit simplex $\sigma_{N-1}$,
\beq
x_0=1-\sum_{l=1}^{N-1}x_l, \qquad 
0<x_j<1-\sum_{l=1}^{j-1}x_l, \qquad j=1,\ldots,N-1,
\eeq
The integral representation \refeq{eq:T02k} is valid both for real and
complex masses.

Specifically, the integrals for $C_{000000}$ and $D_{000000}$,
which are needed for tensors of rank~3, are given by
\beqar
C_{000000} &=& 
\frac{1}{2880}
\left(\Delta+\frac{3}{2}\right)
\Bigl[s_{12}^2+p_1^4+p_2^4+s_{12}(p_1^2+p_2^2)+p_1^2p_2^2
\nl&& \qquad {}
-3(m_0^2s_{12}+p_1^2m_2^2+p_2^2m_1^2)
\nl&& \qquad {}
-6[s_{12}(m_1^2+m_2^2)+p_1^2(m_0^2+m_1^2)+p_2^2(m_0^2+m_2^2)]
\nl&& \qquad {}
+15[m_0^4+m_1^4+m_2^4+m_0^2 m_1^2+m_0^2 m_2^2 +m_1^2m_2^2]\Bigr]
\nn\\
&& {} -\frac{1}{16} \int_{\sigma_2}\rd^2x\, A^2
\ln\left(\frac{A-\ri\eps}{\mu^2}\right),
\label{eq:C000000}
\\    
D_{000000} &=& 
(\Delta+1) \left[
-\frac{1}{960}(s_{12}+s_{13}+s_{23}+p_1^2+p_2^2+p_3^2)
+\frac{1}{192}(m_0^2+m_1^2+m_2^2+m_3^2) \right]
\nn\\
&& {}
-\frac{1}{8} \int_{\sigma_3}\rd^3 x\, A
\ln\left(\frac{A-\ri\eps}{\mu^2}\right), 
\label{eq:D000000}
\eeqar
with the shorthands
\beq
s_{12}=(p_1-p_2)^2, \qquad
s_{13}=(p_1-p_3)^2, \qquad
s_{23}=(p_2-p_3)^2.
\eeq
For an efficient numerical integration of these integrals we use a
fortran code based on the {\sc DCUHRE} algorithm \cite{dcuhre}, as
included in the {\sc CUBA} library \cite{Hahn:2004fe}.  The
UV-divergent parts are integrated analytically in order to ensure
exact cancellation of the singularities.

As mentioned above, the procedure described in this section becomes
unstable if $\det(\Ymod^{(N-1)})$ becomes small.  The basis integrals
$T^N_{0\ldots0}$ are still safely calculated via the numerical
integration, but using the described relations to deduce the remaining
coefficients accumulates an instability in each step that turns an
index pair ``00'' into a non-zero tensor index or that eliminates an
index pair ``00''. This accumulation of an instability can be
suppressed by extending the set of basis integrals. For instance, the
3-point tensor coefficients $C_{i_1 i_2 i_3}$ can be deduced from the
coefficients $C_{00}$, $C_{0000}$, and $C_{000000}$, which all have
logarithmic integrands in their Feynman parametrizations,
upon using the above relations only once.%
\footnote{Note that the Feynman-parameter integral of $D_{00}$ is not
  logarithmic, so that the calculation of $D_{i_1 i_2 i_3}$ from
  $D_{0000}$ and $D_{000000}$ requires the use of the recurrence
  relations twice.}  If $\det(\Ymod^{(N-1)})$ is not small, we prefer
to deduce all tensor coefficients from one basis integral (e.g.,
$D_{000000}$ for $D_{i_1 i_2 i_3}$), because no instabilities
accumulate and the recursion preserves relations among the tensor
coefficients, which are less accurately valid if several coefficients
are calculated numerically.

If $\det(\Ymod^{(N-1)})=0$, the described procedure is not applicable.
For instance, this is the case for 3-point functions that are either
soft or collinear singular. Such cases are much simpler than the case
with general kinematics, so that they can be treated more directly.
For processes with light external fermions only, $\det(\Ymod^{(N-1)})$
is zero only for 3-point functions ($N=3$) where a photon or a gluon
is attached to an external fermion.  A fully analytic treatment of
these cases, which admits a numerically stable evaluation, is
described in \refapp{app:Csing}; this method can be extended to
similar cases that appear in other processes.

Finally, we remark that the method of this section is somewhat related
to the fully numerical procedure advocated in
\citere{Ferroglia:2002mz}. There, a method is described how the
Feynman-parameter representation of one-loop integrals is, upon
partial integrations, transformed into integrals with logarithmic
integrands, which are then treated numerically.  The occurring
algebraic coefficients that express the original integral in terms of
logarithmic integrals are related to the coefficients of the inverse
of the matrix $\Ymod^{(N)}$ introduced in this paper. In fact we have
verified that the reduction of the scalar integral $C_0$ to
logarithmic integrals leads to the same results as our equation
\refeq{eq:D0_cayley} for $N=3$ [see \refeq{eq:N3subst}].  Therefore,
like in our approach, also in the approach of
\citere{Ferroglia:2002mz}, the cases with small or vanishing modified
Cayley determinant $\det(\Ymod^{(N-1)})$ require a special treatment.
Moreover, we emphasize that we treat only one basis integral
numerically, while the procedure of \citere{Ferroglia:2002mz} in
general involves more numerical integrals.

\subsection{Reduction for small Gram determinant}
\label{se:gram}

Let us now derive a reduction scheme that can be used if the Gram
determinant $\Delta^{(3)}$ becomes small, but without changing the set
of basis integrals, which are thus still the standard scalar integrals
$A_0$, $B_0$, $C_0$, $D_0$. Multiplying \refeq{recurDp} with indices
$ni_1\ldots i_P$ by $\Zadj^{(3)}_{jn}$ and summing over $n$ yields
\beqar\label{recurGram1}
\Ymodadj^{(3)}_{0j} D_{i_1\ldots i_P} 
&=& 
-\sum_{n=1}^{N-1} \Zadj^{(3)}_{jn} 
\left(\Shat^{P+1}_{ni_1\ldots i_P} -  2 \sum_{r=1}^P
\de_{ni_r} D_{00i_1\ldots \hat i_r\ldots i_P}\right)
+\Delta^{(3)}  D_{ji_1\ldots i_P} 
\eeqar
for arbitrary $j=1,\ldots,N-1$ and $i_r=0,\ldots,N-1$.  In order to
arrive at this form, \refeq{columnandrowexp} and \refeq{Ymodadj} have
been used.  As long as at least one of the quantities
$\Ymodadj^{(3)}_{0j}$, defined in \refeq{Ymodadj}, is large compared
to $\De^{(3)}$, \refeq{recurGram1} can be used to determine
$D_{i_1\ldots i_P}$ from $D_{00i_1\ldots \hat i_r\ldots i_P}$ up to
terms that are suppressed by the factor $\Delta^{(3)}$.

In order to obtain $D_{00i_1\ldots \hat i_r\ldots i_P}$, we consider
for arbitrary $k,l\ne0$
\beqar
\Delta^{(3)}  D_{kli_3\ldots i_P} &=&
\sum_{i,j=1}^{N-1} \Delta^{(3)} \de_{ki}\de_{lj} D_{iji_3\ldots i_P} 
\nn\\
&=& \sum_{i,j=1}^{N-1} \left( \Zadj^{(3)}_{kl}Z^{(3)}_{ij}
+\sum_{n,m=1}^{N-1} \Zadjadj^{(3)}_{(kn)(lm)} Z^{(3)}_{nj}Z^{(3)}_{im}
\right) D_{iji_3\ldots i_P},
\eeqar
where \refeq{Zadjadjrelation2} has been used.
The first term on the r.h.s.\ can be reduced with
\refeq{recurDg}, the second term on the r.h.s.\ upon using
\refeq{recurDp} twice.
Collecting terms containing $D_{00i_1\ldots i_P}$ and making use of
\refeq{Zadjadjrelation0a} and \refeq{Zadjadjrelation2}, we obtain
\beqar\label{recurGram2}
\lefteqn{
2\left(6+P-N+\sum_{r=1}^P\bar\de_{i_r0}\right)
\Zadj^{(3)}_{kl} D_{00i_1\ldots i_P}  = 
-2(D-4)\Zadj^{(3)}_{kl}D_{00i_1\ldots i_P} -\Delta^{(3)}
D_{kli_1\ldots i_P} }\qquad \nl
&&{} +\Zadj^{(3)}_{kl}  S^{P+2}_{00i_1\ldots i_P} 
+ \sum_{n=1}^{N-1} ( \Zadj^{(3)}_{nl}  \Shat^{P+2}_{nki_1\ldots i_P}
- \Zadj^{(3)}_{kl} \Shat^{P+2}_{nni_1\ldots i_P}) \nl
&&{} -
\sum_{n,m=1}^{N-1} \Zadjadj^{(3)}_{(kn)(lm)}\biggl[
f_n\Shat^{P+1}_{mi_1\ldots i_P} 
+ 2 \sum_{r=1}^P \de_{ni_r} \Shat^{P+2}_{m00i_1\ldots \hat i_r\ldots i_P}
-f_nf_m D_{i_1\ldots i_P} 
\nl&&{} \qquad{}
- 2 \sum_{r=1}^P (f_n\de_{mi_r}+f_m\de_{ni_r}) D_{00i_1\ldots \hat i_r\ldots i_P}
- 4 \sum_{r,s=1\atop r\ne s}^P \de_{ni_r}\de_{mi_s} D_{0000i_1\ldots
  \hat i_r\ldots \hat i_s\ldots i_P}\biggr],\qquad
\eeqar
which holds for arbitrary $k,l=1,\ldots, N-1$ and
$i_1,\ldots,i_P=0,\ldots, N-1$.  Together with \refeq{recurGram1} this
equation allows to iteratively determine the tensor coefficients of
4-point functions in terms of 3-point functions for small Gram
determinant $\Delta^{(3)}$.  If the 3-point functions are known up to
rank $P$, all 4-point tensor coefficients up to this rank can be
determined recursively up to terms of order $\Delta^{(3)}$ from these
equations by putting all terms involving $\Delta^{(3)}$ to zero.
Inserting these results back into the r.h.s.\ of \refeq{recurGram1}
and \refeq{recurGram2} for the terms proportional to $\Delta^{(3)}$,
all 4-point tensor coefficients up to rank $(P-1)$ can be determined
up to terms of order $(\Delta^{(3)})^2$, and so on.  Finally, the
scalar 4-point function is iteratively determined up to terms of order
$(\Delta^{(3)})^{P+1}$.  In order to improve numerical stability, we
can choose $j$ in \refeq{recurGram1} such that $\Ymod^{(3)}_{0j}$ is
maximal, and $k$ and $l$ in \refeq{recurGram2} such that
$\Zadj^{(3)}_{kl}$ is maximal.  For $\Delta^{(3)}=0$ this reduction
scheme essentially corresponds to the one proposed in
\citere{Devaraj:1997es}.

For the lowest tensor coefficients the explicit results read
\beqar
\lefteqn{
\Ymodadj^{(3)}_{0j} D_{0} = 
-\sum_{n=1}^{N-1} \Zadj^{(3)}_{jn} \Shat^{1}_{n}
+\Delta^{(3)}  D_{j},}\qquad
\\
\lefteqn{
2\left(6-N\right)
\Zadj^{(3)}_{kl} D_{00}  = 
-2(D-4)\Zadj^{(3)}_{kl}D_{00} 
-\Delta^{(3)} D_{kl} 
+\Zadj^{(3)}_{kl}  S^2_{00} 
}\qquad \nl && {}
+ \sum_{n=1}^{N-1} ( \Zadj^{(3)}_{nl}  \Shat^2_{nk}
- \Zadj^{(3)}_{kl} \Shat^2_{nn}) 
-\sum_{n,m=1}^{N-1} \Zadjadj^{(3)}_{(kn)(lm)}\biggl[
f_n\Shat^1_{m} -f_nf_m D_{0} 
\biggr],\qquad
\\
\lefteqn{
\Ymodadj^{(3)}_{0j} D_{i_1} =
-\sum_{n=1}^{N-1} \Zadj^{(3)}_{jn} 
\left(\Shat^{2}_{ni_1} 
-  2\de_{ni_1} D_{00}\right)
+\Delta^{(3)}  D_{ji_1},}\qquad
\\
\lefteqn{2\left(8-N\right)
\Zadj^{(3)}_{kl} D_{00i_1}  = 
-2(D-4)\Zadj^{(3)}_{kl}D_{00i_1} 
-\Delta^{(3)} D_{kli_1} 
+ \Zadj^{(3)}_{kl}  S^3_{00i_1} 
}\qquad \nl &&{}
+ \sum_{n=1}^{N-1} ( \Zadj^{(3)}_{nl}  \Shat^3_{nki_1}
- \Zadj^{(3)}_{kl} \Shat^3_{nni_1}) 
-\sum_{n,m=1}^{N-1} \Zadjadj^{(3)}_{(kn)(lm)}\biggl[
f_n\Shat^2_{mi_1} 
+ 2 \de_{ni_1} \Shat^3_{m00}
 \nl && {}
-f_nf_m D_{i_1} 
- 2 (f_n\de_{mi_1}+f_m\de_{ni_1}) D_{00}
\biggr],\qquad
\\
\lefteqn{
\Ymodadj^{(3)}_{0j} D_{i_1i_2} =
-\sum_{n=1}^{N-1} \Zadj^{(3)}_{jn} 
\left[\Shat^{3}_{ni_1i_2} 
-  2(\de_{ni_1} D_{00i_2}+\de_{ni_2} D_{00i_1})\right]
+\Delta^{(3)}  D_{ji_1i_2},}\qquad
\\
\lefteqn{2\left(8-N\right)
\Zadj^{(3)}_{kl} D_{0000}  = 
-2(D-4)\Zadj^{(3)}_{kl}D_{0000} 
  -\Delta^{(3)} D_{00kl}
+ \Zadj^{(3)}_{kl}  S^4_{0000} 
}\qquad\nl&&{} 
+ \sum_{n=1}^{N-1} ( \Zadj^{(3)}_{nl}  \Shat^4_{n00k}
- \Zadj^{(3)}_{kl} \Shat^4_{n00n}) 
-\sum_{n,m=1}^{N-1} \Zadjadj^{(3)}_{(kn)(lm)}\biggl[
f_n\Shat^3_{m00} -f_nf_m D_{00} \biggr],
\\
\lefteqn{2\left(10-N\right)
\Zadj^{(3)}_{kl} D_{00i_1i_2}  = 
-2(D-4)\Zadj^{(3)}_{kl}D_{00i_1i_2} 
-\Delta^{(3)} D_{kli_1i_2} 
+ \Zadj^{(3)}_{kl}  S^4_{00i_1i_2} 
}\qquad \nl && {}
+ \sum_{n=1}^{N-1} ( \Zadj^{(3)}_{nl}  \Shat^4_{nki_1i_2}
- \Zadj^{(3)}_{kl} \Shat^4_{nni_1i_2}) 
\nl && {}
-\sum_{n,m=1}^{N-1} \Zadjadj^{(3)}_{(kn)(lm)}\biggl[
f_n\Shat^3_{mi_1i_2} 
+ 2 (\de_{ni_1} \Shat^4_{m00i_2}+\de_{ni_2} \Shat^4_{m00i_1})
-f_nf_m D_{i_1i_2} 
\nl&&{}
- 2 (f_n\de_{mi_1}+f_m\de_{ni_1}) D_{00i_2}
- 2 (f_n\de_{mi_2}+f_m\de_{ni_2}) D_{00i_1}
\nl&&{}
- 4 (\de_{ni_1}\de_{mi_2}+ \de_{ni_2}\de_{mi_1}) D_{0000}\biggr],\quad
i_1,i_2\ne0,
\\
\lefteqn{
\Ymodadj^{(3)}_{0j} D_{i_1i_2i_3} =
-\sum_{n=1}^{N-1} \Zadj^{(3)}_{jn} 
\left[\Shat^{4}_{ni_1i_2i_3} 
-  2(\de_{ni_1} D_{00i_2i_3}+\de_{ni_2} D_{00i_1i_3}
+\de_{ni_3} D_{00i_1i_2})\right]}\qquad
\nl&&{}
+\Delta^{(3)}  D_{ji_1i_2i_3}. 
\label{D0000smallgram}
\eeqar

Figure~\ref{fig:Gramexp} illustrates a systematic algorithm for this
iteration scheme for 4-point integrals in the plane of tensor
coefficients similar to the previous sections.
\bfi
\centerline{\unitlength 1.0pt
\begin{picture}(140,170)(80,10)%
\SetScale{1.0}
\put(85,150){\small\fbox{$D_0$}}%
\put(85,110){\small $D_{\ina}$}%
\put(85, 70){\small $D_{\ina\inb}$}%
\put(165, 70){\small $D_{00}$}%
\put(85, 30){\small $D_{\ina\inb\inc}$}%
\put(165, 30){\small $D_{00\ina}$}%
\put(95, 10){\small \vdots}%
\put(175, 10){\small \vdots}%
\put(125,170){\small\bf Step 0}%
\Line(205,10)(205,170)
\SetWidth{0.5}
\LongArrow(92,124)(92,140)%
\end{picture}%
\begin{picture}(140,170)(80,10)%
\SetScale{1.0}
\put(85,150){\small $D_0$}%
\put(85,110){\small\fbox{$D_{\ina}$}}%
\put(85, 70){\small $D_{\ina\inb}$}%
\put(165, 70){\small\fbox{$D_{00}$}}%
\put(85, 30){\small $D_{\ina\inb\inc}$}%
\put(165, 30){\small $D_{00\ina}$}%
\put(95, 10){\small \vdots}%
\put(175, 10){\small \vdots}%
\put(125,170){\small\bf Step 1}%
\Line(205,10)(205,170)
\put(125, 73){\small $a$}%
\put(125,125){\small $a$}%
\put(125, 83){\small $b$}%
\put( 95, 88){\small $b$}%
\SetWidth{1.2}\LongArrow(107,143)(158, 80)
\SetWidth{0.5}\LongArrow(110, 70)(158, 70)
\SetWidth{1.2}\LongArrow(158, 75)(112,105)
\SetWidth{0.5}\LongArrow(92, 84)(92,100)
\end{picture}%
\begin{picture}(140,170)(80,10)%
\SetScale{1.0}
\put(85,150){\small $D_0$}%
\put(85,110){\small $D_{\ina}$}%
\put(85, 70){\small\fbox{$D_{\ina\inb}$}}%
\put(165, 70){\small $D_{00}$}%
\put(85, 30){\small $D_{\ina\inb\inc}$}%
\put(165, 30){\small\fbox{$D_{00\ina}$}}%
\put(95, 10){\small \vdots}%
\put(175, 10){\small \vdots}%
\put(125,170){\small\bf Step 2}%
%
\put(125, 33){\small $a$}%
\put(125, 85){\small $a$}%
\put(129, 43){\small $b$}%
\put( 95, 48){\small $b$}%
\put(178, 51){\small $a$}%
\SetWidth{1.2}\LongArrow(107,103)(158, 40)
\SetWidth{0.5}\LongArrow(118, 30)(158, 30)
\SetWidth{1.2}\LongArrow(158, 35)(120, 65)
\SetWidth{0.5}\LongArrow(92, 44)(92, 60)
\SetWidth{1.2}\LongArrow(174, 62)(174, 47)
\end{picture}%
}
\vspace{1em}
\centerline{\unitlength 1.0pt
\begin{picture}(220,170)(70,10)%
\SetScale{1.0}
\put(85,150){\small $D_0$}%
\put(85,130){\small $D_{\ina}$}%
\put(85,110){\small $D_{\ina\inb}$}%
\put(165,110){\small $D_{00}$}%
\put(85, 70){\small\fbox{$D_{\ina\inb\inc}$}}%
\put(165, 70){\small $D_{00\ina}$}%
\put(85, 30){\small $D_{\ina\inb\inc\ind}$}%
\put(165, 30){\small\fbox{$D_{00\ina\inb}$}}%
\put(245, 30){\small\fbox{$D_{0000}$}}%
\put(95, 10){\small \vdots}%
\put(175, 10){\small \vdots}%
\put(255, 10){\small \vdots}%
\put(160,160){\small\bf Step 3}%
\put(205, 85){\small $a$}%
\put(219, 39){\small $a$}%
\put(135, 19){\small $b$}%
\put(219, 19){\small $b$}%
\put(129, 85){\small $b$}%
\put(178, 51){\small $b$}%
\put(129, 45){\small $c$}%
\put( 95, 48){\small $c$}%
\SetWidth{1.2}\LongArrow(187,103)(238, 40)
\SetWidth{1.2}\LongArrow(110,107)(158, 40)
\SetWidth{0.5}\LongArrow(125, 30)(158, 30)
\SetWidth{1.2}\LongArrow(238, 30)(208, 30)
\SetWidth{0.5}\LongArrow(206, 35)(236, 35)
\SetWidth{1.2}\LongArrow(158, 35)(125, 61)
\SetWidth{0.5}\LongArrow(92, 44)(92, 60)
\SetWidth{1.2}\LongArrow(174, 62)(174, 47)
\end{picture}%
}
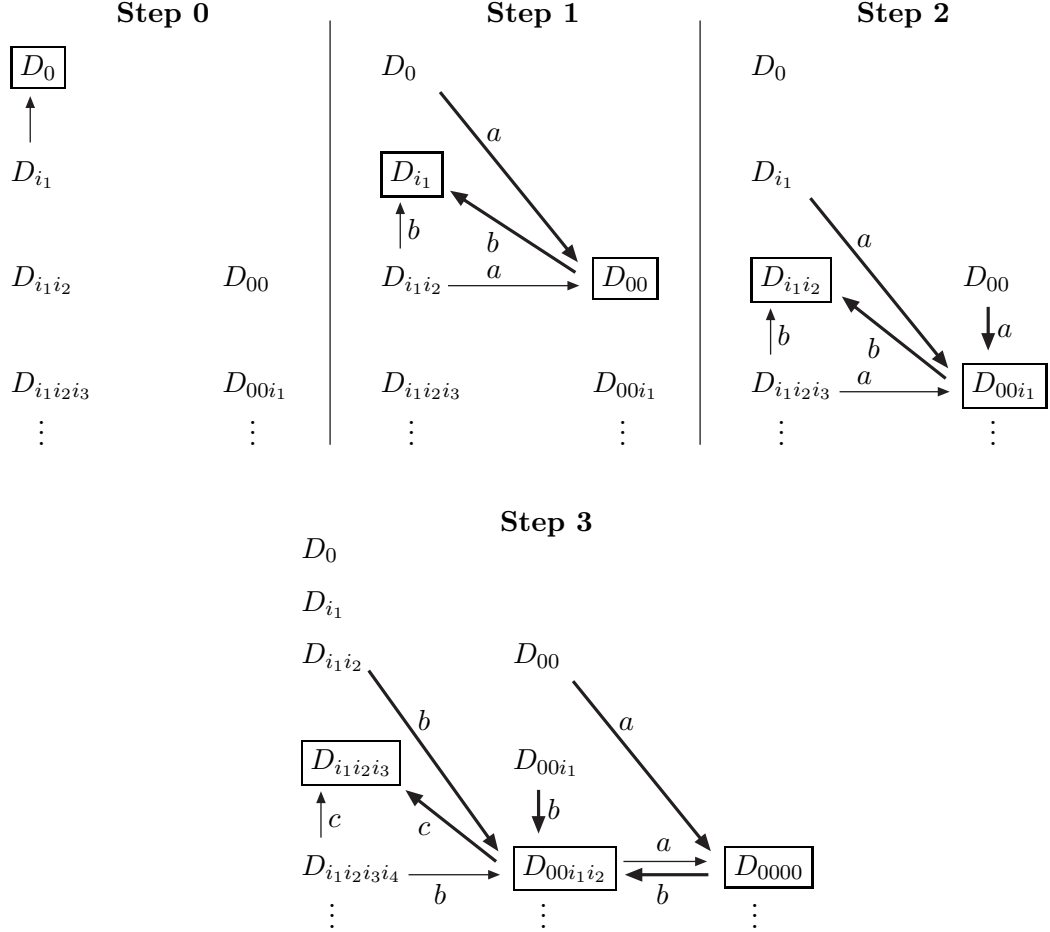
\caption{Schematic illustration of the iteration for small Gram
  determinants, where thin arrows indicate that the relation involves
  a suppression factor $\De^{(3)}$. In each step the boxed
  coefficients are calculated in the order indicated by the labels
  ``$a$'', ``$b$'', etc.  The $n$th iteration consists of the
  following $(n+1)$ steps: $n{\to}(n-1){\to}\dots{\to}1{\to}0$.}
\label{fig:Gramexp}
\efi
Thin arrows indicate that the relation involves a suppression factor
$\De^{(3)}$.  At the beginning of the iteration all 4-point tensor
coefficients as well as the scalar integral $D_0$ are set to zero, \ie
no 4-point basis integral is needed.  The $n$th iteration consists of
the $(n+1)$ steps $n{\to}\mbox{$(n-1)$}{\to}\ldots{\to}1{\to}0$ and
requires all 3-point coefficients of rank $n$.  Step $n$ with $n>0$
starts with the coefficient of rank $(n+1)$ with the highest number of
index pairs ``00'', i.e.\ with the right-most coefficient in the
$(n+1)$th row in the diagrams in \reffi{fig:Gramexp}.  Within a step,
coefficients for rank $(n+1)$ are deduced from the right to the left
in the diagram; only for the last coefficient (which has no index pair
``00'') one has to go one step upwards to rank $n$ in addition.  After
the $n$th iteration the tensor coefficients $D_{\ina\inb\inc\ldots}$
of rank $n$ without index pairs ``00'' and all coefficients
$D_{00\ina\inb\ldots}$ of one rank higher with at least one index pair
``00'' are obtained up to terms that are suppressed by a factor
$\De^{(3)}$.  Coefficients of a rank that is lower by a number $m$ are
known up to terms suppressed by $[\De^{(3)}]^{m+1}$.  Indicating
coefficients that are known up to terms of ${\cal
  O}\left([\De^{(3)}]^{m+1}\right)$ with a superscript ``$(m)$'', the
iteration proceeds as follows:
\begin{itemize}
\item Iteration 0: 
$D_0^{(0)}$ is calculated; all other coefficients are still zero.
\item Iteration 1: 
Step 1 yields $D_{00}^{(0)}$ and $D_{\ina}^{(0)}$; 
step 0 yields $D_0^{(1)}$.
\item Iteration 2: Steps 2 to 0 deliver $D_{00\ina}^{(0)}$,
  $D_{\ina\inb}^{(0)}$, $D_{00}^{(1)}$, $D_{\ina}^{(1)}$, and
  $D_0^{(2)}$.
\item Iteration 3:
Steps 3 to 0 deliver 
$D_{0000}^{(0)}$, $D_{00\ina\inb}^{(0)}$, $D_{\ina\inb\inc}^{(0)}$, 
$D_{00\ina}^{(1)}$, $D_{\ina\inb}^{(1)}$, 
$D_{00}^{(2)}$, $D_{\ina}^{(2)}$, and $D_0^{(3)}$.
\item etc.
\end{itemize}

The reduction method described in this section breaks down if none of
the $\Ymodadj^{(3)}_{0j}$ is large compared to $\De^{(3)}$ or if all
$\Zadj^{(3)}_{kl}$ become small, since in these cases the iteration
does not converge. A reduction for small $\Ymodadj^{(3)}_{0j}$ is
described in \refse{se:gramcayley}.  A reduction for small
$Z^{(3)}_{kl}$ is given in \refse{se:smallmomenta}.  For $N=2$ the
case of small $Z^{(2)}_{kl}$ is equivalent to small
$\Zadj^{(2)}_{kl}$; for $N=3$ the case of small $Z^{(3)}_{kl}$ covers
the case of small $\Zadj^{(3)}_{kl}$ apart from exceptional
configurations.%
\footnote{In an alternative approach, one could disregard
  \refeq{recurGram2} and  use \refeq{recurGram1} also to determine
$D_{00i_1\ldots \hat i_r\ldots i_P}$. This reduction method would also
work if all 
$\Zadj^{(3)}_{kl}$ are small. However, in this case, tensor
integrals of higher rank would be needed. For instance, to calculate
$D_{\ina\inb\inc}$ in leading order in $\De^{(3)}$ one would have to calculate
$D_{000000}$ and  $C_{000000}$ first.\label{fn-gram}}

\subsection{Reduction for small Gram determinant and small modified Cayley
  determinant}
\label{se:gramcayley}

If in addition to the Gram determinant $\De^{(3)}$ also all quantities
$\Ymodadj^{(3)}_{0j}$, $j=1,\ldots, N-1$, become small, the reduction
scheme of \refse{se:gram} breaks down.  As can be seen from
\refeq{eq:Ymodadjadj_rels}, in this case the determinant
$\det(\Ymodadj^{(3)})=\det(Y)$ of \refeq{detCayley} becomes small,
which is a necessary condition for the appearance of leading Landau
singularities. In this situation, we can determine the tensor
coefficients as follows.

For $i_r\ne0$, equation \refeq{recurGram1} can be rewritten as
\beqar\label{recurGramCayley1}
2 \sum_{r=1}^P \Zadj^{(3)}_{ki_r} 
  D_{00i_1\ldots \hat i_r\ldots i_P}
&=& 
\sum_{n=1}^{N-1} \Zadj^{(3)}_{kn} \Shat^{P+1}_{ni_1\ldots i_P} 
+\Ymodadj_{k0}^{(3)} D_{i_1\ldots i_P}
-\Delta^{(3)}  D_{ki_1\ldots i_P}.
\eeqar
This allows to determine $D_{00i_1\ldots \hat i_r\ldots i_P}$ for
$i_1,\ldots,i_P\ne0$ in terms of 3-point functions as:
\beqar\label{recurGramCayley1a}
2P \Zadj^{(3)}_{kl}  D_{00\indexunderbrace{\sst l\ldots l}_{P-1}}
&=& \sum_{n=1}^{N-1}\Zadj^{(3)}_{kn} \Shat^{P+1}_{n\underbrace{\sst l\ldots l}_{P}} 
+\Ymodadj_{k0}^{(3)}
D_{\underbrace{\sst l\ldots l}_{P}} - \Delta^{(3)}
D_{k\underbrace{\sst l\ldots l}_{P}}, \nl
2(P-1) \Zadj^{(3)}_{kl}  D_{00\underbrace{\sst l\ldots l}_{P-2}\!i_1}
&=&  -2 \Zadj^{(3)}_{ki_1}  D_{00\underbrace{\sst l\ldots
    l}_{P-1}}
+\sum_{n=1}^{N-1}\Zadj^{(3)}_{kn} \Shat^{P+1}_{n\underbrace{\sst l\ldots
    l}_{P-1}\!i_1} 
\nl&&{}
+\Ymodadj_{k0}^{(3)}
D_{\underbrace{\sst l\ldots l}_{P-1}\!i_1} - \Delta^{(3)}
D_{k\underbrace{\sst l\ldots l}_{P-1}\!i_1},
\quad i_1\ne 0,l,
 \nl
2(P-2) \Zadj^{(3)}_{kl}  D_{00\underbrace{\sst l\ldots l}_{P-3}\!i_1i_2}
&=&  -2 \Zadj^{(3)}_{ki_1}  D_{00\underbrace{\sst l\ldots
    l}_{P-2}\!i_2}
-2 \Zadj^{(3)}_{ki_2}  D_{00\underbrace{\sst l\ldots
    l}_{P-2}\!i_1}
+\sum_{n=1}^{N-1}\Zadj^{(3)}_{kn} \Shat^{P+1}_{n\underbrace{\sst l\ldots l}_{P-2}\!i_1i_2} 
\nl&&{}
+\Ymodadj_{k0}^{(3)}
D_{\underbrace{\sst l\ldots l}_{P-2}\!i_1i_2} - \Delta^{(3)}
D_{k\underbrace{\sst l\ldots l}_{P-2}\!i_1i_2}, 
\quad i_1,i_2\ne 0,l,
\hspace{4em}
\eeqar
and so on, provided that at least one of the $\Zadj^{(3)}_{kl}$ is not
small.  Again $k\ne0$ and $l\ne0$ can be chosen such that
$\Zadj^{(3)}_{kl}$ is maximal in order to improve the numerical
stability. The tensor coefficients with more index pairs ``00'' can be
determined by equations that are obtained from
\refeq{recurGramCayley1a} by adding additional index pairs ``00'' to
all quantities $S$ and $D$ in \refeq{recurGramCayley1a}.

In order to derive a relation for the calculation of $D_{i_1\ldots i_P}$
we rewrite \refeq{PValt} as
\beqar\label{smallGramCayley}
\lefteqn{
\left(\barr{ccc}
f_1 & f_2 & f_3 \\
2p_1p_1   & 2p_1p_2 & 2p_1p_{3} \\
2p_{2}p_1 & 2p_2p_2 & 2p_2p_{3}  \\
2p_{3}p_1 & 2p_3p_2 & 2p_{3}p_{3} \\
\earr\right) 
\left(\barr{c}
D_{1i_1\ldots i_P} \\
D_{2i_1\ldots i_P} \\
D_{3i_1\ldots i_P} \\
\earr\right) 
}\qquad \nl&&
=\left(\barr{cccc}
2(D+1+P-N) D_{00i_1\ldots i_P} -  C_{i_1\ldots i_P}(0)
- 2m_0^2 D_{i_1\ldots i_P}\\ 
\Shat^{P+1}_{1i_1\ldots i_P} -  2 \sum_{r=1}^P
\de_{1i_r} D_{00i_1\ldots \hat i_r\ldots i_P} 
- f_1 D_{i_1\ldots i_P} \\
\Shat^{P+1}_{2i_1\ldots i_P} -  2 \sum_{r=1}^P
\de_{2i_r} D_{00i_1\ldots \hat i_r\ldots i_P}  
- f_2 D_{i_1\ldots i_P}\\
\Shat^{P+1}_{3i_1\ldots i_P} -   2 \sum_{r=1}^P
\de_{3i_r} D_{00i_1\ldots \hat i_r\ldots i_P}  
- f_3 D_{i_1\ldots i_P}\\
\earr\right).
\eeqar
After discarding the $(j+1)$th of these equations, where $j=1,2$, or 3, 
the remaining three equations have the solution
\beqar
\lefteqn{
\sum_{n=1}^{N-1} f_n \Zadj^{(3)}_{nj} D_{i i_1\ldots i_P}
=  \Zadj^{(3)}_{ij} \left[2(D+1+P-N) D_{00i_1\ldots i_P} 
-  C_{i_1\ldots i_P}(0)
- 2m_0^2 D_{i_1\ldots i_P}\right]}\qquad \nl
&&{}+ \sum_{m,n=1}^{N-1}\Zadjadj^{(3)}_{(in)(jm)}f_n
\left[\Shat^{P+1}_{mi_1\ldots i_P} -  2 \sum_{r=1}^P
\de_{mi_r} D_{00i_1\ldots \hat i_r\ldots i_P} 
- f_m D_{i_1\ldots i_P}\right]. \qquad\qquad\qquad
\eeqar
Using \refeq{Ymodadj} this can be written as 
\beqar\label{recurGramCayley2}
\lefteqn{
\Ymodadj^{(3)}_{ij}
D_{i_1\ldots i_P}
= \Zadj^{(3)}_{ij} \left[2(5+P-N) D_{00i_1\ldots i_P} 
+ 2(D-4) D_{00i_1\ldots i_P} -
  C_{i_1\ldots i_P}(0)\right]  }\qquad\nl
&&{}
+   \sum_{m,n=1}^{N-1}\Zadjadj^{(3)}_{(in)(jm)}f_n
\left[\Shat^{P+1}_{mi_1\ldots i_P} -  2 \sum_{r=1}^P
\de_{mi_r} D_{00i_1\ldots \hat i_r\ldots i_P} \right] 
+\Ymodadj^{(3)}_{0j}
D_{i i_1\ldots i_P},
\eeqar
which holds for arbitrary $i,j=1,\ldots, N-1$ and 
 $i_1,\ldots,i_P=0,\ldots, N-1$.
Together with \refeq{recurGramCayley1} this equation allows to
iteratively determine the tensor coefficients of 4-point functions in
terms of 3-point functions for small Gram determinant $\Delta^{(3)}$
and small $\Ymodadj^{(3)}_{k0}$ and $\Ymodadj^{(3)}_{0j}$ as long as
at least one of the
$\Ymodadj^{(3)}_{ij}$ is not small.  Again $i$ and $j$ can be chosen
suitably in order to improve the numerical accuracy, \eg by choosing
the maximal $\Ymodadj^{(3)}_{ij}$.  If the 3-point functions are known
up to rank $P$, all 4-point tensor coefficients up to rank $(P-1)$ can
be determined up to terms of order $\Delta^{(3)}$,
$\Ymodadj^{(3)}_{k0}$, and $\Ymodadj^{(3)}_{0j}$ from
\refeq{recurGramCayley1} and \refeq{recurGramCayley2} by putting all
terms proportional to these quantities to zero.  Inserting these
results back into the r.h.s.\ of these equations, all 4-point tensor
coefficients up to rank $(P-3)$ can be determined up to terms of order
$[\max(|\Delta^{(3)}|,|\Ymodadj^{(3)}_{k0}|,|\Ymodadj^{(3)}_{0j}|)]^2$,
and so on. Finally, the scalar 4-point function is determined up to
terms of order
$[\max(|\Delta^{(3)}|,|\Ymodadj^{(3)}_{k0}|,|\Ymodadj^{(3)}_{0j}|)]^{[(P+1)/2]}$.

For the tensor coefficients up to rank 3 the reduction formulas explicitly
read
\beqar\label{recurGramCayley2a}
 2 \Zadj^{(3)}_{kl}  D_{00}
&=& \sum_{n=1}^{N-1} \Zadj^{(3)}_{kn} \Shat^{2}_{nl} 
+\Ymodadj_{k0}^{(3)}
D_{l} - \Delta^{(3)}D_{kl},  
\\
%
\Ymodadj^{(3)}_{ij} D_{0}
&=& \Zadj^{(3)}_{ij} \left[2(5-N) D_{00} + 2(D-4) D_{00} -  C_{0}(0)\right] 
\nl&&
+   \sum_{m,n=1}^{N-1}\Zadjadj^{(3)}_{(in)(jm)}f_n\Shat^{1}_{m} 
+\Ymodadj^{(3)}_{0j} D_{i}, \\
4 \Zadj^{(3)}_{kl}  D_{00l}
&=& \sum_{n=1}^{N-1}\Zadj^{(3)}_{kn} \Shat^{3}_{nll} 
+\Ymodadj_{k0}^{(3)}
D_{ll} - \Delta^{(3)}D_{kll}, 
\nl
2 \Zadj^{(3)}_{kl}  D_{00i_1}
&=&  -2 \Zadj^{(3)}_{ki_1}  D_{00l}
+\sum_{n=1}^{N-1}\Zadj^{(3)}_{kn} \Shat^{3}_{nli_1} 
+\Ymodadj_{k0}^{(3)}
D_{li_1} - \Delta^{(3)}D_{kli_1}, \qquad i_1\ne 0,l, 
\\
%
\Ymodadj^{(3)}_{ij} D_{i_1}
&=& \Zadj^{(3)}_{ij} \left[2(6-N) D_{00i_1} + 2(D-4) D_{00i_1} -
  C_{i_1}(0)\right]  
\nl
&&{}
+   \sum_{m,n=1}^{N-1}\Zadjadj^{(3)}_{(in)(jm)}f_n
\left[\Shat^{2}_{mi_1} - 2\de_{mi_1} D_{00}  
\right] 
+\Ymodadj^{(3)}_{0j} D_{i i_1},\\
6 \Zadj^{(3)}_{kl}  D_{00ll}
&=& \sum_{n=1}^{N-1}\Zadj^{(3)}_{kn} \Shat^{4}_{nlll} 
+\Ymodadj_{k0}^{(3)}
D_{lll} - \Delta^{(3)}D_{klll}, 
\nl
4 \Zadj^{(3)}_{kl}  D_{00li_1}
&=&  -2 \Zadj^{(3)}_{ki_1}  D_{00l l}
+\sum_{n=1}^{N-1}\Zadj^{(3)}_{kn} \Shat^{4}_{nlli_1} 
+\Ymodadj_{k0}^{(3)}
D_{lli_1} - \Delta^{(3)}D_{klli_1},   \quad i_1\ne0,l,
\nl
2 \Zadj^{(3)}_{kl}  D_{00i_1i_2}
&=&  -2 \Zadj^{(3)}_{ki_1}  D_{00li_2}
-2 \Zadj^{(3)}_{ki_2}  D_{00li_1}
\nl&&{}
+\sum_{n=1}^{N-1} \Zadj^{(3)}_{kn} \Shat^{4}_{nli_1i_2} 
+\Ymodadj_{k0}^{(3)}
D_{li_1i_2} - \Delta^{(3)} D_{kli_1i_2},   \quad i_1,i_2\ne0,l,
\\
\Ymodadj^{(3)}_{ij} D_{i_1i_2}
&=& \Zadj^{(3)}_{ij} \left[2(7-N) D_{00i_1i_2} + 2(D-4) D_{00i_1i_2} -
  C_{i_1i_2}(0)\right]  
\nl
&&{}
+   \sum_{m,n=1}^{N-1}\Zadjadj^{(3)}_{(in)(jm)}f_n
\left[\Shat^{3}_{mi_1i_2} - 2\de_{mi_1} D_{00i_2} 
- 2\de_{mi_2} D_{00i_1} 
\right] 
+\Ymodadj^{(3)}_{0j} D_{i i_1i_2},\\
8 \Zadj^{(3)}_{kl}  D_{00lll}
&=& \sum_{n=1}^{N-1}\Zadj^{(3)}_{kn} \Shat^{5}_{nllll} 
+\Ymodadj_{k0}^{(3)}
D_{llll} - \Delta^{(3)}D_{kllll}, 
\nl
6 \Zadj^{(3)}_{kl}  D_{00lli_1}
&=& -2 \Zadj^{(3)}_{ki_1}  D_{00lll}
+\sum_{n=1}^{N-1}\Zadj^{(3)}_{kn} \Shat^{5}_{nllli_1} 
+\Ymodadj_{k0}^{(3)}
D_{llli_1} - \Delta^{(3)}D_{kllli_1}, \quad i_1\ne0,l, 
\nl
4 \Zadj^{(3)}_{kl}  D_{00li_1i_2}
&=&  -2 \Zadj^{(3)}_{ki_1}  D_{00lli_2}-2 \Zadj^{(3)}_{ki_2}  D_{00lli_1}
\nl&&{}
+\sum_{n=1}^{N-1}\Zadj^{(3)}_{kn} \Shat^{5}_{nlli_1i_2} 
+\Ymodadj_{k0}^{(3)}
D_{lli_1i_2} - \Delta^{(3)}D_{klli_1i_2},   \quad i_1,i_2\ne0,l,
\nl
2 \Zadj^{(3)}_{kl}  D_{00i_1i_2i_3}
&=&  -2 \Zadj^{(3)}_{ki_1}  D_{00li_2i_3}
-2 \Zadj^{(3)}_{ki_2}  D_{00li_1i_3}
-2 \Zadj^{(3)}_{ki_3}  D_{00li_1i_2}
\nl&&{}
+\sum_{n=1}^{N-1} \Zadj^{(3)}_{kn} \Shat^{5}_{nli_1i_2i_3} 
+\Ymodadj_{k0}^{(3)}
D_{li_1i_2i_3} - \Delta^{(3)} D_{kli_1i_2i_3},   \quad i_1,i_2,i_3\ne0,l,
\\
%
\Ymodadj^{(3)}_{ij} D_{i_1i_2i_3}
&=& \Zadj^{(3)}_{ij} \left[2(8-N) D_{00i_1i_2i_3} + 2(D-4) D_{00i_1i_2i_3}
  - C_{i_1i_2i_3}(0)\right]  
\nl
&&{}
+   \sum_{m,n=1}^{N-1}\Zadjadj^{(3)}_{(in)(jm)}f_n
\left[\Shat^{4}_{mi_1i_2i_3} - 2\de_{mi_1} D_{00i_2i_3} 
- 2\de_{mi_2} D_{00i_1i_3} -  2\de_{mi_3} D_{00i_1i_2} 
\right] 
\nl&&{}
+\Ymodadj^{(3)}_{0j} D_{i i_1i_2i_3}.
\eeqar

Figure~\ref{fig:GramCayleyexp} illustrates a systematic algorithm for
the iteration scheme for 4-point integrals in the plane of tensor
coefficients similar to the previous sections.
\bfi
\centerline{\unitlength 1.0pt
\begin{picture}(240,245)(70,10)%
\SetScale{1.0}
\put( 85,230){\small\fbox{$D_0$}}%
\put( 85,190){\small\fbox{$D_{\ina}$}}%
\put( 85,150){\small $D_{\ina\inb}$}%
\put(165,150){\small\fbox{$D_{00}$}}%
\put( 85,110){\small $D_{\ina\inb\inc}$}%
\put(165,110){\small\fbox{$D_{00\ina}$}}%
\put( 85, 70){\small $D_{\ina\inb\inc\ind}$}%
\put(165, 70){\small $D_{00\ina\inb}$}%
\put(245, 70){\small $D_{0000}$}%
\put( 85, 30){\small $D_{\ina\inb\inc\ind\ine}$}%
\put(165, 30){\small $D_{00\ina\inb\inc}$}%
\put(245, 30){\small $D_{0000\ina}$}%
\put(95, 10){\small \vdots}%
\put(175, 10){\small \vdots}%
\put(255, 10){\small \vdots}%
\put(160,240){\small\bf Step 0}%
\put( 85,209){\small $b$}%
\put( 85,169){\small $b$}%
\put(111,206){\small $b$}%
\put(119,194){\small $b$}%
\put(111,166){\small $b$}%
\put(125,102){\small $a$}%
\put(125,125){\small $a$}%
\put(125,141){\small $a$}%
\put(141,155){\small $a$}%
\SetWidth{0.5}\LongArrow(92,204)(92,220)
\SetWidth{0.5}\LongArrow(92,164)(92,180)
\SetWidth{1.2}\LongArrow(160,125)(112,183)
\SetWidth{1.2}\LongArrow(160,165)(109,223)
\SetWidth{1.2}\LongArrow(160,158)(112,196)
\SetWidth{0.5}\LongArrow(110,190)(158,152)
\SetWidth{0.5}\LongArrow(110,145)(158,117)
\SetWidth{0.5}\LongArrow(110,150)(158,145)
\SetWidth{0.5}\LongArrow(119,110)(158,110)
\Line(295,10)(295,240)
\end{picture}%
\begin{picture}(220,245)(70,10)%
\SetScale{1.0}
\put( 85,230){\small $D_0$}%
\put( 85,190){\small $D_{\ina}$}%
\put( 85,150){\small\fbox{$D_{\ina\inb}$}}%
\put(165,150){\small $D_{00}$}%
\put( 85,110){\small\fbox{$D_{\ina\inb\inc}$}}%
\put(165,110){\small $D_{00\ina}$}%
\put( 85, 70){\small $D_{\ina\inb\inc\ind}$}%
\put(165, 70){\small\fbox{$D_{00\ina\inb}$}}%
\put(245, 70){\small\fbox{$D_{0000}$}}%
\put( 85, 30){\small $D_{\ina\inb\inc\ind\ine}$}%
\put(165, 30){\small\fbox{$D_{00\ina\inb\inc}$}}%
\put(245, 30){\small\fbox{$D_{0000i}$}}%
\put(95, 10){\small \vdots}%
\put(175, 10){\small \vdots}%
\put(255, 10){\small \vdots}%
\put(160,240){\small\bf Step 1}%
\put( 85,129){\small $b$}%
\put( 85, 89){\small $b$}%
\put(117,129){\small $b$}%
\put(139,139){\small $b$}%
\put(132,108){\small $b$}%
\put(125, 83){\small $b$}%
\put(135, 22){\small $a$}%
\put(132, 46){\small $a$}%
\put(132, 61){\small $a$}%
\put(144, 75){\small $a$}%
\put(220, 22){\small $c$}%
\put(220, 38){\small $c$}%
\put(220, 62){\small $c$}%
\put(220, 78){\small $c$}%
\SetWidth{0.5}\LongArrow( 92,124)( 92,140)
\SetWidth{0.5}\LongArrow( 92, 84)( 92,100)
\SetWidth{1.2}\LongArrow(160,118)(119,153)
\SetWidth{1.2}\LongArrow(160, 85)(119,146)
\SetWidth{1.2}\LongArrow(160, 45)(125,100)
\SetWidth{1.2}\LongArrow(160, 78)(125,113)
\SetWidth{0.5}\LongArrow(124,107)(158, 72)
\SetWidth{0.5}\LongArrow(124, 65)(158, 37)
\SetWidth{0.5}\LongArrow(123, 70)(158, 65)
\SetWidth{0.5}\LongArrow(130, 30)(158, 30)
\SetWidth{0.5}\LongArrow(190,106)(238, 77)
\SetWidth{0.5}\LongArrow(205, 60)(238, 37)
\SetWidth{0.5}\LongArrow(206, 70)(238, 70)
\SetWidth{0.5}\LongArrow(213, 30)(238, 30)
\end{picture}%
}
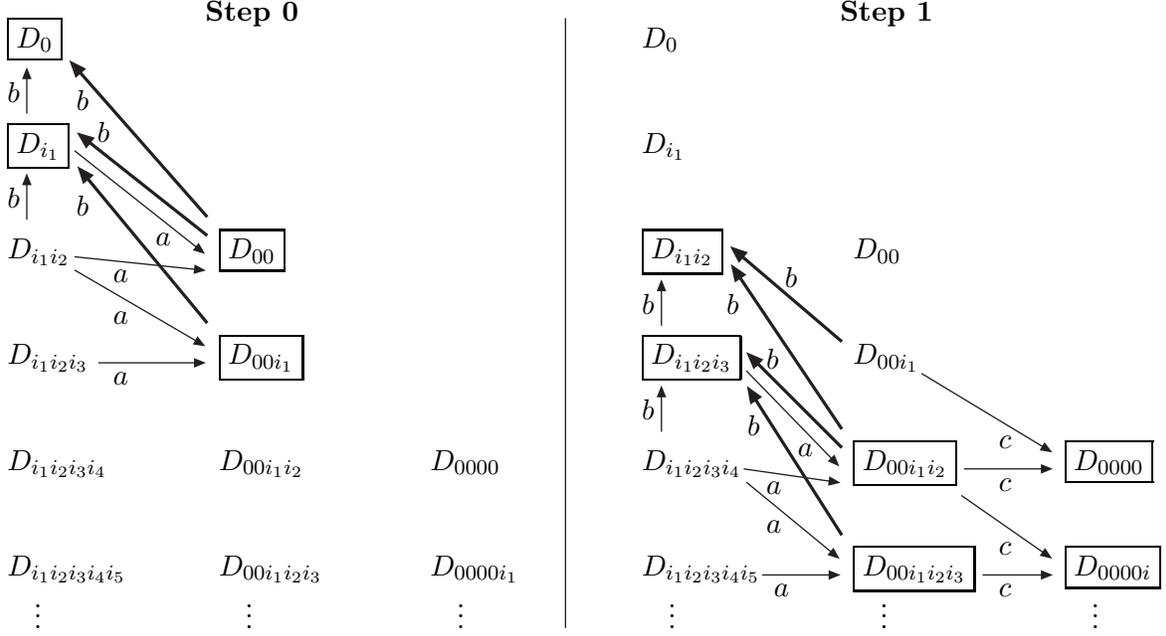
\caption{Schematic illustration of the iteration for small Gram
  and modified Cayley determinants, where thin arrows indicate that
  the relation involves a suppression factor $\De^{(3)}$,
  $\protect\Ymodadj^{(3)}_{k0}$, or $\protect\Ymodadj^{(3)}_{0j}$.  In
  each step the boxed coefficients are calculated in the order
  indicated by the labels ``$a$'', ``$b$'', etc.  The $n$th iteration
  consists of the following $(n+1)$ steps:
  $n{\to}(n-1){\to}\dots{\to}1{\to}0$.}
\label{fig:GramCayleyexp}
\efi
Thin arrows indicate that the relation involves a suppression factor
$\De^{(3)}$, $\Ymodadj^{(3)}_{k0}$, or $\Ymodadj^{(3)}_{0j}$.  At the
beginning of the iteration all 4-point tensor coefficients as well as
the scalar integral $D_0$ are set to zero, \ie no 4-point basis
integral is needed.  The $n$th iteration consists of the $(n+1)$ steps
$n{\to}(n-1){\to}\ldots{\to}1{\to}0$ and requires all 3-point
coefficient functions up to rank $2(n+1)$.  Step $n$ starts with the
two coefficients of ranks $(2n+2)$ and $(2n+3)$ that have exactly one
index pair ``00'', i.e.\ which belong to the second column in the
respective rows.  in the diagrams in \reffi{fig:GramCayleyexp}.
Within a step, first the two coefficients are calculated that are
reached upon omitting the index pair ``00'' from the starting
coefficients; they are located in the first column two rows above the
starting rows in the diagram.  Then all coefficients that lie to the
right of the starting coefficients are calculated column by column.
After the $n$th iteration the tensor coefficients
$D_{\ina\inb\inc\ldots}$ of ranks $2n$ and $(2n+1)$ without index
pairs ``00'' and all coefficients $D_{00\ina\inb\ldots}$ of two ranks
higher with at least one index pair ``00'' are obtained up to terms
that are suppressed by a factor $\De^{(3)}$, 
$\Ymodadj^{(3)}_{k0}$, or
$\Ymodadj^{(3)}_{0j}$.  Coefficients of a rank that is lower by a
number $2m$ are known up to terms suppressed by
$[\max(|\Delta^{(3)}|,|\Ymodadj^{(3)}_{k0}|,|\Ymodadj^{(3)}_{0j}|)]^{m+1}$.
The iteration proceeds as follows:
\begin{itemize}
\item Iteration 0: 
$D_{00}^{(0)}$, $D_{00\ina}^{(0)}$, $D_0^{(0)}$, and $D_{\ina}^{(0)}$ 
are calculated; all other coefficients are still zero.
\item Iteration 1: 
Step 1 yields 
$D_{00\ina\inb}^{(0)}$, $D_{00\ina\inb\inc}^{(0)}$, $D_{\ina\inb}^{(0)}$, 
$D_{\ina\inb\inc}^{(0)}$,
$D_{0000}^{(0)}$, and $D_{0000\ina}^{(0)}$, 
step 0 yields 
$D_{00}^{(1)}$, $D_{00\ina}^{(1)}$, $D_0^{(1)}$, and $D_{\ina}^{(1)}$.
\item etc.
\end{itemize}

The reduction described in this section breaks down if none of
$\Ymodadj^{(3)}_{ij}$ is large compared to $\De^{(3)}$ and
$\Ymodadj^{(3)}_{0j}$, or if all $\Zadj^{(3)}_{kl}$ become small,
since in these cases the iteration does not converge. A reduction for
small $Z^{(3)}_{kl}$, and thus for small $\Zadj^{(3)}_{kl}$ in
non-exceptional configurations, is described in
\refse{se:smallmomenta}.  If both $\Delta^{(3)}$ and all
$\Ymodadj^{(3)}_{k0}$
and $\Ymodadj^{(3)}_{ij}$
become small, in some cases the alternative Passarino--Veltman
reduction of \refse{se:PValt} works. In other cases, none of the
discussed reduction methods is really 
good. However, this happens only in
exceptional cases, and one of the discussed methods yields at least
crude results.%
\footnote{An alternative reduction could be derived, by considering
  $\sum_{r=1}^{P+1}\Ymodadj^{(3)}_{ii_r} D_{i_1\ldots \hat i_r\ldots
    i_{P+1}}$, using \refeq{recurGramCayley2} and inserting
  \refeq{recurGramCayley1} on the r.h.s.\ of the resulting equation to
  eliminate $D_{00i_1\ldots \hat i_r\ldots i_{P+1}}$. From the
  obtained relation, all tensor coefficients could be calculated.
  This reduction method would also work if all $\Zadj^{(3)}_{kl}$ are
  small. However, in this case, tensor integrals of higher rank would
  be needed. \label{fn-GramCayley}}

\subsection{Reduction for small momenta}
\label{se:smallmomenta}

Finally, we provide a reduction scheme for the case where all
$Z^{(3)}_{kl}$ and thus all momenta become small. Note that in
this case also all of the quantities $\Delta^{(3)}$, $\Zadj^{(3)}_{kl}$,
$\Ymodadj^{(3)}_{0k}$, and $\Ymodadj^{(3)}_{kl}$
become small.  If the $f_k$ are not small as well, we can proceed as
follows.  We rewrite \refeq{recurDp} as
\beqar\label{recurDpsmallmom}
 f_k D_{i_1\ldots i_P} &=&
\Shat^{P+1}_{ki_1\ldots i_P} 
- 2 \sum_{r=1}^P\de_{ki_r} D_{00i_1\ldots \hat i_r\ldots i_P}
-\sum_{m=1}^{N-1} Z^{(3)}_{km} D_{mi_1\ldots i_P},
\nl
&& k=1,\ldots,N-1, \qquad i_1,\ldots,i_P=0,\ldots, N-1,
\eeqar
and \refeq{recurDg} as
\beqar\label{recurDgsmallmom}
2 \Bigl(4+P + \sum_{r=1}^P \bar\de_{i_r 0} \Bigr)
D_{00i_1\ldots i_P} &=& -2(D-4) D_{00i_1\ldots i_P}
+2 C_{i_1\ldots i_P}(0) + 2 m_0^2 D_{i_1\ldots i_P}
\hspace{3em}
\\ &&{}
- \sum_{n,m=1}^{N-1}Z^{(3)}_{nm} D_{nmi_1\ldots i_P},
\quad i_1,\ldots,i_P=0,\ldots, N-1.
\nn
\eeqar
By using these equations iteratively, we can determine $D_{i_1\ldots
  i_P}$ and $D_{00i_1\ldots i_P}$ for given 3-point functions for
small $Z^{(3)}_{kl}$. If the 3-point functions are known up to rank
$P$, we can determine the coefficients of the 4-point functions with
rank $P$ up to terms of order $Z^{(3)}_{kl}$, those of rank $(P-1)$ up
to terms of order $[Z^{(3)}_{kl}]^2$, \ldots, and those of order $0$
up to terms of order $[Z^{(3)}_{kl}]^{P+1}$.  In order to improve
numerical stability, we can choose $k$ such that $f_k$ is maximal.
Note that the structure of \refeq{recurDpsmallmom} and
\refeq{recurDgsmallmom} is similar to the one of \refeq{recurGram1}
and \refeq{recurGram2}. 
In fact, a systematic algorithm for this iteration scheme for 4-point
integrals is given by \reffi{fig:Gramexp}, if the arrows that point
vertically downwards or horizontally to the left are omitted.

Up to tensor rank 3 the explicit formulas read:
\beqar\label{recurDpsmallmom2}
f_k D_{0} &=&
\Shat^{1}_{k} 
-\sum_{m=1}^{N-1} Z^{(3)}_{km} D_{m},\\
 8 D_{00} &=&
-2(D-4) D_{00}+
2 C_{0}(0) + 2 m_0^2 D_{0} 
- \sum_{n,m=1}^{N-1}Z^{(3)}_{nm} D_{nm},
\\ 
 f_k D_{i_1} &=&
\Shat^{2}_{ki_1} 
- 2 \de_{ki_1} D_{00}
-\sum_{m=1}^{N-1} Z^{(3)}_{km} D_{mi_1},\\
 12 D_{00i_1} &=&
-2(D-4) D_{00i_1}+
2 C_{i_1}(0) + 2 m_0^2 D_{i_1} 
- \sum_{n,m=1}^{N-1}Z^{(3)}_{nm} D_{nmi_1},
\\ 
 f_k D_{i_1i_2} &=&
\Shat^{3}_{ki_1i_2} 
- 2 \de_{ki_1} D_{00i_2}- 2 \de_{ki_2} D_{00i_1}
-\sum_{m=1}^{N-1} Z^{(3)}_{km} D_{mi_1i_2},\\
 16 D_{00i_1i_2} &=&
-2(D-4) D_{00i_1i_2}+
2 C_{i_1i_2}(0) + 2 m_0^2 D_{i_1i_2} 
- \sum_{n,m=1}^{N-1}Z^{(3)}_{nm} D_{nmi_1i_2},\\ 
 f_k D_{i_1i_2i_3} &=&
\Shat^{4}_{ki_1i_2i_3} 
- 2 \de_{ki_1} D_{00i_2i_3}- 2 \de_{ki_2} D_{00i_1i_3}
- 2 \de_{ki_3} D_{00i_1i_2}
\nl&&{}
-\sum_{m=1}^{N-1} Z^{(3)}_{km} D_{mi_1i_2i_3}.
\eeqar

If also all the $f_k$ become small we can rewrite
\refeq{recurDpsmallmom} and \refeq{recurDgsmallmom} as 
\beqar\label{recurDpsmallmomandf}
2 \sum_{r=1}^P\de_{ki_r} D_{00i_1\ldots \hat i_r\ldots i_P} &=&
\Shat^{P+1}_{ki_1\ldots i_P} 
-  f_k D_{i_1\ldots i_P}
-\sum_{m=1}^{N-1} Z^{(3)}_{km} D_{mi_1\ldots i_P},
\nl
&& k=1,\ldots,N-1, \qquad i_1,\ldots,i_P=0,\ldots, N-1,
\eeqar
and
\beqar\label{recurDgsmallmomandf}
2 m_0^2 D_{i_1\ldots i_P}  &=& 
2 \Bigl(4+P + \sum_{r=1}^P \bar\de_{i_r 0} \Bigr)
D_{00i_1\ldots i_P} 
 + 2 (D-4) D_{00i_1\ldots i_P}
\nl && {}
-2 C_{i_1\ldots i_P}(0) + \sum_{n,m=1}^{N-1} Z^{(3)}_{nm} D_{nmi_1\ldots i_P},
\quad i_1,\ldots,i_P=0,\ldots, N-1.
\hspace{2em}
\eeqar
By using these equations iteratively, we can determine $D_{i_1\ldots
  i_P}$ and $D_{00i_1\ldots i_P}$ for given 3-point functions for
small $Z^{(3)}_{kl}$ and small $f_k$. The structure of
\refeq{recurDpsmallmomandf} and \refeq{recurDgsmallmomandf} is similar
to the one of \refeq{recurGramCayley1} and \refeq{recurGramCayley2}.
If the 3-point functions are known up to rank $P$, we can determine
the coefficients of the 4-point functions with rank $(P-1)$ up to terms
of order $\max(|Z^{(3)}_{kl}|,|f_n|)$, those of rank $(P-3)$ up to terms
of order $[\max(|Z^{(3)}_{kl}|,|f_k|)]^2$, and so on.  Finally, the
scalar 4-point function is determined up to terms of order
$[\max(|Z^{(3)}_{kl}|,|f_{k}|)]^{[(P+1)/2]}$.

\subsection{Summary of reduction schemes and
application to {$\Pep\Pem\to4f$} at one loop}
\label{se:ee4f}

Table~\ref{tab:methods} briefly summarizes some of the features of the
described reduction schemes for 3- and 4-point tensor integrals.
\begin{table}
\centerline{
\begin{tabular}{c|cccccc}
Section & \ref{se:PV} & \ref{se:PValt} & \ref{se:Cayley} & \ref{se:gram} 
& \ref{se:gramcayley} & \ref{se:smallmomenta}
\\
\hline
method & PV & PV$'$ & Cayley & Gram & Gram/Cayley & momenta
\\[-.2em]
type & red. & red. & red. & exp. & exp. & exp.
\\
\hline
applicability & $|Z|{\ne}0$ & 
\parbox{1.3cm}{$|Z|{\ne}0 \\[-.0em] |X|{\ne}0$} \vphantom{\rule{0em}{1.8em}} &
$|X|{\ne}0$ & 
\parbox{1.4cm}{$|Z|{\to}0 \\[-.0em] \Ymodadj_{0j}{\ne}0\\[-.0em]  
               \Zadj_{kl}\ne0$ } \vphantom{\rule{0em}{1.8em}} &
\parbox{2.2cm}{$|Z|{\to}0 \\[-.0em]  \Ymodadj_{0k},\Ymodadj_{0j}{\to}0 \\[-.0em]  
               \Ymodadj_{ij}\ne0  \\[-.0em] \Zadj_{kl}\ne0$} 
\vphantom{\rule[-2.4em]{0em}{5.5em}} &
\parbox{1.3cm}{$Z{\to}0 \\[-.0em] f_k{\ne}0$}
\\
\hline
\parbox{1.8cm}{stable for \\[-.2em] $|Z|{\to0}$ ?} \vphantom{\rule{0em}{1.8em}} &
no & no & yes & yes & yes & yes 
\\
\parbox{1.8cm}{stable for \\[-.2em] $|X|{\to0}$ ?} \vphantom{\rule{0em}{1.8em}} &
yes & no & no &       & yes & yes 
\\
\parbox{1.8cm}{stable for \\[-.2em] $Z{\to0}$ ?} \vphantom{\rule{0em}{1.8em}} &
no & no & no & no & no & yes 
\\[.5em]
fast ? & yes & yes & no & yes & yes & yes
\end{tabular} }
\caption{Summary of features of the reduction schemes for 3- and
  4-point tensor integrals. 
The type of the method is either 
``reduction (red.)'' or ``expansion (exp.)'';
Gram and Cayley determinants are generically indicated by
$|Z|$ and $|X|$, respectively.}
\label{tab:methods}
\end{table}
The type of the method, ``reduction (red.)'' or ``expansion (exp.)''
is indicated in the third row.  In the fourth row we summarize the
conditions for the applicability of the schemes. Conditions that
depend on indices $i,j,k,l$ have to be fulfilled for at least one
choice of these indices.  The ``yes'' and ``no'' in the last rows
indicate whether a method is stable or unstable in the corresponding
limits or if the method is fast in terms of CPU time.  A blank entry 
means that the method can be stable or unstable in the
considered limit.

The reduction schemes described above have been successfully applied
in the calculation of the complete one-loop corrections to the
charged-current processes $\Pep\Pem\to4f$ as presented in
\citere{Denner:2005es}. As described there, actually two independent
calculations of the corrections have been carried out employing two
different procedures (called ``rescue systems'' there) for the
evaluation of the one-loop tensor integrals in the numerically
delicate kinematical configurations. Both procedures make use of the
conventional Passarino--Veltman reduction (see \refse{se:PV}) as long
as internal consistency checks prove this method to be reliable. If
this is not the case, the procedures differ:

\paragraph{Procedure 1: reduction with modified Cayley determinants
and further exception handling}

If conventional Passarino--Veltman reduction seems not to be trustworthy,
since consistency relations among the tensor coefficients are
valid only to very few digits or even violated,
the method with modified Cayley determinants is used as described in
\refse{se:Cayley}. Because of the vanishing modified Cayley determinant
this is not possible for the IR-singular 
(i.e.\ soft or collinear divergent)
3-point functions. Therefore,
these cases are evaluated as described in \refapp{app:Csing}, yielding
perfectly stable results. 

The described procedure fails if both the Gram and the modified Cayley
determinants are very small. In practice, this happens only at a small
fraction of events that hardly contribute to the $\Pep\Pem\to4f$ cross
section. However, if this limitation of the procedure becomes serious
in other cases, the double limit of small Gram and modified Cayley
determinants can be covered using the method of \refses{se:gramcayley}
etc., as it is done in Procedure~2.

\paragraph{Procedure 2: expansions for small Gram determinants etc.}

\begin{sloppypar}
The second procedure is based on the two versions of Passarino--Veltman 
reduction of \refses{se:PV} and \ref{se:PValt} and
on the expansion methods described
in \refses{se:gram}--\ref{se:smallmomenta}.
If the Passarino--Veltman reduction fails, at least the Gram determinant
of the corresponding integral (or an integral related to a subgraph)
is small. The question which of the different expansions is most appropriate 
is decided by estimating the number of valid digits in each of the
expansion variants; the variant promising the highest precision is
taken. 
\end{sloppypar}

For the application to the processes $\Pep\Pem\to4f$, it turned out to be
sufficient to implement the expansions for  small Gram determinant
(\refse{se:gram}), for  small Gram and modified Cayley determinants 
(\refse{se:gramcayley}), and for small momenta
(\refse{se:smallmomenta})  up to tensor rank~4 for 4-point functions and the
corresponding formulas for 3-point functions up to tensor rank~5.
The implementation of the modified procedure for small $f_k$ 
was not required. We also did not yet implement the schemes mentioned in
footnotes \ref{fn-gram} and \ref{fn-GramCayley}.

Note that in the one-loop diagrams for $\Pep\Pem\to4f$ 3- and 4-point
functions appear only up to rank~3, i.e.\ the implemented reductions
go beyond taking pure limits of vanishing determinants.  For these
processes, the exceptional cases where none of the expansions is good
appeared only for a very small fraction of events and did not yield
sizeable contributions to integrated physical quantities.

\subsection{UV and IR divergences in dimensional regularization
and terms of order $(D-4)$}
\label{se:divs}

In the preceding equations we have kept all terms of order $(D-4)$ that
multiply one-loop coefficient integrals. These terms give rise to
finite terms in dimensional regularization if these integrals are
divergent in four dimensions. It is convenient to discuss UV~divergences,
which formally result from loop momenta $q$ tending to infinity,
and IR~divergences, which arise from finite loop momenta but specific
kinematical configurations, separately:

\setcounter{paragraph}{0}
\paragraph{UV divergences}
UV~divergences are universal in the sense that the divergent terms
in an integral are regular functions of the external momenta $p_k$ and 
internal masses $m_k$, but these terms do not change if these kinematical 
quantities approach exceptional configurations (zero limits,
on-shell configurations, etc.). At one loop, UV~divergences 
generally have the form $1/(D-4)$ times a polynomial in
$p_k$ and $m_k$. Therefore, the terms $(D-4)T^N_{i_1\ldots i_P}$
contained in the above formulas are finite polynomials in
$p_k$ and $m_k$.
We have listed these $(D-4)T^N_{i_1\ldots i_P}$ terms for
1-point functions of arbitrary rank, 2-point functions up to rank~5,
3-point functions up to rank~7, 
4-point functions up to rank~7, and
five point functions up to rank~6 in \refapp{se:tendiv}. 

\paragraph{IR divergences}
IR~divergences at one loop originate either from soft or collinear
configurations of a loop momentum \cite{Kinoshita:ur}.  These type of
divergences have the property that they do not show up in tensor
coefficients with at least one index pair ``00'', i.e.\ all tensor
structures containing at least one factor of the metric tensor are
IR~finite.

These fact can be seen by inspecting the Feynman-parameter integrals
of the tensor coefficients or by the following arguments.  A soft
singularity results from the limit of zero-momentum transfer of a
massless particle ($q\to0$) between two on-shell particles.  Assuming
that the massless particle correspond to the propagator denominator
$N_0$, power counting in $q$ shows that soft divergences can appear
only in the scalar integral, but not in tensor integrals, because loop
momenta in the numerator render the limit $q\to0$ in the integral
non-singular.  Thus, in the general case, where the massless particle
corresponds to any propagator denominator $N_k$, soft divergent parts
of tensor integrals are always proportional to powers of the momenta
$p_k$, as can be seen by performing a shift $q\to q-p_k$, which maps
$N_k$ to $N_0$.  A collinear singularity results from the range where
the loop momentum $q$ is parallel to the momentum $p_k$ of a light
external on-shell particle that splits into two light particles. If a
tensor $q_{\mu_1}\dots q_{\mu_P}$ is present in the loop integral, the
divergence can only show up in covariants that are built up in the
singular region. Thus, collinear divergences of tensor integrals
appear in covariants containing only the momentum $p_k$.

In the reduction formulas given above the factor $(D-4)$ appears only
in front of tensor coefficients $T^N_{00i_3\ldots i_P}$ containing at
least one index pair ``00'', which have been shown to be IR~finite.
Therefore, all the reduction formulas are valid without modification
if IR singularities are regularized dimensionally.  All terms
$(D-4)T^N_{00i_3\ldots i_P}$ can be taken from \refapp{se:tendiv}; if
more of these terms are needed, they can be easily derived from the
reduction formulas themselves.

\section{Reduction of 5-point integrals}
\label{se:5pf}

In four space-time dimensions, 5-point integrals can be reduced to
4-point integrals. In \citere{Denner:2002ii} we have given relations
that express 5-point tensor integrals of rank $P$ by 4-point tensor
integrals of rank $P$ (see also \refapp{se:5pf-alt}).  This method
follows the strategy proposed in \citere{Me65} for the reduction of
scalar integrals and was actually used in the calculation of one-loop
corrections to $\Pep\Pem\to4f$ \cite{Denner:2005es}.  Here we derive
formulas that directly reduce 5-point tensor integrals of rank $P$ to
4-point tensor integrals of rank $(P-1)$. While similar results have
been presented in \citere{Binoth:2005ff}, our derivation is more
transparent.

We start by considering the determinant
{\arraycolsep 4pt
\beqar\label{Eredstart}
{\cal E} &=&
\left\vert
\barr{ccccc}
q^\mu       & -2 q^2 & 2qp_{1}     & \ldots & 2qp_{4} \\
0           & \,2m_0^2 & f_1         & \ldots & f_{4} \\
p_{1}^{\mu} & -2p_1q & 2p_{1}p_{1} & \ldots & 2p_{1}p_{4} \\
\vdots      & \vdots & \vdots      & \ddots & \vdots     \\
p_{4}^{\mu} & -2p_4q & 2p_{4}p_{1} & \ldots & 2p_{4}p_4
\earr\right\vert
\nl[2ex]
&=&{}
\left\vert
\barr{ccccc}
q^\mu       & -N_0-2m_0^2  & 2qp_{1}     & \ldots & 2qp_{4} \\
0           & 2m_0^2       & f_1         & \ldots & f_{4} \\
p_{1}^{\mu} & f_1          & 2p_{1}p_{1} & \ldots & 2p_{1}p_{4} \\
\vdots      & \vdots       & \vdots      & \ddots & \vdots     \\
p_{4}^{\mu} & f_4          & 2p_{4}p_{1} & \ldots & 2p_{4}p_4
\earr\right\vert
+
\left\vert
\barr{ccccc}
q^\mu       & -N_0    & 2qp_{1}     & \ldots & 2qp_{4} \\
0           &  0      & f_1         & \ldots & f_{4} \\
p_{1}^{\mu} & N_0-N_1 & 2p_{1}p_{1} & \ldots & 2p_{1}p_{4} \\
\vdots    & \vdots    & \vdots      & \ddots & \vdots    \\
p_{4}^{\mu} & N_0-N_4 & 2p_{4}p_{1} & \ldots & 2p_{4}p_4
\earr\right\vert
\nl[2ex]&=&{}
\left\vert
\barr{ccccc}
q^\mu       & -N_0   & N_1-N_0     & \ldots & N_4-N_0 \\
0           & 2m_0^2 & f_1         & \ldots & f_{4} \\
p_{1}^{\mu} & f_1    & 2p_{1}p_{1} & \ldots & 2p_{1}p_{4} \\
\vdots      & \vdots & \vdots      & \ddots & \vdots   \\
p_{4}^{\mu} & f_4    & 2p_{4}p_{1} & \ldots & 2p_{4}p_4
\earr\right\vert
+
\left\vert
\barr{ccccc}
g^\mu_\al   & -N_0    & 2p_{1,\al}  & \ldots & 2qp_{4,\al} \\
0           &  0       & f_1         & \ldots & f_{4} \\
p_{1}^{\mu} & q^\al (N_0-N_1) & 2p_{1}p_{1} & \ldots & 2p_{1}p_{4} \\
\vdots    & \vdots    & \vdots      & \ddots & \vdots    \\
p_{4}^{\mu} & q^\al (N_0-N_4) & 2p_{4}p_{1} & \ldots & 2p_{4}p_4
\earr\right\vert.
\nn\\
\eeqar}%
In the first manipulation, we have split the determinant in the second
column, and in the second we have added the second row of the first
determinant to its first row and we have moved $q^\al$ from the first
row to the second column in the second determinant. Moreover, we have
used the definitions \refeq{D0Di} and \refeq{def-fi}.

In four dimensions, the determinant ${\cal E}$ vanishes, as can be
seen from its defining form, because $q$ is linearly dependent on the
four momenta $p_i$, $i=1,\ldots,4$.
Since we want to derive a relation that also holds in dimensional
regularization we do not use this fact, but translate the integral
over ${\cal E}$ into a form that has a factor of ${\cal O}(D-4)$
rendering the whole contribution zero for finite integrals.
Inserting the first form of ${\cal E}$ in \refeq{Eredstart} into the
integrand of the tensor integral $E^{\mu_1\ldots\mu_P}$ results in
{\arraycolsep 4pt
\beqar\label{Eredtensor}
\int {\cal E} &\!\equiv\!&
\frac{(2\pi\mu)^{4-D}}{\ri\pi^{2}}\int \rd^{D}q\,
\frac{q^{\mu_{1}}\cdots q^{\mu_{P}}}
{N_0N_1\ldots N_4} \, {\cal E} 
\nn\\
&\!=\!&
2m_0^2E^{\al \mu_1\ldots\mu_P}\left\vert
\barr{ccccc}
g^\mu_\al   & 2p_{1,\al}  & \ldots & 2p_{4,\al} \\
p_{1}^{\mu} & 2p_{1}p_{1} & \ldots & 2p_{1}p_{4} \\
 \vdots     &\vdots       & \ddots & \vdots   \\
p_{4}^{\mu} & 2p_{4}p_{1} & \ldots & 2p_{4}p_4
\earr\right\vert + 
E^{\al\be\mu_1\ldots\mu_P}\left\vert
\barr{ccccc}
g^\mu_\al   & -2 g_{\al\be} & 2p_{1,\al}  & \ldots & 2p_{4,\al}  \\
0           &   0            & f_1         & \ldots & f_{4}       \\
p_{1}^{\mu} & -2p_{1,\be}   & 2p_{1}p_{1} & \ldots & 2p_{1}p_{4} \\
\vdots    & \vdots           & \vdots      & \ddots & \vdots      \\
p_{4}^{\mu} & -2p_{4,\be}   & 2p_{4}p_{1} & \ldots & 2p_{4}p_4
\earr\right\vert.
\nn\\
\eeqar
This form can be written more compactly by introducing the
four-dimensional metric tensor
\beq\label{4Dmetric}
g_{(4)}^{\mu\nu}= \sum_{j,k=1}^4 2p_j^{\mu}p_k^{\nu} (Z^{(4)})^{-1}_{kj}
=-\frac{1}{\De^{(4)}}\left\vert
\barr{ccccc}
 0          & 2p_{1}^\nu  & \ldots & 2p_{4}^\nu \\
p_{1}^{\mu} & 2p_{1}p_{1} & \ldots & 2p_{1}p_{4} \\
 \vdots     &\vdots       & \ddots & \vdots   \\
p_{4}^{\mu} & 2p_{4}p_{1} & \ldots & 2p_{4}p_4
\earr\right\vert,
\eeq
leading to the result
\beqar\label{Eredtensor0a}
\int {\cal E} &=&
2m_0^2\De^{(4)}(g^\mu_\al-{g_{(4)}}^\mu_\al)E^{\al\mu_1\ldots\mu_P}
\nl&&{}
+2\sum_{n=1}^4 \Ymodadj^{(4)}_{n0}
\Bigl[p_n^\mu(g_{\al\be}-{g_{(4)}}_{\al\be})
- p_{n,\be} (g^\mu_\al-{g_{(4)}}^\mu_\al)\Bigr]E^{\al\be \mu_1\ldots\mu_P} .
\eeqar
The second term is obtained by expanding the second determinant
in \refeq{Eredtensor} along the first two rows and the first two
columns according to \refeq{columnandrowexp} and using
\refeq{Zadjadjrelation1}.  

Alternatively integrating over the last form of ${\cal E}$ in
\refeq{Eredstart}, we obtain 
\beqar\label{Eredtensor0}
\arraycolsep 3pt
\int {\cal E} &=&
\left\vert
\barr{ccccc}
E^{\mu\mu_1\ldots\mu_P}  & -D^{\mu_1\ldots\mu_P}(0) & 
D^{\mu_1\ldots\mu_P}(1)-D^{\mu_1\ldots\mu_P}(0)     & \ldots & 
D^{\mu_1\ldots\mu_P}(4)-D^{\mu_1\ldots\mu_P}(0)         \\
0           & 2m_0^2 & f_1         & \ldots & f_{4}      \\
p_{1}^{\mu} & f_1    & 2p_{1}p_{1} & \ldots & 2p_{1}p_{4} \\
\vdots    & \vdots   & \vdots      & \ddots & \vdots      \\
p_{4}^{\mu} & f_4    & 2p_{4}p_{1} & \ldots & 2p_{4}p_4
\earr\right\vert
\nl[2ex]&&{}
+ \left\vert
\barr{ccccc}
g^\mu_\al   & -D^{\mu_1\ldots\mu_P}(0) & 2p_{1,\al}     & \ldots & 2p_{4,\al} \\
0           &   0     & f_1         & \ldots & f_{4} \\
p_{1}^{\mu} & D^{\al\mu_1\ldots\mu_P}(0)-D^{\al\mu_1\ldots\mu_P}(1) & 2p_{1}p_{1} & \ldots & 2p_{1}p_{4} \\
\vdots    & \vdots    & \vdots      & \ddots &  \vdots   \\
p_{4}^{\mu} & D^{\al\mu_1\ldots\mu_P}(0)-D^{\al\mu_1\ldots\mu_P}(4) & 2p_{4}p_{1} & \ldots & 2p_{4}p_4
\earr\right\vert.
\eeqar}%
The last determinant can be written as 
{\arraycolsep 4pt
\beqar\label{det3}
\lefteqn{
\left\vert
\barr{ccccc}
g^\mu_\al   & -D^{\mu_1\ldots\mu_P}(0) & 2p_{1,\al}     & \ldots & 2p_{4,\al} \\
0           &   0     & f_1         & \ldots & f_{4} \\
p_{1}^{\mu} & D^{\al\mu_1\ldots\mu_P}(0)-D^{\al\mu_1\ldots\mu_P}(1) & 2p_{1}p_{1} & \ldots & 2p_{1}p_{4} \\
\vdots    & \vdots    & \vdots      & \ddots &  \vdots   \\
p_{4}^{\mu} & D^{\al\mu_1\ldots\mu_P}(0)-D^{\al\mu_1\ldots\mu_P}(4) & 2p_{4}p_{1} & \ldots & 2p_{4}p_4
\earr\right\vert
}\qquad\nl[2ex]
&=&
\left\vert
\barr{ccccc}
g^\mu_\al   &   0     & 2p_{1,\al}     & \ldots & 2p_{4,\al} \\
0           &   0     & f_1         & \ldots & f_{4} \\
p_{1}^{\mu} &
D^{\al\mu_1\ldots\mu_P}(0)+p_{1}^{\al}D^{\mu_1\ldots\mu_P}(0)
-D^{\al\mu_1\ldots\mu_P}(1) 
& 2p_{1}p_{1} & \ldots & 2p_{1}p_{4} \\
\vdots    & \vdots    & \vdots      & \ddots &  \vdots   \\
p_{4}^{\mu} &
D^{\al\mu_1\ldots\mu_P}(0)+p_{1}^{\al}D^{\mu_1\ldots\mu_P}(0)
-D^{\al\mu_1\ldots\mu_P}(4) 
& 2p_{4}p_{1} & \ldots & 2p_{4}p_4
\earr\right\vert
\nl[2ex]
&=&
\left\vert
\barr{ccccc}
g^\mu_\al   &   0     & 2p_{1,\al}     & \ldots & 2p_{4,\al} \\
0           &   0     & f_1         & \ldots & f_{4} \\
p_{1}^{\mu} & -\D^{\al\mu_1\ldots\mu_P}(1)
& 2p_{1}p_{1} & \ldots & 2p_{1}p_{4} \\
\vdots    & \vdots    & \vdots      & \ddots &  \vdots   \\
p_{4}^{\mu} & -\D^{\al\mu_1\ldots\mu_P}(4) 
& 2p_{4}p_{1} & \ldots & 2p_{4}p_4
\earr\right\vert.
\eeqar}%
The first equality in \refeq{det3} can be easily checked by expanding
along the second column. In order to explain the second equality,
we introduce the Lorentz-covariant decompositions
\beqar\label{decomp}
D^{\al\mu_1\ldots\mu_P}(i) &=& [D^{\al\mu_1\ldots\mu_P}(i)]^{(p)}
+ [D^{\al\mu_1\ldots\mu_P}(i)]^{(g)}, \qquad i=0,\ldots,4,\nl{}
 [D^{\al\mu_1\ldots\mu_P}(i)]^{(p)}&=&\sum_{n=1\atop n\ne i}^4 
p_n^\al x_n^{\mu_1\ldots\mu_P}(i),\nl{}
 [D^{\al\mu_1\ldots\mu_P}(i)]^{(g)}&=&\sum_{r=1}^P 
g^{\al\mu_r}y_r^{\mu_1\ldots\hat\mu_r\ldots\mu_P}(i),\nl{}
\lefteqn{[D^{\al\mu_1\ldots\mu_P}(0)
+p_1^\al D^{\mu_1\ldots\mu_P}(0)]^{(p)}=
\sum_{n=2}^4 
(p_{n}-p_{1})^\al z_n^{\mu_1\ldots\mu_P}.}
\phantom{[D^{\al\mu_1\ldots\mu_P}(i)]^{(p)}}
\eeqar
The operation ``$(g)$'' isolates all tensor structures in which the
first Lorentz index appears at a metric tensor; the remaining part of
the tensor furnishes the ``$(p)$'' contribution.  The last
decomposition in \refeq{decomp} becomes obvious after performing a
shift $q\to q-p_1$ in the integral. From \refeq{decomp} it follows
immediately that the terms in the second line of \refeq{det3} that
involve $[D_{\al\mu_1\ldots\mu_P}(i)]^{(p)}$, $i=1,\ldots,4$
drop out when expanding the determinant along the second column,
because the resulting determinants vanish. Similarly, the contribution
proportional to $[D_{\al\mu_1\ldots\mu_P}(0)
+p_{1\al}D_{\mu_1\ldots\mu_P}(0)]^{(p)}$ vanishes after
summation over all contributions.  The remaining terms involving
$[D]^{(g)}$ are collected in the quantity
\beq
\D^{\al\mu_1\ldots\mu_P}(i) =
[D^{\al\mu_1\ldots\mu_P}(i)-D^{\al\mu_1\ldots\mu_P}(0)]^{(g)}, 
\qquad i=1,\dots,4.
\eeq
Inserting \refeq{det3} into \refeq{Eredtensor0} and expanding the
determinants we find
\beqar
\int {\cal E} &=&
\det(\Ymod^{(4)}) E^{\mu \mu_1\ldots\mu_P}
-\sum_{n,m=1}^4  \Ymodadj^{(4)}_{mn}p_m^\mu
[D^{\mu_1\ldots\mu_P}(n)- D^{\mu_1\ldots\mu_P}(0)]
\\ && {}
-\sum_{n=1}^4  \Ymodadj^{(4)}_{n0}[-p_n^\mu
D^{\mu_1\ldots\mu_P}(0)+
\D^{\mu\mu_1\ldots\mu_P}(n)]
+\sum_{n=1}^4 \D^{\al\mu_1\ldots\mu_P}(n)
\sum_{m,l=1}^4  2p_{m,\al}p_l^\mu \Ymodadjadj^{(4)}_{(ln)(0m)},
\nn
\eeqar
where  $\Ymodadjadj^{(4)}_{(ln)(0m)}$ is given in \refeq{eq:Ymodadjadj}.
Setting this equal to \refeq{Eredtensor0a}, we obtain
\beqar\label{Eredtensor1}
\det(\Ymod^{(4)}) E^{\mu \mu_1\ldots\mu_P} &=&
\sum_{n,m=1}^4  \Ymodadj^{(4)}_{mn}p_m^\mu
[D^{\mu_1\ldots\mu_P}(n)- D^{\mu_1\ldots\mu_P}(0)]
\nl&&{}
+\sum_{n=1}^4  \Ymodadj^{(4)}_{n0}[-p_n^\mu
D^{\mu_1\ldots\mu_P}(0)+
\D^{\mu\mu_1\ldots\mu_P}(n)]
\nl&&{}
-\sum_{n=1}^4 \D^{\al\mu_1\ldots\mu_P}(n) 
\sum_{m,l=1}^4  2p_{m,\al}p_l^\mu
\Ymodadjadj^{(4)}_{(ln)(0m)}
\nl&&{}
+2m_0^2\De^{(4)}(g^\mu_\al-{g_{(4)}}^\mu_\al)E^{\al\mu_1\ldots\mu_P}
\nl&&{}
+2\sum_{n=1}^4 \Ymodadj^{(4)}_{n0}
\Bigl[p_n^\mu(g_{\al\be}-{g_{(4)}}_{\al\be})
- p_{n,\be} (g^\mu_\al-{g_{(4)}}^\mu_\al)\Bigr]E^{\al\be \mu_1\ldots\mu_P} .
\hspace{2em}
\eeqar
In this result, all inverse Gram determinants have been absorbed in
the four-dimensional metric tensor, which appears only in the
difference $(g-g_{(4)})$. In four dimensions, all these terms vanish
identically.  In dimensional regularization they contribute only if
$E^{\al\be \mu_1\ldots\mu_P}$ involves singularities, \ie only the
singular terms in $E^{\al\be \mu_1\ldots\mu_P}$ are relevant.  As
explained in \refse{se:divs}, IR singularities of $E^{\al
  \mu_1\ldots\mu_P}$ appear only in contributions that are
proportional to a momentum $p_k^\al$.  These contributions vanish
exactly in \refeq{Eredtensor1} as long as the external momenta have
only non-vanishing components in the four-dimensional subspace.  UV
singularities appear only if $P\ge4$.  Therefore, we can omit the
last two terms in \refeq{Eredtensor1} for $P<4$.%
\footnote{This result is in agreement with the observation made in
  \citere{Dittmaier:2003bc} that in the absence of UV divergences
  reduction formulas valid in 4 dimensions remain valid in $D$
  dimensions up to terms of ${\cal O}(D-4)$, independent of the
  possible occurrence of IR singularities.}
For $P\ge4$ the inverse Gram determinant that is implicitly
contained in $g_{(4)}$ can always be cancelled by a prefactor 
$\De^{(4)}$. In the last-but-one term of \refeq{Eredtensor1} this
prefactor is already explicit; for the last contribution it is
straightforward to check%
\footnote{Contributions to $E^{\al\be \mu_1\ldots\mu_P}$ involving
  $p^\alpha_k$ vanish, and those involving
  $g^{\alpha\mu_i}g^{\be\mu_j}$ cancel after symmetrizing w.r.t.\ the
  indices $\mu$, $\mu_1$, $\ldots$, $\mu_P$. In terms involving
  $g^{\alpha\be}$ the surviving  $(g-g_{(4)})$ turns
  into $(D-4)$. Finally, terms involving $p^\be_k$ get a factor
  $\sum_{n=1}^4 Z^{(4)}_{in} \Ymodadj^{(4)}_{n0} = -f_i \Delta^{(4)} $ owing to
  \refeq{Ymodadj} and \refeq{Zinv}.}
that this factor always arises after symmetrizing the
r.h.s.\ of \refeq{Eredtensor1} w.r.t.\ the indices $\mu$, $\mu_1$,
$\ldots$, $\mu_P$.

The next step consists in the insertion of the decompositions of
tensor integrals into Lorentz covariants.  Here and in the following
we omit the terms involving $(g-{g_{(4)}})$ if $P<4$.  The general
tensor decompositions up to rank~5 explicitly read
\beqar
E^{\mu}&=&\sum_{\ina=1}^{4} p_{\ina}^{\mu}E_{\ina}, \qquad
E^{\mu\nu}=\sum_{\ina,\inb=1}^{4} p_{\ina}^{\mu}p_{\inb}^{\nu}E_{\ina\inb}
+g^{\mu\nu}E_{00},\nls
E^{\mu\nu\rho}&=&\sum_{\ina,\inb,\inc=1}^{4} p_{\ina}^{\mu}p_{\inb}^{\nu}p_{\inc}^{\rho}E_{\ina\inb\inc}
+\sum_{\ina=1}^{4}\{g p\}_{\ina}^{\mu\nu\rho} E_{00\ina}, 
\nl
E^{\mu\nu\rho\si} &=& 
\sum_{\ina,\inb,\inc,\ind=1}^{4} p_{\ina}^{\mu}p_{\inb}^{\nu}p_{\inc}^{\rho}p_{\ind}^\si E_{\ina\inb\inc\ind}
+\sum_{\ina,\inb=1}^{4}
\{g pp\}_{\ina\inb}^{\mu\nu\rho\si}E_{00\ina\inb}
+\{g g\}^{\mu\nu\rho\si} E_{0000},\nl
E^{\mu\nu\rho\si\tau} &=& 
\sum_{\ina,\inb,\inc,\ind,\ine=1}^{4}
p_{\ina}^{\mu}p_{\inb}^{\nu}p_{\inc}^{\rho}p_{\ind}^\si p_{\ine}^\tau E_{\ina
\inb\inc\ind\ine}
+\sum_{\ina,\inb,\inc=1}^{4}
\{g ppp\}_{\ina\inb\inc}^{\mu\nu\rho\si\tau}
E_{00\ina\inb\inc}
\nl&&{}
+\sum_{\ina=1}^{4} 
\{g g p\}_{\ina}^{\mu\nu\rho\si\tau} 
E_{0000\ina}.
\eeqar
In four dimensions, the covariants involving metric tensors are
redundant in these decompositions, since the metric tensor could be
replaced by \refeq{4Dmetric}.  By keeping these coefficients we can
avoid the appearance of explicit inverse Gram determinants in the
reduction formulas.

Inserting the Lorentz decompositions of the tensor integrals into
\refeq{Eredtensor1}, we find the following reduction equations for the
tensor coefficients upon comparing coefficients of covariants,
\beqar\label{Eredcoeffa}
\lefteqn{\det(\Ymod^{(4)}) \Ebar_{k i_1\ldots i_P} =\sum_{n=1}^4
\Ymodadj^{(4)}_{kn} \Bigl[
D_{(i_1)_n\ldots (i_P)_n}(n)\debar_{i_1n}\ldots\debar_{i_Pn}
- D_{i_1\ldots i_P}(0)\Bigr]
-\Ymodadj^{(4)}_{k0} D_{i_1\ldots i_P}(0)}
 \nl&&{}
-2 \sum_{n=1}^4 \sum_{r=1}^P  \Ymodadjadj^{(4)}_{(kn)(0 i_r)}\Bigl[
D_{00(i_1)_n\ldots \widehat{(i_r)_n} \ldots (i_P)_n}(n)
\debar_{i_1n}\ldots\debar_{i_{r-1}n}\debar_{i_{r+1}n}\ldots\debar_{i_Pn}
 -D_{00i_1\ldots \hat i_r \ldots i_P}(0) \Bigr],
\nn\\[.3em]
&&\qquad 
k=1,\ldots,4, \qquad P<4,
\\\label{Eredcoeffb}
\lefteqn{\det(\Ymod^{(4)}) \Ebar_{00i_2\ldots i_P} =
\sum_{n=1}^4  \Ymodadj^{(4)}_{n0} \Bigl[D_{00(i_2)_n\ldots (i_P)_n}(n)
\debar_{i_2n}\ldots\debar_{i_Pn}
-D_{00i_2\ldots i_P}(0) \Bigr],\qquad P<4.}
\nln
\eeqar
Since we have distinguished the index $k$ in the derivation of
\refeq{Eredcoeffa}, the resulting tensor coefficients $\Ebar_{\ldots}$
are not symmetric under the exchange of $k$ with one of the indices
$i_r$, $r=1,\ldots,P$. In order to distinguish them from the symmetric
tensor coefficients $E_{\ldots}$, we marked them with a bar.
Symmetric tensor coefficients can be easily obtained by adding all $P$
results with $k$ exchanged with one of the $i_r$ and dividing the sum
by $P$, \eg,
\beqar\label{Eijksym}
E_{i_1i_2i_3} &=&
\frac{1}{3}(\Ebar_{i_1i_2i_3}+\Ebar_{i_2i_1i_3}+\Ebar_{i_3i_2i_1}),
\\\label{E00isym}
E_{00i_1} &=&
\frac{1}{3}(\Ebar_{00i_1}+\Ebar_{0i_10}+\Ebar_{i_100}).
\eeqar
In \refeq{E00isym}, $\Ebar_{00i_1}$ and $\Ebar_{0i_10}$
are determined from \refeq{Eredcoeffb}, while $\Ebar_{i_100}$ is
determined from \refeq{Eredcoeffa}.  

For $P\ge 4$ extra terms of order $(D-4)E_{00\ldots}$ have to be added
to the equations \refeq{Eredcoeffa} and \refeq{Eredcoeffb}.  For $P=4$
the last-but-one contribution in \refeq{Eredtensor1} is of ${\cal
  O}(D-4)$, but the last term yields a finite contribution for
$D\to4$, because the coefficient $E_{000000}$ is UV divergent.  We
calculate this contribution upon inserting $E^{\al\be
  \mu_1\ldots\mu_4}|_{\mathrm{div}} =\{ggg\}^{\al\be
  \mu_1\ldots\mu_4}E_{000000}|_{\mathrm{div}}$ into
\refeq{Eredtensor0a} and using $(D-4)E_{000000}$ from
\refeq{eq:E000000div}. After symmetrizing in the Lorentz indices, we
get
\beq
\int {\cal E} = -\frac{1}{240} \sum_{n=1}^4 \Ymodadj^{(4)}_{n0}
\{ggp\}_n^{\mu\mu_1\ldots\mu_4}.
\eeq
This contribution to the coefficients $E_{i0000}$
can be included by replacing
$-\Ymodadj^{(4)}_{i0} D_{0000}(0)$ in \refeq{Eredcoeffa}
by $-\Ymodadj^{(4)}_{i0} \Bigl[D_{0000}(0) +\frac{1}{48}\Bigr]$.
The cases $P>4$ can be treated analogously, but 
usually do not appear in renormalizable quantum field theories.

After the symmetrization, we thus find 
for the tensor coefficients up to rank~5:
\beqar\label{Eredcoeff}
\det(\Ymod^{(4)}) E_{i_1} &=&
\sum_{n=1}^4
\Ymodadj^{(4)}_{i_1n} \Bigl[D_{0}(n)- D_{0}(0)\Bigr]
-\Ymodadj^{(4)}_{i_10} D_{0}(0),
 \\
\det(\Ymod^{(4)}) E_{00} &=&
\sum_{n=1}^4  \Ymodadj^{(4)}_{n0} \Bigl[D_{00}(n)
-D_{00}(0)\Bigr], 
\nl
2\det(\Ymod^{(4)}) E_{i_1i_2} &=&
\biggl\{\sum_{n=1}^4
\Ymodadj^{(4)}_{i_1n} \Bigl[D_{(i_2)_n}(n)\debar_{i_2n}- D_{i_2}(0)\Bigr]
-\Ymodadj^{(4)}_{i_1 0} D_{i_2}(0)
 \nl&&{}
-2 \sum_{n=1}^4 \Ymodadjadj^{(4)}_{(i_1n)(0i_2)}
\Bigl[D_{00}(n) -D_{00}(0)\Bigr]
\biggr\}
+ (i_1\leftrightarrow i_2),
 \\
3\det(\Ymod^{(4)}) E_{00i_1} &=&
2\sum_{n=1}^4  \Ymodadj^{(4)}_{n0} \Bigl[ D_{00(i_1)_n}(n)\debar_{i_1n}
-D_{00i_1}(0)\Bigr] 
\nl&&
+
\sum_{n=1}^4
\Ymodadj^{(4)}_{i_1n} \Bigl[D_{00}(n)- D_{00}(0)\Bigr]
-\Ymodadj^{(4)}_{i_10} D_{00}(0),
\nl
3\det(\Ymod^{(4)}) E_{i_1i_2i_3} &=&
\biggl\{\sum_{n=1}^4
\Ymodadj^{(4)}_{i_1n} \Bigl[D_{(i_2)_n(i_3)_n}(n)\debar_{i_2n}\debar_{i_3n}
 - D_{i_2i_3}(0)\Bigr]
- \Ymodadj^{(4)}_{i_10} D_{i_2i_3}(0)
 \nl&&{}
- 2 \sum_{n=1}^4 \Ymodadjadj^{(4)}_{(i_1n)(0i_2)}
\Bigl[D_{00(i_3)_n}(n)\debar_{i_3n} -D_{00i_3}(0) \Bigr]
\nn\\
&&{}
- 2 \sum_{n=1}^4 \Ymodadjadj^{(4)}_{(i_1n)(0i_3)}
\Bigl[D_{00(i_2)_n}(n)\debar_{i_2n} -D_{00i_2}(0) \Bigr]
\biggr\}
\nn\\ &&{}
+ (i_1\leftrightarrow i_2) + (i_1\leftrightarrow i_3),
\\
\det(\Ymod^{(4)}) E_{0000} &=&
\sum_{n=1}^4  \Ymodadj^{(4)}_{n0} \Bigl[D_{0000}(n)
-D_{0000}(0)\Bigr], 
\nl
4\det(\Ymod^{(4)}) E_{00i_1i_2} &=&
2\sum_{n=1}^4  \Ymodadj^{(4)}_{n0} \Bigl[ D_{00(i_1)_n(i_2)_n}(n)\debar_{i_1n}
\debar_{i_2n}-D_{00i_1i_2}(0)\Bigr]
\nl&&{}
+ \biggl\{\sum_{n=1}^4
\Ymodadj^{(4)}_{i_1n} \Bigl[D_{00(i_2)_n}(n)\debar_{i_2n}- D_{00i_2}(0)\Bigr]
-\Ymodadj^{(4)}_{i_1 0} D_{00i_2}(0)
 \nl&&{}
-2 \sum_{n=1}^4 \Ymodadjadj^{(4)}_{(i_1n)(0i_2)}
\Bigl[D_{0000}(n) -D_{0000}(0)\Bigr]
+ (i_1\leftrightarrow i_2)\biggr\},
\nl
4\det(\Ymod^{(4)}) E_{i_1i_2i_3i_4} &=&
\biggl\{\sum_{n=1}^4
\Ymodadj^{(4)}_{i_1n} \Bigl[D_{(i_2)_n(i_3)_n(i_4)_n}(n)\debar_{i_2n}\debar_{i_3n}\debar_{i_4n}
 - D_{i_2i_3i_4}(0)\Bigr]
- \Ymodadj^{(4)}_{i_10} D_{i_2i_3i_4}(0)
 \nl&&{}
- 2 \sum_{n=1}^4
\Ymodadjadj^{(4)}_{(i_1n)(0i_2)}
\Bigl[D_{00(i_3)_n(i_4)_n}(n)\debar_{i_3n}\debar_{i_4n} 
-D_{00i_3i_4}(0) \Bigr]
 \nl&&{}
- 2 \sum_{n=1}^4 \Ymodadjadj^{(4)}_{(i_1n)(0i_3)}
\Bigl[D_{00(i_2)_n(i_4)_n}(n)\debar_{i_2n}\debar_{i_4n} 
-D_{00i_2i_4}(0) \Bigr]
 \nl&&{}
- 2 \sum_{n=1}^4 \Ymodadjadj^{(4)}_{(i_1n)(0i_4)}
\Bigl[D_{00(i_2)_n(i_3)_n}(n)\debar_{i_2n}\debar_{i_3n} 
-D_{00i_2i_3}(0) \Bigr] 
\biggr\}
 \nl&&{}
+ (i_1\leftrightarrow i_2) + (i_1\leftrightarrow i_3)+ (i_1\leftrightarrow i_4)
,
 \\
5\det(\Ymod^{(4)}) E_{0000i_1} &=&
4\sum_{n=1}^4  \Ymodadj^{(4)}_{n0} \Bigl[D_{0000(i_1)_n}(n)\debar_{i_1n}
-D_{0000i_1}(0)\Bigr] 
\nl
&&{}+\sum_{n=1}^4
\Ymodadj^{(4)}_{i_1n} \Bigl[D_{0000}(n)- D_{0000}(0)\Bigr]
-\Ymodadj^{(4)}_{i_10}  \Bigl[D_{0000}(0) +\textstyle \frac{1}{48}\Bigr],
\nl
5\det(\Ymod^{(4)}) E_{00i_1i_2i_3} &=&
2\sum_{n=1}^4  \Ymodadj^{(4)}_{n0} \Bigl[
D_{00(i_1)_n(i_2)_n(i_3)_n}(n)\debar_{i_1n}\debar_{i_2n}\debar_{i_3n}
-D_{00i_1i_2i_3}(0)\Bigr] 
\nl&&{}
+ \biggl\{\sum_{n=1}^4
\Ymodadj^{(4)}_{i_1n} \Bigl[D_{00(i_2)_n(i_3)_n}(n)\debar_{i_2n}\debar_{i_3n}- D_{00i_2i_3}(0)\Bigr]
-\Ymodadj^{(4)}_{i_1 0} D_{00i_2i_3}(0)
 \nl&&{}
-2 \sum_{n=1}^4 \Ymodadjadj^{(4)}_{(i_1n)(0i_2)}
\Bigl[D_{0000(i_3)_n}(n)\debar_{i_3n} -D_{0000i_3}(0)\Bigr]
 \nl&&{}
-2 \sum_{n=1}^4 \Ymodadjadj^{(4)}_{(i_1n)(0i_3)}
\Bigl[D_{0000(i_2)_n}(n)\debar_{i_2n} -D_{0000i_2}(0)\Bigr]
\nn\\&&{}
+ (i_1\leftrightarrow i_2)+ (i_1\leftrightarrow i_3)\biggr\},
\nl
5\det(\Ymod^{(4)}) E_{i_1i_2i_3i_4i_5} &=&
\biggl\{\sum_{n=1}^4
\Ymodadj^{(4)}_{i_1n} \Bigl[D_{(i_2)_n(i_3)_n(i_4)_n(i_5)_n}(n)\debar_{i_2n}\debar_{i_3n}\debar_{i_4n}\debar_{i_5n}
 - D_{i_2i_3i_4i_5}(0)\Bigr]
 \nl&&{}
- \Ymodadj^{(4)}_{i_10} D_{i_2i_3i_4i_5}(0)
 \nl&&{}
- 2 \sum_{n=1}^4 \Ymodadjadj^{(4)}_{(i_1n)(0i_2)}
\Bigl[D_{00(i_3)_n(i_4)_n(i_5)_n}(n)\debar_{i_3n}\debar_{i_4n}\debar_{i_5n} -D_{00i_3i_4i_5}(0) \Bigr]
 \nl&&{}
- 2 \sum_{n=1}^4 \Ymodadjadj^{(4)}_{(i_1n)(0i_3)}
\Bigl[D_{00(i_2)_n(i_4)_n(i_5)_n}(n)\debar_{i_2n}\debar_{i_4n}\debar_{i_5n} -D_{00i_2i_4i_5}(0) \Bigr]
 \nl&&{}
- 2 \sum_{n=1}^4 \Ymodadjadj^{(4)}_{(i_1n)(0i_4)}
\Bigl[D_{00(i_2)_n(i_3)_n(i_5)_n}(n)\debar_{i_2n}\debar_{i_3n}\debar_{i_5n} -D_{00i_2i_3i_5}(0) \Bigr]
 \nl&&{}
- 2 \sum_{n=1}^4 \Ymodadjadj^{(4)}_{(i_1n)(0i_5)}
\Bigl[D_{00(i_2)_n(i_3)_n(i_4)_n}(n)\debar_{i_2n}\debar_{i_3n}\debar_{i_4n} -D_{00i_2i_3i_4}(0) \Bigr]
\biggr\}
 \nl&&{}
+ (i_1\leftrightarrow i_2) + (i_1\leftrightarrow i_3)+ (i_1\leftrightarrow i_4)+ (i_1\leftrightarrow i_5)
.
\eeqar

For the 4-point tensor coefficients that result from omitting $N_0$
in the 5-point integrals, we have introduced the auxiliary quantities
\beqar\label{auxD}
\Dcomb_{\ina}(0) &=& \Dtilde_{\ina-1}(0), \quad \ina=2,3,4,
\nl
\Dcomb_{1}(0) &=& -\sum_{\sina=2}^4D_{\sina}(0)-D_{0}(0),
\\
\Dcomb_{\ina\inb}(0) &=& \Dtilde_{\ina-1,\inb-1}(0),
\quad \ina,\inb=2,3,4,
\nl
\Dcomb_{1\ina}(0) &=& -\sum_{\sina=2}^4D_{\sina\ina}(0)-D_{\ina}(0),
\quad \ina=1,\ldots,4, 
\\
\Dcomb_{\ina\inb\inc}(0) &=& \Dtilde_{\ina-1,\inb-1,\inc-1}(0),
\quad \ina,\inb,\inc=2,3,4,
\nl
\Dcomb_{1\ina\inb}(0) &=&  -\sum_{\sina=2}^4D_{\sina\ina\inb}(0)-D_{\ina\inb}(0),
\quad \ina,\inb=1,\ldots,4,
\\
\Dcomb_{\ina\inb\inc\ind}(0) &=& \Dtilde_{\ina-1,\inb-1,\inc-1,\ind-1}(0),
\quad \ina,\inb,\inc,\ind=2,3,4,
\nl
\Dcomb_{1\ina\inb\inc}(0) &=&  
-\sum_{\sina=2}^4D_{\sina\ina\inb\inc}(0)-D_{\ina\inb\inc}(0),
\quad \ina,\inb,\inc=1,\ldots,4,
\eeqar
and similar quantities resulting from these relations with index pairs
``00'' added to the $D_{\ldots}(0)$ functions on both sides.

\section{Reduction of 6-point integrals}
\label{se:6pf}

Following the guideline of the reduction of the scalar 6-point integral
to six scalar 5-point integrals \cite{Me65}, the 6-point tensor
integrals of rank~$P$ can be reduced to six 5-point tensor integrals
of rank~$P$ as described in \citere{Denner:1993kt}.
This method, which was used in the calculation of one-loop corrections
to $\Pep\Pem\to4f$ \cite{Denner:2005es}, is more explicitly worked
out in \refapp{se:6pf-alt}.

In the following we describe a method that reduces 6-point tensor
integrals of rank~$P$ to 5-point tensor integrals of rank $(P-1)$.
The scalar 6-point integral should be treated following
\citeres{Me65,Denner:2005es} as explicitly described in
\refapp{se:6pf-alt}.  The tensor reduction can be derived by
considering the determinant
\beq\label{Fredstart}\arraycolsep 6pt
{\cal F}=
\left\vert
\barr{cccc}
q^\mu  & 2qp_{1}    & \ldots & \;2qp_{5} \\
p_{1}^{\mu} & 2p_{1}p_1 & \ldots & \;2p_{1}p_{5} \\
\vdots    & \vdots     & \ddots     &\;\vdots     \\
p_{k-1}^{\mu} & 2p_{k-1}p_1 & \ldots & \;2p_{k-1}p_{5} \\
0      &  f_1       & \ldots & \; f_5 \\
p_{k+1}^{\mu} & 2p_{k+1}p_1 & \ldots & \;2p_{k+1}p_{5} \\
\vdots    & \vdots     & \ddots     &\;\vdots     \\
p_{5}^{\mu} & 2p_{5}p_{1} & \ldots &\; 2p_{5}p_5
\earr\right\vert
=
\left\vert
\barr{cccc}
q^\mu   & N_1-N_0    & \ldots & \;N_5-N_0 \\
p_{1}^{\mu} & 2p_{1}p_1 & \ldots & \;2p_{1}p_{5} \\
\vdots    & \vdots     & \ddots     &\;\vdots     \\
p_{k-1}^{\mu} & 2p_{k-1}p_1 & \ldots & \;2p_{k-1}p_{5} \\
0      & f_1       & \ldots & \;f_5 \\
p_{k+1}^{\mu} & 2p_{k+1}p_1 & \ldots & \;2p_{k+1}p_{5} \\
\vdots    & \vdots     & \ddots     &\;\vdots     \\
p_{5}^{\mu} & 2p_{5}p_{1} & \ldots &\; 2p_{5}p_5
\earr\right\vert.
\eeq
The r.h.s.\ is obtained by adding
the $(k+1)$th row to the first row and using \refeq{contrtensor},

In four dimensions, this determinant vanishes, as can be seen from the
first form in \refeq{Fredstart}, because $q$ is linearly dependent on
the four (non-exceptional) momenta $p_i$, $i=1,\ldots,5$, $i\ne k$.
We again do not use this fact, but translate the integral
over ${\cal F}$ into a form that has a factor of ${\cal O}(D-4)$
rendering the whole contribution zero for finite integrals.  Inserting
\refeq{Fredstart} into the integrand of the tensor integral
$F^{\mu_1\ldots\mu_P}$ results in
\beq
\int {\cal F} \equiv
\frac{(2\pi\mu)^{4-D}}{\ri\pi^{2}}\int \rd^{D}q\,
\frac{q^{\mu_{1}}\cdots q^{\mu_{P}}}
{N_0N_1\ldots N_5} \, {\cal F} =
F^{\al \mu_1\ldots\mu_P}\left\vert
\barr{cccc}
g^{\mu}_{\al}& 2p_{1\al}    & \ldots & \;2p_{5\al} \\
p_1^\mu & 2p_{1}p_1 & \ldots & \;2p_{1}p_{5} \\
\vdots    & \vdots     & \ddots     &\;\vdots     \\
p_{k-1}^\mu & 2p_{k-1}p_1 & \ldots & \;2p_{k-1}p_{5} \\
0      & f_1       & \ldots & \; f_5 \\
p_{k+1}^\mu & 2p_{k+1}p_1 & \ldots & \;2p_{k+1}p_{5} \\
\vdots    & \vdots     & \ddots     &\;\vdots     \\
p_5^\mu & 2p_{5}p_{1} & \ldots &\; 2p_{5}p_5
\earr\right\vert.
\eeq
We expand the determinant along the $(k+1)$th row and use the fact
that the four-dimensional metric tensor can be written as
\beq\label{4Dmetric2}
g_{(4)}^{\mu\nu}\left\vert
\barr{cccc}
 2k_{1}p_{1} & \ldots & 2k_{1}p_{4} \\
 \vdots      & \ddots & \vdots   \\
 2k_{4}p_{1} & \ldots & 2k_{4}p_4
\earr\right\vert
=-\left\vert
\barr{ccccc}
 0          & 2p_{1}^\nu  & \ldots & 2p_{4}^\nu \\
k_{1}^{\mu} & 2k_{1}p_{1} & \ldots & 2k_{1}p_{4} \\
 \vdots     &\vdots       & \ddots & \vdots   \\
k_{4}^{\mu} & 2k_{4}p_{1} & \ldots & 2k_{4}p_4
\earr\right\vert 
\eeq
for two arbitrary sets of linear independent momenta $p_1,p_2,p_3,p_4$
and $k_1,k_2,k_3,k_4.$ This yields
\beq\label{Fredtensor1a}
\int {\cal F} =
-\Ymodadj^{(5)}_{k0}F^{\al \mu_1\ldots\mu_P}(g_\al^\mu-{g_{(4)}}_{\al}^\mu).
\eeq

Inserting the r.h.s.\ of \refeq{Fredstart} into the integrand of the
tensor integral $F^{\mu_1\ldots\mu_P}$ results in
{\arraycolsep 6pt
\beq
\int {\cal F} 
= \left\vert
\barr{ccc@{\hspace{\arraycolsep}\;}c}
F^{\mu \mu_1\ldots\mu_P} 
& E^{\mu_1\ldots\mu_P}(1)- E^{\mu_1\ldots\mu_P}(0) & \ldots 
& E^{\mu_1\ldots\mu_P}(5)- E^{\mu_1\ldots\mu_P}(0) \\
p_1^\mu & 2p_{1}p_1 & \ldots & \;2p_{1}p_{5} \\
\vdots    & \vdots     & \ddots     &\;\vdots     \\
p_{k-1}^\mu & 2p_{k-1}p_1 & \ldots & \;2p_{k-1}p_{5} \\
0      & f_1       & \ldots & \;f_5 \\
p_{k+1}^\mu & 2p_{k+1}p_1 & \ldots & \;2p_{k+1}p_{5} \\
\vdots    & \vdots     & \ddots     &\;\vdots     \\
p_5^\mu & 2p_{5}p_{1} & \ldots &\; 2p_{5}p_5
\earr\right\vert.
\eeq}%
Expanding the determinant along the first row
and the first column according to the analogue of \refeq{Zadjadjrelation3},
yields     
\beq\label{Fredtensor1b}
\int {\cal F} =
-\Ymodadj^{(5)}_{k0}F^{\mu \mu_1\ldots\mu_P} 
-\sum_{n,m=1}^5  \Ymodadjadj^{(5)}_{(km)(0n)}p_m^\mu
\Bigl[E^{\mu_1\ldots\mu_P}(n)- E^{\mu_1\ldots\mu_P}(0)\Bigr],
\eeq
where  $\Ymodadjadj^{(5)}_{(km)(0n)}$ is given in \refeq{eq:Ymodadjadj}.

From \refeq{Fredtensor1a} and \refeq{Fredtensor1b} we obtain
\beq\label{Fredtensor}
\Ymodadj^{(5)}_{k0}F^{\mu \mu_1\ldots\mu_P} = 
-\sum_{n,m=1}^5  \Ymodadjadj^{(5)}_{(km)(0n)}p_m^\mu
\Bigl[E^{\mu_1\ldots\mu_P}(n)- E^{\mu_1\ldots\mu_P}(0)\Bigr]
+\Ymodadj^{(5)}_{k0}F^{\al \mu_1\ldots\mu_P}(g_\al^\mu-{g_{(4)}}_{\al}^\mu).
\eeq
The last term in \refeq{Fredtensor} only contributes in dimensional
regularization if $F^{\al \mu_1\ldots\mu_P}$ is singular. For UV
singularities this is the case if $P\ge7$, which is usually not needed
in renormalizable theories.  As explained in \refse{se:divs}, IR (soft
and collinear) singularities of $F^{\al \mu_1\ldots\mu_P}$ only appear
in contributions that are proportional to a momentum $p_i^\al$.  These
contributions vanish exactly in \refeq{Fredtensor}. Therefore, the
terms involving $(g-g_{(4)})$ in \refeq{Fredtensor} can be omitted for
$P<7$.%
\footnote{This is again in agreement with the observation
  \cite{Dittmaier:2003bc} that in the absence of UV divergences
  reduction formulas valid in 4 dimensions remain valid in $D$
  dimensions up to terms of ${\cal O}(D-4)$, independent of possible
  IR singularities.}
For $P\ge7$ the inverse determinant that is implicitly
contained in $g_{(4)}$ can always be cancelled.%
\footnote{According to \refeq{Ymodadj}, $\Ymodadj^{(5)}_{k0} =
  -\sum_{n=1}^{5} \Zadj^{(5)}_{kn} f_n$. For each of these terms,
  $\Zadj^{(5)}_{kn} g_{(4)}$ can be expressed via \refeq{4Dmetric2} by
  a determinant without denominator.}

Introducing the matrix
\beq\arraycolsep 6pt
M_{(k)}=
\left(
\barr{ccc@{\hspace{\arraycolsep}\;}c}
2p_{1}p_1 & \ldots & \;2p_{1}p_{5} \\
 \vdots     & \ddots     &\;\vdots     \\
 2p_{k-1}p_1 & \ldots & \;2p_{k-1}p_{5} \\
 f_1       & \ldots & \;f_5 \\
 2p_{k+1}p_1 & \ldots & \;2p_{k+1}p_{5} \\
 \vdots     & \ddots     &\;\vdots     \\
2p_{5}p_{1} & \ldots &\; 2p_{5}p_5
\earr\right)\,,
\eeq
\refeq{Fredtensor} can be written as
\beq \label{Fredtensor3}
F^{\mu \mu_1\ldots\mu_P} =
\sum_{n=1}^5 
\sum_{m=1 \atop m\ne k}^5 
\left(M_{(k)}^{-1}\right)_{nm} \, p_m^\mu
\Bigl[E^{\mu_1\ldots\mu_P}(n)- E^{\mu_1\ldots\mu_P}(0)\Bigr]
+F^{\al \mu_1\ldots\mu_P}(g_\al^\mu-{g_{(4)}}_{\al}^\mu),
\eeq
which expresses the 6-point tensor integral of rank~$P$ in terms of
six 5-point tensor integrals of rank $(P-1)$. The inverse of
$M_{(k)}$ is given by
\beq
\left(M_{(k)}^{-1}\right)_{ij} = -\Ymodadjadj^{(5)}_{(kj)(0i)} / 
\Ymodadj^{(5)}_{k0}, \qquad i,j,k=1,\ldots,N.
\eeq

In the form \refeq{Fredtensor3} our result can easily be extended to
the reduction of $N$-point functions with $N>6$ by simply forming a
matrix similar to $M_{(k)}$ by selecting five momenta for the columns
and four momenta for the rows out of the $(N-1)$ available momenta of
the $N$-point function.

Equation \refeq{Fredtensor} can also be used to derive an alternative
reduction of tensor 6-point integrals.  Multiplying it with $X_{k0}$,
summing over $k=1,\ldots,N$, and using \refeq{eq:Ymodadjadj_rels},
for $\Ymodadjadj^{(5)}_{(km)(0n)}$ yields
\beqar\label{Fredtensor2}
\det(\Ymod^{(5)})F^{\mu \mu_1\ldots\mu_P} &=&
\sum_{n,m=1}^5  \Ymodadj^{(5)}_{nm}p_m^\mu
\Bigl[E^{\mu_1\ldots\mu_P}(n)- E^{\mu_1\ldots\mu_P}(0)\Bigr]
\nl&&{}
+\det(\Ymod^{(5)})F^{\al \mu_1\ldots\mu_P}(g_\al^\mu-{g_{(4)}}_{\al}^\mu).
\eeqar
Here, as in \refeq{Eredtensor1}, all inverse Gram determinants have
been absorbed in the four-dimensi\-on\-al metric tensor, which appears
only in the difference $(g-g_{(4)})$.
The result \refeq{Fredtensor2} is equivalent to Eq.~(64) of
\citere{Binoth:2005ff}.

Finally, we insert the decompositions of tensor 6-point integrals into
Lorentz covariants in order to derive explicit reduction formulas for
the tensor coefficients. Since we consider only tensors up to rank~3,
we can omit the terms involving $(g-g_{(4)})$.
The tensor decompositions explicitly read
\beqar\label{LorentzdecF}
F^{\mu}&=&\sum_{\ina=1}^{5} p_{\ina}^{\mu}F_{\ina}, \qquad
F^{\mu\nu}=\sum_{\ina,\inb=1}^{5} p_{\ina}^{\mu}p_{\inb}^{\nu}F_{\ina\inb}
+g^{\mu\nu}F_{00},\nls
F^{\mu\nu\rho}&=&\sum_{\ina,\inb,\inc=1}^{5} p_{\ina}^{\mu}p_{\inb}^{\nu}p_{\inc}^{\rho}F_{\ina\inb\inc}
+\sum_{\ina=1}^{5}\{g p\}_{\ina}^{\mu\nu\rho} F_{00\ina}.
\eeqar
In four dimensions, some covariants in these decompositions are
redundant in the sense that they can be expressed by the others.  For
instance, in the decomposition of $F^\mu$ one of the five covariants
$p_{\ina}^{\mu}F_{\ina}$ is redundant, because one of the momenta
$p_{\ina}$ can be expressed by the other four linearly independent
vectors.  Similarly, all covariants involving metric tensors are
redundant.  However, by keeping these coefficients we can
avoid the appearance of explicit inverse Gram determinants in the
reduction formulas.

Inserting the Lorentz decompositions of the tensor integrals in the
reduction formulas given above, we can read off the reduction formulas
for the tensor coefficients upon comparing coefficients of covariants
on both sides.  Generically we find
\beqar\label{Fredcoeff}
\Fbar_{j i_1\ldots i_P} &=&
\sum_{n=1}^5 c_{jn} 
\Bigl[E_{(i_1)_n\ldots(i_P)_n}(n)\debar_{i_1n}\ldots\debar_{i_Pn}
- E_{i_1\ldots i_P}(0)\Bigr], \qquad P<7,
\eeqar
with 
\beq\label{Fredcoeff3}
c_{0n} = c_{kn} = 0,\qquad c_{jn} = \left(M_{(k)}^{-1}\right)_{nj},
\quad j,n=1,\ldots,5,\quad j\ne k
\eeq
for the reduction given in \refeq{Fredtensor3} and with
\beq\label{Fredcoeff2}
c_{0n} = 0,\qquad c_{jn} = \Ymodadj^{(5)}_{nj} / \det(X^{(5)})
=\left(X^{(5)}\right)^{-1}_{jn},
\quad j,n=1,\ldots,5
\eeq
for the reduction given in \refeq{Fredtensor2}.
In the numerical reduction we can select the equation that is
numerically most stable. For example, in \refeq{Fredcoeff3} we can
choose $k$ such that the modulus of $\Ymodadj^{(5)}_{k0}=-\det
M_{(k)}$ is maximal.

Since we have distinguished one momentum in the derivation of
\refeq{Fredtensor3} the resulting tensor coefficients $\Fbar_{\ldots}$ 
are not symmetric under the exchange of $j$ with one of the indices
$i_r$.  This can be easily cured as in the case of 5-point functions
[see \refeq{Eijksym} and 
\refeq{E00isym}] by adding all $P$ results with $j$ exchanged
with one of the $i_r$, and dividing the sum by $P$.

Thus, we find from \refeq{Fredcoeff} for the tensor
coefficients up to rank~3
\beqar
F_{i_1} &=& \sum_{n=1}^5 c_{i_1 n}
 \Bigl[E_{0}(n)- E_{0}(0)\Bigr], \quad i_1=1,\dots,5,
\\
F_{00} &=& 0,
\nl
F_{i_1i_2} &=& \frac{1}{2}\sum_{n=1}^5 \biggl\{  
  c_{i_1 n} \Bigl[E_{(i_2)_n}(n)\debar_{i_2n}- E_{i_2}(0)\Bigr]
+ (i_1\leftrightarrow i_2) \biggr\},
\quad 
i_1,i_2=1,\dots,5,
\\
F_{00i_1} &=&
\frac{1}{3}\sum_{n=1}^5 c_{i_1 n} \Bigl[E_{00}(n)- E_{00}(0)\Bigr],
\nl
F_{i_1i_2i_3} &=&
\frac{1}{3}\sum_{n=1}^5 \biggl\{  
  c_{i_1 n} \Bigl[E_{(i_2)_n(i_3)_n}(n)\debar_{i_2n}\debar_{i_3n}- E_{i_2i_3}(0)\Bigr]
+ (i_1\leftrightarrow i_2) + (i_1\leftrightarrow i_3) 
\biggr\}, 
\nl&&{}
i_1,i_2,i_3=1,\dots,5.
\hspace{2em}
\eeqar
For the 5-point tensor coefficients that result from omitting $N_0$
in the 6-point integrals, we have again used the auxiliary quantities
\beqar\label{auxE}
\Ecomb_{\ina}(0) &=& \Etilde_{\ina-1}(0), \quad \ina=2,\dots,5,
\nl
\Ecomb_{1}(0) &=& -\sum_{\sina=2}^5E_{\sina}(0)-E_{0}(0),
\\
\Ecomb_{\ina\inb}(0) &=& \Etilde_{\ina-1,\inb-1}(0),
\quad \ina,\inb=2,\dots,5,
\nl
\Ecomb_{1\ina}(0) &=& -\sum_{\sina=2}^5E_{\sina\ina}(0)-E_{\ina}(0),
\quad \ina=1,\dots,5, 
\\
\Ecomb_{\ina\inb\inc}(0) &=& \Etilde_{\ina-1,\inb-1,\inc-1}(0),
\quad \ina,\inb,\inc=2,\dots,5,
\nl
\Ecomb_{1\ina\inb}(0) &=&  -\sum_{\sina=2}^5E_{\sina\ina\inb}(0)-E_{\ina\inb}(0),
\quad \ina,\inb=1,\dots,5.
\eeqar

\section{Summary}
\label{se:summary}

Methods for a systematic evaluation of one-loop tensor integrals
have been described for graphs with up to six external legs.
The results are presented in a form that can be directly translated
into a computer code; only the scalar 3- and 4-point integrals
have to be taken from elsewhere.

While UV divergences are treated in dimensional regularization,
possible IR (soft or collinear) divergences can be regularized either
dimensionally or with small mass parameters; the described results
are valid in either IR regularization scheme.
Moreover, the results hold if internal masses are complex parameters,
which naturally appear for unstable internal particles.
The generalization of the proposed methods to functions with more 
than six external lines is straightforward.

Particular attention is paid to the issue of numerical stability.  For
1- and 2-point integrals of arbitrary tensor rank, general numerically
stable results are presented.  For 3- and 4-point tensor integrals,
serious numerical instabilities are known to arise in the frequently
used Passarino--Veltman reduction if Gram determinants built of
external momenta become small. For these cases we have developed
dedicated reduction techniques. One of the techniques replaces the
standard scalar integral by a specific tensor coefficient that can be
safely evaluated numerically and reduces the remaining tensor
coefficients as well as the standard scalar integral to the new set of
basis integrals. In this scheme no dangerous inverse Gram determinants
occur, but inverse modified Cayley determinants instead.  In a second
class of techniques we keep the basis set of standard scalar integrals
and iteratively deduce the tensor coefficients up to terms that are
systematically suppressed by small Gram determinants or by other
kinematical determinants in specific kinematical configurations.  The
convergence of the iteration can be systematically improved upon
including higher tensor ranks.  For 5- and 6-point tensor integrals,
we describe reductions to 5- and 4-point integrals, respectively, that
do not involve inverse Gram determinants either. Compared to some
other existing methods, the described methods are distinguished by the
fact that the reduction from 6-(5-) to 5-(4-)point integrals decreases
the tensor rank at the same time.

We finally emphasize that the presented methods have already been
successfully applied in the calculation of a complete one-loop
correction to a $2\to4$ scattering reaction, viz.\ the 
electroweak corrections to the charged-current processes
$\Pep\Pem\to4f$.
The described methods, thus, have proven their reliability in 
practice and will certainly be used in future loop calculations
for interesting many-particle production processes at the
LHC and ILC.

\par
\vskip 1cm

\section*{Acknowledgements}

This work was supported in part by the Swiss National Science
Foundation.  We are grateful to L. Wieders for performing some checks
of our results.  We thank him and M.~Roth for the successful
collaboration on electroweak corrections to 4-fermion production which
triggered this work. A.D. is also thankful to T. Binoth for useful
discussions. 

\appendix
\section*{Appendix}

\section{UV-divergent parts of tensor integrals}
\label{se:tendiv}

In the reduction formulas given above, products of $(D-4)$ with tensor
integrals appear. These give rise to finite terms originating from
UV singularities in the loop integrals. As mentioned above, no IR-singular 
integrals multiplied with $(D-4)$ appear in the reduction
formulas. The UV-singular parts of the loop integrals can be derived
easily from the Feynman-parameter representation or by using
\refeq{PVrecursionD1} for these parts only.  In the following,
we list results for $(D-4)$ times one-loop integrals omitting 
terms of order $\ord(D-4)$.  For the 1-point functions
$A_{\ldots}(m_{0})$ we get
\beq\label{TNdiv}
(D-4)\,A_{0} = -2m_0^{2}, \qquad
(D-4)A_{\underbrace{\sst 0\ldots0}_{2n}} = -\frac{m_0^{2n+2}}{2^{n-1}(n+1)!},
\qquad n=1,2\ldots\,.
\eeq
For the IR-finite 2-point functions $B_{\ldots}(p_{1},m_{0},m_{1})$, 
i.e.\ excluding the case $p_1^2=m_0^2=m_1^2=0$,
we obtain
\beqar
(D-4)\,B_{0} &=& -2 ,\nl[1ex]
(D-4)\,B_{1} &=&  1 ,\nl[1ex]
(D-4)\,B_{00} &=&
\textfrac{1}{6}(p_{1}^{2}-3m_{0}^{2}-3m_{1}^{2}) ,\qquad
(D-4)\,B_{11} =  -\textfrac{2}{3} ,\nl[1ex]
(D-4)\,B_{001} &=&  -\textfrac{1}{12}(p_{1}^{2}-2m_{0}^{2}-4m_{1}^{2}) ,\qquad  
(D-4)\,B_{111} =  \textfrac{1}{2} ,\nl[1ex]
(D-4)\,B_{0000} &=&
-\textfrac{1}{120}\Bigl[p_{1}^{4}-5p_1^2(m_{0}^{2}+m_1^2)
+10(m_{0}^{4}+m_0^2m_1^2+m_1^4)\Bigr] ,\nl[1ex]
(D-4)\,B_{0011} &=&  \textfrac{1}{60}(3p_{1}^{2}-5m_{0}^{2}-15m_{1}^{2}) ,\qquad
(D-4)\,B_{1111} =  -\textfrac{2}{5} ,\nl[1ex]
(D-4)\,B_{00001} &=&  \textfrac{1}{240}
\Bigl[p_{1}^{4}-4p_1^2m_{0}^{2}-6p_1^2m_{1}^{2}
+5m_0^4+10m_0^2m_1^2+15m_1^4\Bigr] ,\!\!\nl[1ex]
(D-4)\,B_{00111} &=&  -\textfrac{1}{60}(2p_{1}^{2}-3m_{0}^{2}-12m_{1}^{2}) ,\qquad
(D-4)\,B_{11111} =  \textfrac{1}{3} .
\eeqar

\begin{sloppypar}
For the 3-point functions $C_{\ldots}(p_{1},p_{2},m_{0},m_{1},m_{2})$
we obtain, denoting $(p_1-p_2)^2=s_{12}$,
\beqar
(D-4)\,C_{00} &=&  -\textfrac{1}{2} ,\nl[1ex]
(D-4)\,C_{00i} &=&  \textfrac{1}{6} ,
\nl[1ex]
(D-4)\,C_{0000} &=&
\textfrac{1}{48}[s_{12}+p_1^2+p_2^2]
-\textfrac{1}{12}(m_0^2+m_1^2+m_2^2) ,
\nl[1ex]
(D-4)\,C_{00ii} &=&  -\textfrac{1}{12} , \qquad
(D-4)\,C_{00ij} =  -\textfrac{1}{24} ,
\nl[1ex]
(D-4)\,C_{0000i} &=&
-\textfrac{1}{240}
\biggl[2s_{12}-5m_0^2+\sum_{n=1}^2(p_n^2-5m_n^2)(1+\de_{in})\biggr],\nl[1ex]
(D-4)\,C_{00iii} &=&  
\textfrac{1}{20}, \qquad
(D-4)\,C_{00iij} =
\textfrac{1}{60},
\nl[1ex]
(D-4)\,C_{000000} &=&
-\textfrac{1}{2880}
\Bigl[2s_{12}^2-6s_{12}m_0^2+30m_0^4
+2s_{12}\sum_{n=1}^2 (p_n^2-6m_n^2)
\nl&&{}
-6m_0^2\sum_{n=1}^2 (2p_n^2-5m_n^2)
\nl&&{}
+\sum_{m,n=1}^2 (p_m^2 p_n^2-6p_m^2 m_n^2+15m_m^2 m_n^2)(1+\de_{mn})
\Bigr],
\nl[1ex]
(D-4)\,C_{0000ii} &=&
\textfrac{1}{720}\biggr[3s_{12}-6m_0^2
+\sum_{n=1}^2(p_n^2-6m_n^2)(1+2\de_{in})\biggl],
\nl
(D-4)\,C_{0000ij} &=&
\textfrac{1}{720}\biggr[2s_{12}-3m_0^2
+\sum_{n=1}^2(p_n^2-6m_n^2)\biggl],
\nl[1ex]
(D-4)\,C_{00iiii} &=&  
-\textfrac{1}{30}, \qquad
(D-4)\,C_{00iiij} =  
-\textfrac{1}{120}, \qquad
(D-4)\,C_{00iijj} =  
-\textfrac{1}{180},
\nl[1ex]
(D-4)\,C_{000000i} &=& \textfrac{1}{10080} \biggr[ 3s_{12}^2-7s_{12}m_0^2
+21m_0^4+s_{12}\sum_{n=1}^2(p_n^2-7m_n^2)(2+\de_{in})
\nn\\ && {}
-7m_0^2\sum_{n=1}^2(p_n^2-3m_n^2)(1+\de_{in})
\nn\\ && {}
+\sum_{m,n=1}^2(p_m^2 p_n^2-7p_m^2 m_n^2+21m_m^2 m_n^2)(1+2\de_{im}\de_{in})
\biggl],
\nl[1ex]
(D-4)\,C_{0000iii} &=& -\textfrac{1}{1680} \biggr[ 4s_{12}-7m_0^2
+\sum_{n=1}^2(p_n^2-7m_n^2)(1+3\de_{in})\biggl],
\nl[1ex]
(D-4)\,C_{0000iij} &=& -\textfrac{1}{5040} \biggr[ 6s_{12}-7m_0^2
+\sum_{n=1}^2(p_n^2-7m_n^2)(2+\de_{in})\biggl],
\nl[1ex]
(D-4)\,C_{00iiiii} &=&
\textfrac{1}{42},
\qquad
(D-4)\,C_{00iiiij} =  
\textfrac{1}{210},
\qquad     
(D-4)\,C_{00iiijj} =  
\textfrac{1}{420}, 
\hspace{2em}
\eeqar
where $i,j=1,2$ but 
$i\ne j$.  All other 3-point tensor coefficients up to rank~7
are UV finite, so that for them $(D-4)C_{\dots}=0$
if they are IR finite.  For the 4-point
functions $D_{\ldots}(p_{1},p_{2},p_{3},m_{0},m_{1},m_{2},m_{3})$ we
find, denoting $(p_1-p_2)^2=s_{12}$, $(p_1-p_3)^2=s_{13}$, and
$(p_2-p_3)^2=s_{23}$:
\beqar
(D-4)\,D_{0000} &=&
-\textfrac{1}{12} ,
\nl[1ex]
(D-4)\,D_{0000i} &=&
\textfrac{1}{48} .
\nl[1ex]
(D-4)\,D_{000000} &=&
\textfrac{1}{480}[s_{12}+s_{13}+s_{23}+p_1^2+p_2^2+p_3^2]
-\textfrac{1}{96}(m_0^2+m_1^2+m_2^2+m_3^2) ,
\nl[1ex]
(D-4)\,D_{0000ii} &=&  -\textfrac{1}{120} ,
\qquad
(D-4)\,D_{0000ij} =  -\textfrac{1}{240} ,
\nl[1ex]
(D-4)\,D_{000000i} &=&
-\textfrac{1}{2880}\Biggl[\sum_{n=1}^3 p_n^2(1+\de_{in})
+\sum_{m,n=1\atop m>n}^3 s_{mn}(1+\de_{in}+\de_{im})\Biggr]
+\textfrac{1}{480}\sum_{n=0}^3m_n^2,
\nl[1ex]
(D-4)\,D_{0000iii} &=&  
\textfrac{1}{240},
\qquad
(D-4)\,D_{0000iij} =  
\textfrac{1}{720},
\qquad     
(D-4)\,D_{0000ijk} =  
\textfrac{1}{1440}, \hspace{2em}
\eeqar
where $i,j,k=1,2,3$ but are pairwise different.
All other 4-point tensor coefficients up to rank~7 are UV finite.
\end{sloppypar}

For the 5-point functions
$E_{\ldots}(p_{1},p_{2},p_{3},p_{4},m_{0},m_{1},m_{2},m_{3},m_{4})$,
there is only one UV-singular tensor coefficient up to rank~6,
\beqar\label{eq:E000000div}
(D-4)\,E_{000000} &=&
-\textfrac{1}{96}. 
\eeqar

\section{Tensor coefficients of singular 3-point functions}
\label{app:Csing}

The vanishing of the modified Cayley determinant $\det (X^{(N)})$, as
defined via \refeq{detCayley}, is a necessary condition for the
existence of a leading Landau singularity in a one-loop $N$-point
integral. For 3-point integrals this means that $\det X^{(3)}=0$ for
IR-singular (either soft or collinear) integrals, so that the
reduction methods of \refses{se:PValt} and \ref{se:Cayley} are not
applicable in this case.  If in addition the Gram determinant is
small, for IR-singular 3-point integrals also the $\Ymodadj_{0j}$ are
small, and the reduction method of \refse{se:gram} cannot be used
either.  One could still, however, use the method of
\refse{se:gramcayley}.  In the following we describe a way of
evaluating these specific 3-point functions that does not make use of
an iteration technique, but is based on analytical simplifications
that are admitted by the simple structure of the special cases.

The simplifications are achieved by directly using the analytical results
for the standard scalar integrals and for the tensor coefficients, as
obtained with the Passarino--Veltman reduction, and by rewriting
them in such a way that the limit of vanishing Gram determinant does
not involve numerical cancellations. To this end, the scalar integrals
are split into two parts: one contains the asymptotic behaviour of
the integral in the limit of vanishing Gram determinant $\De^{(3)}$
up to a specific order $n$ and a corresponding remainder which is of
${\cal O}\left([\De^{(3)}]^{n+1}\right)$.
We symbolize this splitting by introducing the asymptotic operators
${\cal T}^{(n)}_{x\to x_0}$ and ${\cal R}^{(n)}_{x\to x_0}$, which
define the asymptotic behaviour of a function $f(x)$ for $x\to x_0$
by
\beq
f(x) = {\cal T}^{(n)}_{x\to x_0}\left[f(x)\right] 
+ {\cal R}^{(n)}_{x\to x_0}\left[f(x)\right], \quad
{\cal R}^{(n)}_{x\to x_0}\left[f(x)\right] = 
{\cal O}\left((x-x_0)^{n+1}\right), \quad n=0,1,\ldots\, .
\eeq
If the function $f(x)$ is analytical at $x=x_0$, ${\cal T}^{(n)}_{x\to x_0}$
is the usual operator for a Taylor expansion up to order $n$.

Making use of these definitions, we now describe the treatment
of the IR-singular 3-point tensor integrals that were needed in the
calculation of the one-loop corrections to
$\Pep\Pem\to4f$ \cite{Denner:2005es}.
It is convenient to switch from the original definition \refeq{tensorint}
of arguments on tensor coefficients to the new notation
\beqar
B_{\dots}(p_1^2,m_0,m_1) &\equiv& B_{\dots}(p_1,m_0,m_1), \nn\\
C_{\dots}(p_1^2,(p_2-p_1)^2,p_2^2,m_0,m_1,m_2) &\equiv& 
C_{\dots}(p_1,p_2,m_0,m_1,m_2). 
\eeqar

\setcounter{paragraph}{0}
\paragraph{Collinear-singular case with two off-shell legs:
$C_{\dots}(m^2,s,s',0,m,M)$}
\label{par:case1}

Here $m$ denotes a small real mass, which will be neglected whenever possible.
In this limit the relevant scalar integrals read
\beqar\label{eq:scalint1}
B_0(0) &=& B_0(s,m,M) = \De+\ln\biggl(\frac{\mu^2}{M^2}\biggr)+2
+\biggl(\frac{M^2}{s}-1\biggr)\ln\biggl(\frac{M^2-s}{M^2}\biggr),
\nn\\
B_0(1) &=& B_0(s',0,M) = \De+\ln\biggl(\frac{\mu^2}{M^2}\biggr)+2
+\biggl(\frac{M^2}{s'}-1\biggr)\ln\biggl(\frac{M^2-s'}{M^2}\biggr),
\nn\\
B_0(2) &=& B_0(m^2,0,m) = \De+\ln\biggl(\frac{\mu^2}{m^2}\biggr)+2,
\nn\\
C_0 &=&
\frac{1}{s-s'} \Biggl\{
 \ln\biggl(\frac{M^2-s}{m^2}\biggr)
 \ln\biggl(\frac{M^2-s}{M^2}\biggr)
-\ln\biggl(\frac{M^2-s'}{m^2}\biggr)
 \ln\biggl(\frac{M^2-s'}{M^2}\biggr)
\nn\\
&& {}
-2\Li\biggl(\frac{s-s'}{M^2-s'}\biggr)
+\Li\biggl(\frac{s}{M^2}\biggr)
-\Li\biggl(\frac{s'}{M^2}\biggr)
\Biggr\},
\eeqar
where $M^2$ is complex with a finite or infinitesimal negative
imaginary part, which is also present for vanishing $M^2$.
The Gram determinant is given by
\beq
\De^{(2)} = -(s-s')^2,
\eeq
so that the delicate limit is $\delta s\equiv s'-s\to 0$.
The asymptotic expansions of the scalar integrals in \refeq{eq:scalint1}
for this limit can be worked out easily; the first few terms read
\beqar\label{eq:scal_collexpand1}
B_0(1) &=& B_0(0) 
-\frac{\de s}{s}
\biggl[1+\frac{M^2}{s}\ln\biggl(\frac{M^2-s}{M^2}\biggr)\biggr]
+{\cal R}^{(1)}_{\de s\to0}\left[B_0(1)\right],
\nn\\[.5em]
C_0 &=&
\frac{1}{s-M^2}\Biggl\{ 
\biggl[1+\frac{\de s}{2(M^2-s)}\biggr]
\ln\biggl(\frac{M^2-s}{m^2}\biggr) 
+\frac{M^2}{s}\biggl[1-\frac{\de s(M^2-2s)}{2s(M^2-s)}\biggr]
\ln\biggl(\frac{M^2-s}{M^2}\biggr)
\nn\\
&&{}
+2-\frac{\de s(M^2-2s)}{2s(M^2-s)}
\Biggr\}
+{\cal R}^{(1)}_{\de s\to0}\left[C_0\right],
\eeqar
where we have kept $s$ fixed.
Inserting these or forms with more explicit terms of the asymptotic
expansion for the scalar integrals into the explicit
formulas for the tensor coefficients, one obtains expressions like
\beqar
C_1 &=&
\frac{M^2[(M^2-s)s-\de s(4M^2-5s)]}{2s^2(M^2-s)^2}
+ \frac{M^2-s+\de s}{2(M^2-s)^2} \ln\biggl(\frac{M^2-s}{m^2}\biggr) 
\nn\\          
&& {}
+ \frac{M^2[M^2s(M^2-s)-\de s(4M^4-7M^2s+2s^2)]}{2(M^2-s)^2s^3}
\ln\biggl(\frac{M^2-s}{M^2}\biggr)
\nn\\          
&& {}
- \frac{2(s+\de s)}{(\de s)^2} {\cal R}^{(2)}_{\de s\to0}\left[B_0(1)\right]
+ \frac{M^2-s-\de s}{\de s} {\cal R}^{(1)}_{\de s\to0}\left[C_0\right],
\nn\\
C_2 &=&
-\frac{M^2}{s^2} \ln\biggl(\frac{M^2-s}{M^2}\biggr)
-\frac{1}{s}
+\frac{1}{\de s}
{\cal R}^{(1)}_{\de s\to0}\left[B_0(1)\right],
\nn\\[.3em]
C_{00} &=& \frac{1}{4}\Delta
+\frac{1}{4}\ln\biggl(\frac{\mu^2}{M^2-s}\biggr)
+\frac{M^2(M^2-\de s)}{4s^2} \ln\biggl(\frac{M^2-s}{M^2}\biggr)
+\frac{M^2+2s-\de s}{4s} 
\nn\\          
&& {}
-\frac{M^2-s-\de s}{4\de s}
{\cal R}^{(1)}_{\de s\to0}\left[B_0(1)\right].
\eeqar
Here the orders $n$ in the ${\cal R}^{(n)}$ operators are chosen in
such a way that all terms involving ${\cal R}^{(n)}$ contribute only
in ${\cal O}(\de s)$ in spite of the enhancement factors $1/\de s^m$.
Note that no delicate cancellations for $\de s\to0$ appear in the
other terms, although the original Passarino--Veltman results contain
plenty of terms involving $1/\de s^m$ in front of linear combinations
of scalar integrals.  Thus, the above forms are numerically stable as
long as the remainder terms ${\cal R}^{(n)}$ can be evaluated in a
stable way. This task is, however, easily achieved upon expanding the
scalar integrals as in \refeq{eq:scal_collexpand1} to a high order,
e.g., with computer-algebraic methods, and dropping the first $n$
orders. The resulting series are easy to evaluate, and an arbitrarily
high precision can be achieved by including sufficiently high orders
in the expansions.  On the other hand, if $\de s$ is not small, the
${\cal R}^{(n)}$ terms can safely be 
evaluated upon numerically
subtracting the ${\cal T}^{(n)}$ terms from the scalar integrals.  In
this way an arbitrarily high precision can be achieved as long as
$s,s'\ne 0$ and $s\ne M^2$. The case $s=M^2$ does not occur in our
application, the cases $s=0$ and $s'=0$ are treated below.

\paragraph{Collinear-singular case with one off-shell leg:
$C_{\dots}(m^2,0,s',0,m,M)$}

Specializing the previous case to $s=0$, the scalar integrals read 
\beqar\label{eq:scalint2}
B_0(0) &=& B_0(0,m,M) = \De+\ln\biggl(\frac{\mu^2}{M^2}\biggr)+1,
\nn\\
C_0 &=&
\frac{1}{s'} \Biggl\{
\ln\biggl(\frac{M^2}{m^2}\biggr)
 \ln\biggl(\frac{M^2-s'}{M^2}\biggr)
-\Li\biggl(\frac{s'}{M^2}\biggr)
\Biggr\},
\eeqar
with $B_0(1)$ and $B_0(2)$ still as given in \refeq{eq:scalint1}.
The limit of vanishing Gram determinant is now reached for $s'\to0$,
where the scalar integrals can be expanded according to
\beqar\label{eq:scal_collexpand2}
B_0(1) &=& B_0(0) 
+\frac{s'}{2M^2}
+{\cal R}^{(1)}_{s'\to0}\left[B_0(1)\right],
\nn\\[.5em]
C_0 &=&
-\frac{1}{M^2}\biggl[ 
\biggl(1+\frac{s'}{2M^2}\biggr)
\ln\biggl(\frac{M^2}{m^2}\biggr) 
+1+\frac{s'}{4M^2}
\biggr]
+{\cal R}^{(1)}_{s'\to0}\left[C_0\right],
\eeqar
or to higher orders if needed.
Making use of these expansions, the first few tensor coefficients can 
be written as
\beqar
C_1 &=&
\frac{M^2+s'}{2M^4} \ln\biggl(\frac{M^2}{m^2}\biggr) 
-\frac{M^2-s'}{4M^4}
- \frac{2}{s'} {\cal R}^{(1)}_{s'\to0}\left[B_0(1)\right]
+ \frac{M^2-s'}{s'} {\cal R}^{(1)}_{s'\to0}\left[C_0\right],
\nn\\
C_2 &=&
\frac{1}{2M^2}
+\frac{1}{s'}
{\cal R}^{(1)}_{s'\to0}\left[B_0(1)\right],
\nn\\[.3em]
C_{00} &=& \frac{1}{4}\Delta
+\frac{1}{4}\ln\biggl(\frac{\mu^2}{M^2}\biggr)
+\frac{3M^2+s'}{8M^2} 
-\frac{M^2-s'}{4s'}
{\cal R}^{(1)}_{s'\to0}\left[B_0(1)\right].
\eeqar
The ${\cal R}^{(n)}$ terms, which are suppressed by a factor $(s')^{n+1}$,
can be evaluated to arbitrary precision for all values of $s'$ as described
above.

\paragraph{Collinear-singular case with one off-shell leg:
$C_{\dots}(m^2,s,0,0,m,M)$}

Specializing case \ref{par:case1}
to $s'=0$, the scalar integrals read 
\beqar\label{eq:scalint3}
B_0(1) &=& B_0(0,0,M) = \De+\ln\biggl(\frac{\mu^2}{M^2}\biggr)+1,
\nn\\
C_0 &=&
\frac{1}{s} \Biggl\{
 \ln\biggl(\frac{M^2-s}{m^2}\biggr)
 \ln\biggl(\frac{M^2-s}{M^2}\biggr)
-\Li\biggl(\frac{s}{M^2}\biggr)
\Biggr\},
\eeqar
with $B_0(0)$ and $B_0(2)$ still as given in \refeq{eq:scalint1}.
The limit of vanishing Gram determinant is reached for $s\to0$,
where the scalar integrals can be expanded according to
\beqar\label{eq:scal_collexpand3}
B_0(0) &=& B_0(1) 
+\frac{s}{2M^2}
+{\cal R}^{(1)}_{s\to0}\left[B_0(0)\right],
\nn\\[.5em]
C_0 &=&
-\frac{1}{M^2}\biggl[ 
\biggl(1+\frac{s}{2M^2}\biggr)
\ln\biggl(\frac{M^2}{m^2}\biggr) 
+1-\frac{3s}{4M^2}
\biggr]
+{\cal R}^{(1)}_{s\to0}\left[C_0\right],
\eeqar
or to higher orders if needed.  Making use of these expansions, the
first few tensor coefficients can be written as
\beqar
C_1 &=&
\frac{1}{2M^2} \ln\biggl(\frac{M^2}{m^2}\biggr) 
-\frac{1}{4M^2}
+ \frac{1}{s} {\cal R}^{(1)}_{s\to0}\left[B_0(0)\right]
- \frac{M^2}{s} {\cal R}^{(1)}_{s\to0}\left[C_0\right],
\nn\\
C_2 &=&
\frac{1}{2M^2}
+\frac{1}{s}
{\cal R}^{(1)}_{s\to0}\left[B_0(0)\right],
\nn\\[.3em]
C_{00} &=& \frac{1}{4}\Delta
+\frac{1}{4}\ln\biggl(\frac{\mu^2}{M^2}\biggr)
+\frac{3M^2+s}{8M^2} 
-\frac{M^2-s}{4s}
{\cal R}^{(1)}_{s\to0}\left[B_0(0)\right].
\eeqar
The ${\cal R}^{(n)}$ terms, which are suppressed by a factor $s^{n+1}$,
can be evaluated to arbitrary precision for all values of $s$ as described
above.

\paragraph{Soft-singular case:
$C_{\dots}(m_1^2,s,m_2^2,\lambda,m_1,m_2)$}

For processes with external fermions in the massless limit
($m_i\to0$), the Passarino--Veltman reduction of this case turns out
to be less delicate than the previous ones.  In fact, no special
treatment was necessary for $\Pep\Pem\to4f$ \cite{Denner:2005es},
although one could also improve the stability as described in the
previous sections.  We attribute the robustness of 
this case to the
following reasons.  Firstly, because $\la$ is an infinitesimal photon
mass $f_1=f_2=0$ and all 3-point tensor coefficients are directly
obtained from 2-point coefficients without further recursions. Thus,
instabilities do not accumulate. Secondly, for massless fermions the
Gram determinant $\De^{(2)}=-s^2$ vanishes only for $s\to0$, and this
case appears for $\Pep\Pem\to4f$ only in regions of phase space that
are suppressed by $\GW/\MW$.

\section{Alternative reduction of 5-point integrals}
\label{se:5pf-alt}

In \citere{Denner:2002ii} we have worked out a reduction of 5-point
tensor integrals that follows the strategy proposed
in \citere{Me65} for scalar integrals in four space--time dimensions.
Here we briefly describe the derivation of this method in $D$
dimensions, to make closer contact to the methods used in this paper.

The reduction is based on different ways of evaluating the determinant
{\arraycolsep 4pt
\beqar\label{oldEredstart}
{\cal E}' &=&
\left\vert
\barr{cccc}
2q^2   & 2qp_{1}     & \ldots & 2qp_{4} \\
2p_1q & 2p_{1}p_{1} & \ldots & 2p_{1}p_{4} \\
\vdots& \vdots      & \ddots & \vdots     \\
2p_4q & 2p_{4}p_{1} & \ldots & 2p_{4}p_4
\earr\right\vert,
\eeqar}%
which vanishes in four dimensions owing to the linear dependence of
any five momenta. In $D$ dimensions the integral over ${\cal E}'$
can be easily evaluated to
{\arraycolsep 4pt
\beqar\label{oldEredtensor}
\frac{(2\pi\mu)^{4-D}}{\ri\pi^{2}}\int \rd^{D}q\,
\frac{q^{\mu_{1}}\cdots q^{\mu_{P}}}
{N_0N_1\ldots N_4} \, {\cal E}'
&=&
E^{\al\be\mu_1\ldots\mu_P}\left\vert
\barr{cccc}
2 g_{\al\be} & 2p_{1,\al}  & \ldots & 2p_{4,\al}  \\
2p_{1,\be}   & 2p_{1}p_{1} & \ldots & 2p_{1}p_{4} \\
\vdots       & \vdots      & \ddots & \vdots      \\
2p_{4,\be}   & 2p_{4}p_{1} & \ldots & 2p_{4}p_4
\earr\right\vert
\nn\\
&=&
2E^{\al\be\mu_1\ldots\mu_P} \De^{(4)} \left(g_{\al\be}-g_{(4),\al\be}\right), 
\eeqar}%
where we have identified the form \refeq{4Dmetric} of the metric tensor
$g_{(4),\al\be}$ in four dimensions.

On the other hand, the integral over ${\cal E}'$ can be evaluated in
terms of 4-point functions as described in Section~2 of
\citere{Denner:2002ii} with the only difference that no additional UV
regularization is needed, because we now keep the dimension $D$
general.  In detail, this means that the factor
$-\La^2/(q^2-\Lambda^2)$ introduced in (2.5) of \citere{Denner:2002ii}
is absent, and the result analogous to (2.19) of
\citere{Denner:2002ii} becomes
\beqar\label{oldEfinal}
\det(Y) E^{\mu_1\ldots\mu_P} &=& - \sum_{n=0}^4
 \,\det(Y_n) \,D^{\mu_1\ldots\mu_P}(n)
+\sum_{n,m=1}^4
\Zadj^{(4)}_{nm}\,2p_{m,\al}
\D^{\al\mu_1\ldots\mu_P}(n)
\nn\\
&& {}
+2E^{\al\be\mu_1\ldots\mu_P} \De^{(4)} \left(g_{\al\be}-g_{(4),\al\be}\right), 
\eeqar
where $Y=(Y_{ij})$, $i,j=0,\ldots4$, was defined in \refeq{eq:Y}, and
$Y_n$ is obtained from the 5-dimensional modified Cayley matrix $Y$ by
replacing all entries in the $n$th column by~1.  The last term of
\refeq{oldEfinal}, which results from \refeq{oldEredtensor},
contributes only if $E^{\al\be\mu_1\ldots\mu_P}$ involves a divergent
coefficient $E_{00\dots}$ corresponding to a covariant containing a
metric tensor.  As explained in \refse{se:divs}, such coefficients are
free of IR divergences, and power counting shows that UV divergences
only occur for $P\ge4$. Therefore, the last term in \refeq{oldEfinal}
is of ${\cal O}(D-4)$, and thus irrelevant, for $P\le3$.  For $P=4$,
this term can be explicitly evaluated using \refeq{eq:E000000div}
yielding
\beqar
\label{eq:Edecompold}
\det(Y) E^{\mu_1\ldots\mu_4} &=& - \sum_{n=0}^4
 \,\det(Y_n) \,D^{\mu_1\ldots\mu_4}(n)
+\sum_{n,m=1}^4
\Zadj^{(4)}_{nm}\,2p_{m,\al}
\D^{\al\mu_1\ldots\mu_4}(n)
\nn\\*
&& {}
-\frac{1}{48}\De^{(4)}\{gg\}^{\mu_1\ldots\mu_4}
-\frac{1}{24(D-4)}\De^{(4)}\{(g-g_{(4)})g\}^{\mu_1\ldots\mu_4},
\nln
\eeqar
where $\{(g-g_{(4)})g\}^{\mu_1\ldots\mu_4}$ is a symmetric tensor of
rank~4 constructed according to the rules explained in 
\refse{se:conventions},
\beqar
\{(g-g_{(4)})g\}^{\mu_1\ldots\mu_4} &=& 
(g-g_{(4)})^{\mu_1\mu_2}g^{\mu_3\mu_4}
+(g-g_{(4)})^{\mu_1\mu_3}g^{\mu_2\mu_4}
+(g-g_{(4)})^{\mu_1\mu_4}g^{\mu_2\mu_3}
\nn\\ && {}
+(g-g_{(4)})^{\mu_2\mu_3}g^{\mu_1\mu_4}
+(g-g_{(4)})^{\mu_2\mu_4}g^{\mu_1\mu_3}
+(g-g_{(4)})^{\mu_3\mu_4}g^{\mu_1\mu_2}.
\nn\\
\eeqar
The first term in the last line of \refeq{eq:Edecompold} is just the
finite contribution $U^{\mu_1\ldots\mu_4}$ defined in (2.15) of
\citere{Denner:2002ii}, and the UV-divergent terms of the second term
in the last line exactly cancel the UV divergences of the 4-point
integrals in the first line. Thus, the result for
$E^{\mu_1\ldots\mu_4}$ exactly receives the form of (2.19) of
\citere{Denner:2002ii},
\beqar
\det(Y) E^{\mu_1\ldots\mu_4} &=& - \sum_{n=0}^4
 \,\det(Y_n) \,D^{(\mathrm{fin})\mu_1\ldots\mu_4}(n)
+\sum_{n,m=1}^4
\Zadj^{(4)}_{nm}\,2p_{m,\al}
\D^{(\mathrm{fin})\al\mu_1\ldots\mu_4}(n)
\nn\\*
&& {}
-\frac{1}{48}\De^{(4)}\{gg\}^{\mu_1\ldots\mu_4},
\eeqar
where the superscript ``(fin)'' indicates that the UV parts have to be 
consistently omitted, as e.g.\ following the $\overline{\mathrm{MS}}$ 
prescription.

\section{Alternative reduction of 6-point integrals}
\label{se:6pf-alt}

Here we describe the reduction of 6-point tensor integrals of rank~$P$
(including the scalar case $P=0$) to six 5-point tensor integrals of
equal rank that is based on the strategy of \citere{Denner:1993kt}.
This reduction is related to the reduction of 5-point functions as
given in \citere{Denner:2002ii} and \refapp{se:5pf-alt} and has been
used in the calculation of the electroweak corrections to
$\Pep\Pep\to4f$ \cite{Denner:2005es}. Moreover, it is needed to reduce
the scalar 6-point function to 5-point functions \cite{Me65}.

It starts from the observation that 
\beq \label{eq:Fred2}
\frac{(2\pi\mu)^{4-D}}{\ri\pi^{2}}\int\!\rd^Dq\,
\frac{q^{\mu_1}\ldots q^{\mu_P}}{N_{0}N_1\cdots N_5}
\left\vert
\barr{cccc}
N_{0}+Y_{00}  & 2qp_{1}    & \ldots & \;2qp_{5} \\
Y_{10}-Y_{00} & 2p_{1}p_1 & \ldots & \;2p_{1}p_{5} \\
\vdots    & \vdots     & \ddots     &\;\vdots     \\
Y_{50}-Y_{00} & 2p_{5}p_{1} & \ldots &\; 2p_{5}p_5
\earr\right\vert
= 0,
\eeq
which is correct in any space--time dimension $D$ as long as the five
four-momenta $p_i$ $(i=1,\dots,5)$ are linearly dependent, and thus
for four-dimensional $p_i$, because then the five last columns of the
determinant are linearly dependent for an arbitrary $D$-dimensional
momentum $q$.  The l.h.s.\ of this relation is practically the same as
in Eq.~(2.10) of \citere{Denner:2002ii}, where the reduction of
5-point integrals is described. The same manipulations as described
there lead to the result
\beqar \label{eq:Fred}
\left\vert \barr{cccccc}
F^{\mu_1\ldots\mu_P}
&\:-E^{\mu_1\ldots\mu_P}(0)&\:-E^{\mu_1\ldots\mu_P}(1)&
\cdots&
\:-E^{\mu_1\ldots\mu_P}(5)\\
  1   &  Y_{00}   &  Y_{01}   &  \cdots   &  Y_{05}   \\
  1   &  Y_{10}   &  Y_{11}   &  \cdots   &  Y_{15}   \\
\vdots&  \vdots   &  \vdots   &  \ddots   &  \vdots   \\
  1   &  Y_{50}   &  Y_{51}   &  \cdots   &  Y_{55}
\earr \right\vert
&=&0.
\eeqar
Equation~\refeq{eq:Fred} expresses
$F^{\mu_1\ldots\mu_P}$ in terms of six 5-point integrals,
\beqar\label{eq:Fredf}
 F^{\mu_1\ldots\mu_P} &=& - \sum_{n=0}^5
 \,\eta_n \,E^{\mu_1\ldots\mu_P}(n)
\qquad \mbox{with} \quad \eta_n = \frac{\det(Y_n)}{\det(Y)},
\eeqar
where $Y=(Y_{ij})$, $i,j=0,\ldots5$, and $Y_n$ is obtained from the
6-dimensional modified Cayley matrix $Y$ by replacing all entries in
the $n$th column by~1.  For the scalar integral $F_0$, this result is
identical with the one of \citere{Me65}.

By inserting the Lorentz decompositions as given in
\refeq{LorentzdecF}, we can derive explicit formulas for the scalar
6-point function and the coefficients of tensor 6-point integrals from
\refeq{eq:Fredf}:
\beqar
F_{0} &=& - \sum_{n=0}^5\detY{n}E_{0}(n),
\\
F_{\ina}&=& - \sum_{\sinN=1}^5\detY{\sinN} E_{(\ina)_\sinN}(\sinN)\debar_{\ina\sinN}
-\detY{0}\Ecomb_{\ina}(0),
\quad \ina=1,\dots,5,
\\
F_{00}&=& - \sum_{\sinN=0}^5\detY{\sinN} E_{00}(\sinN),
\nn\\
F_{\ina\inb}&=&
 - \sum_{\sinN=1}^5\detY{\sinN} E_{(\ina)_\sinN(\inb)_\sinN}(\sinN)\debar_{\ina\sinN}\debar_{\inb\sinN}
-\detY{0}\Ecomb_{\ina\inb}(0),\quad \ina,\inb=1,\dots,5,
\\
F_{00\ina}&=& - \sum_{\sinN=1}^5\detY{\sinN} E_{00(\ina)_\sinN}(\sinN)\debar_{\ina\sinN}
-\detY{0}\Ecomb_{00\ina}(0), \quad \ina=1,\dots,5,
\nn\\
F_{\ina\inb\inc}&=& 
- \sum_{\sinN=1}^5
 \detY{\sinN}
 E_{(\ina)_\sinN(\inb)_\sinN(\inc)_\sinN}(\sinN)\debar_{\ina\sinN}\debar_{\inb\sinN}\debar_{\inc\sinN }
-\detY{0}\Ecomb_{\ina\inb\inc}(0),
\quad 
\ina,\inb,\inc=1,\dots,5.
\hspace{2em}
\eeqar
The 5-point tensor coefficients that result from omitting $N_0$
in the 6-point integrals have been given in \refeq{auxE}.


\end{document}